\documentclass[twocolumn]{aa} 

\usepackage{graphicx}
\usepackage[varg]{txfonts}
\usepackage[percent]{overpic}
\usepackage{color}
\usepackage{multirow}
\usepackage{natbib,twoopt} 
\usepackage[colorlinks=true,urlcolor=blue,linkcolor=red]{hyperref} 
\begin{document}
   
   \title{Convective blueshifts in the solar atmosphere}
   \subtitle{{\rm I.} Absolute measurements with LARS of the spectral lines at 6302\,\AA}
   
   \author{J. L\"ohner-B\"ottcher\inst{\ref{inst_kis}} \and W. Schmidt\inst{\ref{inst_kis}} \and F. Stief\inst{\ref{inst_kis}} \and T. Steinmetz\inst{\ref{inst_mpq},\ref{inst_menlo}} \and R. Holzwarth\inst{\ref{inst_mpq},\ref{inst_menlo}}}
    \institute{
    	Kiepenheuer-Institut f\"ur Sonnenphysik, Sch\"oneckstr. 6, 79104 Freiburg, Germany\\ \email{jlb@leibniz-kis.de}\label{inst_kis} \and
	Max-Planck-Institut f\"ur Quantenoptik, Hans-Kopfermann-Strasse 1, 85748 Garching, Germany\label{inst_mpq} \and
	Menlo Systems GmbH, Am Klopferspitz 19, 82152 Martinsried, Germany\label{inst_menlo}}
   \date{Received 16 Oct 2017 / Accepted xx Nov 2017}

  \abstract 
  {The solar convection manifests as granulation and intergranulation at the solar surface. In the photosphere, convective motions induce differential Doppler shifts to spectral lines. The observed convective blueshift varies across the solar disk.}
  {We focus on the impact of solar convection on the atmosphere and aim to resolve its velocity stratification in the photosphere.} 
  {We performed high-resolution spectroscopic observations of the solar spectrum in the 6302\,\AA\ range with the Laser Absolute Reference Spectrograph (LARS) at the Vacuum Tower Telescope. A laser frequency comb enabled the calibration of the spectra to an absolute wavelength scale with an accuracy of ${\rm 1\,m\,s^{-1}}$. We systematically scanned the Quiet Sun from disk center to the limb at ten selected heliocentric positions. The analysis included 99 time sequences of up to 20\,min in length. By means of ephemeris and reference corrections, we translated wavelength shifts into absolute line-of-sight velocities. A bisector analysis on the line profiles yielded the shapes and convective shifts of seven photospheric lines.}
  {At disk center, the bisector profiles of the iron lines feature a pronounced C-shape with maximum convective blueshifts of up to ${\rm -450\,m\,s^{-1}}$ in the spectral line wings. Toward the solar limb, the bisectors change into a \textbackslash-shape {with} a {saturation in the line core at a redshift of ${\rm +100\,m\,s^{-1}}$. The} center-to-limb variation of the line core velocities shows a slight increase in blueshift when departing the disk center for larger heliocentric angles. {This increase in blueshift is more pronounced for the magnetically less active meridian than for the equator. Toward the solar limb, the blueshift decreases and can turn into a redshift}. In general, weaker lines exhibit stronger blueshifts.}
  {Best spectroscopic measurements enabled the accurate determination of absolute convective shifts in the solar photosphere. We convolved the results to lower spectral resolution to permit a comparison with observations from other instruments.}

  \maketitle
  \titlerunning{Convective blueshifts in the solar atmosphere. I} 
  \authorrunning{L\"ohner-B\"ottcher et al.}

\section{Introduction}\label{sec1_intro}
In the convection zone, hot plasma rises up to the solar surface by buoyancy forces and constitutes as the well-known granulation pattern. When reaching the surface and lower photosphere, the hot gas cells appear as bright granules with a typical size between 0.5\arcsec\ and 2\arcsec\ on the sky, or 1.4\,Mm in average at an absolute scale \citep{1977SoPh...54..319B,1986SoPh..107...11R}. At the surface thermal boundary layer, the upward-moving convective plasma in the granular center cools by radiation, its vertical motion decelerates and spreads horizontally towards the granular periphery. Thereafter, the flow takes an overturning motion and converges into the intergranular lanes where it sinks back to deeper layers to be heated up again \citep[e.g.,][]{1967sogr.book.....B}. 

At the surface, the intergranular lanes appear substantially darker than the granules caused by the temperature decrease of the plasma by several hundred to one thousand Kelvin \citep{1998ApJ...499..914S}. Reconstructed images arrive at a root-mean-square intensity contrast of around 13\% at 550\,nm \citep{1982ARA&A..20...61D}. While upflows within the tip of granules can reach speeds of ${\rm 2\,km\,s^{-1}}$ \citep{1979SoPh...61..251D}, intergranular downflow velocities are of the same order (or slightly faster) in the opposite direction. \citet{1981A&A....96...96M} found the velocity of the horizontal component to be of the same magnitude as the vertical component. At disk center, granular upflows in the line-of-sight cause blueshifts of spectral lines to shorter wavelengths, translated into negative Doppler velocities. Line-of-sight downflows in the intergranulum evoke redshifts, or positive Doppler velocities. 

The observation of an integrated Quiet Sun region at disk center yield an overall convective blueshift of photospheric spectral lines \citep{1978SoPh...58..243B}. 
{In total, the blueshifted bright granules} contribute more light to the average line profile than the redshifted dark lanes. The average blueshift of a few hundred ${\rm \,m\,s^{-1}}$ depends on the line depth, excitation potential, and wavelength region of the photospheric line \citep{1981A&A....96..345D}. Since lines with a higher excitation potential of the lower level are predominantly formed in hotter gas (inside granules), they exhibit a stronger blueshift. Furthermore, the difference in light contribution introduces a line asymmetry into the average profile \citep[e.g.,][]{1975A&A....43...45D}. Redshifts from intergranular downflows affect the outer line wings, while granular upflows induce blueshifts especially of the inner line wings and core. In addition, the velocity stratification of the convective overshooting into the atmosphere strongly contributes to the differential line shift. Owing to excess momentum, the convective motion reaches a few hundred km into the photosphere. With increasing height the upward motion of the plasma decreases by several hundred ${\rm \,m\,s^{-1}}$ \citep{1979SoPh...61..251D,2000A&A...355..381K}. Since the highest formation layers are sampled deep in the spectral line core, the convective blueshift decreases as a function of depth from the line wings to the minimum position. A conventional tool to analyze the line asymmetry is the determination of the bisector curve, defined as the median or connection of the midpoints of horizontal line segments at certain depth of the absorption profile. In combination, all effects (granulation, intergranulum and height-dependence) lead to a differential convex asymmetry of photospheric line profiles with bisectors often resembling a "C"-shape \citep{1956ZA.....40..157V,1957ZA.....41..141S}. Numerous spectroscopic studies \citep[e.g.,][]{1976MNRAS.177..687A,1981A&A....96..345D,1982A&A...109..233C} investigated the bisectors of solar (predominantly iron) lines measured at solar disk center. In a comprehensive study, \citet{1984SoPh...93..219B} compared the bisector shapes of 143 photospheric lines in the visible range and generated an atlas with a classification into three typical bisector profiles, also with regard to their evolution from disk center to the solar limb. 

More than a century ago, \citet{1907AN....173..273H} was the first to discover that solar spectral lines shift to longer wavelength when measured at the solar limb. The shifts amounted to Doppler velocities of several hundred ${\rm m\,s^{-1}}$. Later confirmed by \citet{1910ApJ....31...30A} and, after the postulating of the theory of general relativity and the uniform gravitational redshift of the Sun, the limb effect was ascribed to the solar convection \citep{1928ApJ....67..195S}.  \citet{Beckers1977} proved the convective origin of the effect by verifying that sunspot umbrae which lack granular convection omit the limb variation. From then on, the so called "center-to-limb variation of the convective blueshift" was a topic of frequent research \citep[e.g.,][]{1982SoPh...79....3B,1984SoPh...93..219B,1985SoPh...99...31B,1984SoPh...94...49A,1991A&A...245..279K}. It has been found that the variation toward the solar limb reveals more than a plain decrease of blueshifts. Firstly, moving from disk center toward a heliocentric position of around $\mu=\cos\theta=0.8$ (where $\theta$ is the line-of-sight angle to the solar surface vertical), many lines show an increasing blueshift before the decrease in blueshift toward the limb sets in. Secondly, at the solar limb, the convective shift often turns into a slight redshift (commonly called "supergravitational redshift"). \citet{1978SoPh...58..243B} suggested the horizontal motion of the granular gas and the effective line-of-sight to be the cause of both phenomena. Later on, \citet{1985SoPh...99...31B} confirmed and illustrated the impact of the overall convective motion on the observed center-to-limb variation and changing line shape. Further information on solar granulation and the convective blueshift can be found in the reviews by \citet{1982ARA&A..20...61D} and \citet{1984cup..book.....B}.

In this work, we will present a systematic investigation of the spectral lines around 6302\,\AA with regard to their convective blueshift and center-to-limb variation. The key advance compared to other comparable or former studies comes from the instrument being used for the measurements. The Laser Absolute Reference Spectrograph \citep[LARS,][]{Doerr2015,2017A&A...607A..12L} allows for spectroscopic observations of the Sun with an unprecedented spectral quality. The combination of a high-resolution echelle spectrograph ($\lambda/\Delta\lambda>700\,000$) with a laser frequency comb for the absolute wavelength calibration enables the measurement of spectral line profiles with a spectral accuracy of $1\,{\rm m\,s^{-1}}$ \citep{Steinmetz+etal2008}. In Section \ref{sec2_data}, we will introduce the LARS instrument in more detail and present the systematic observations. {Enabled by the multiplicity of calibration modes provided by a frequency comb, the spectral investigation presented in this work outperforms earlier studies by almost two orders of magnitude in terms of intrinsic instrumental accuracy.} 
In this context, we will point out our results in Section \ref{sec3_results} and compare them with former observational studies and a synthetic analysis by \citet{2011A&A...528A.113D}. Therein, we focus our attention on the center-to-limb variation of the spectral line shapes in the 6302\,\AA\ region. To increase the benefit of our reference measurements for future solar spectroscopy, we complete this study with a translation of our results to the conditions of other upcoming instruments.

\begin{figure}[htpb]
\includegraphics[trim=0.7cm 0cm 0.4cm 1.0cm,clip,width=\columnwidth]{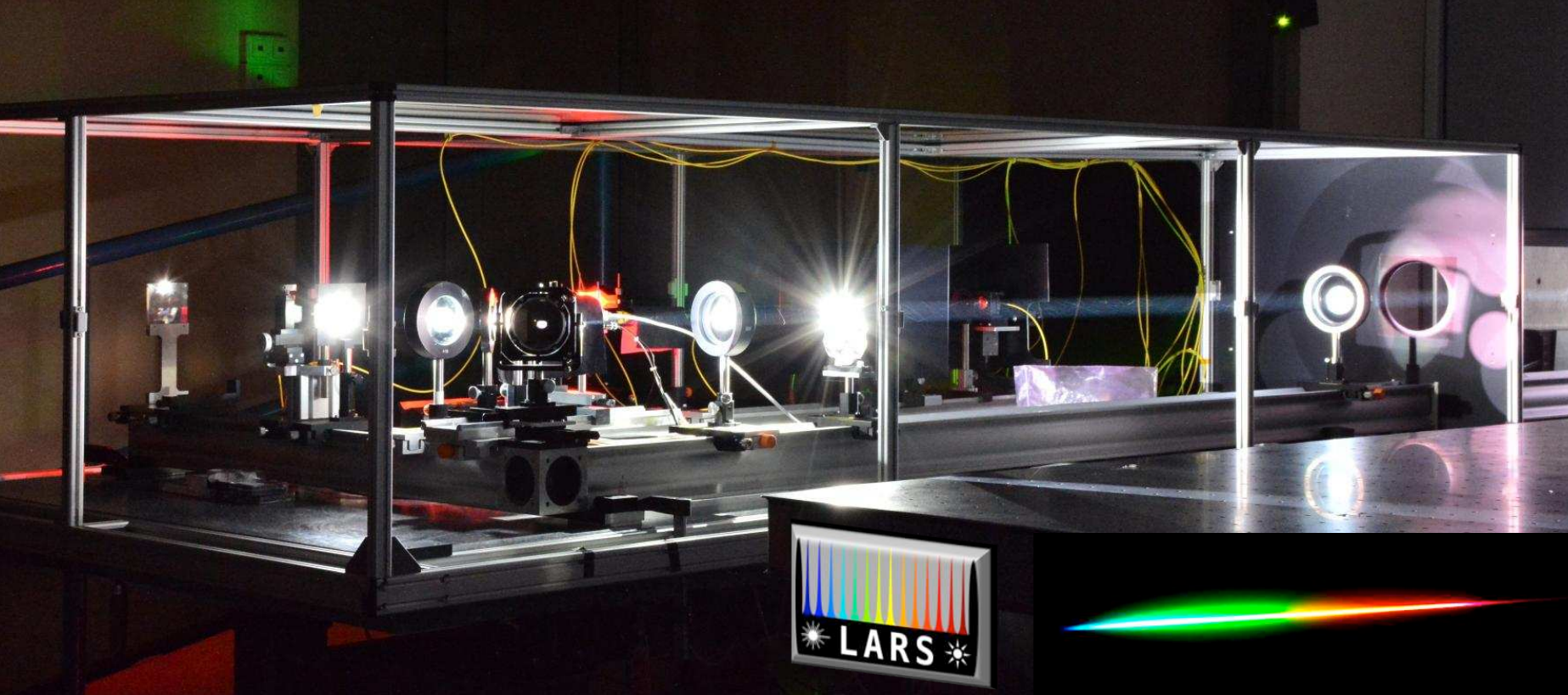}
\caption{Optical setup of LARS in 2016. The beam from the telescope enters the instrument on the upper right-hand side. A beamsplitter guides the sunlight to the Context Imager (10\%) and the fiber-coupling unit (90\%). A single-mode fiber (yellow cable) guides the integrated light from a circular 10\arcsec-wide solar region to the spectrograph. The logo of LARS and the first light spectrum of the laser frequency comb (upgraded in May 2016) are overplotted in the lower right-hand side.}
\label{fig_sec2_lars}
\end{figure}

\section{Observations}\label{sec2_data}
The extensive spectroscopic study of absolute convective shifts is based on a number of important preconditions. High spectral resolution is essential to obtain the required precision. Long-term measurements are needed to reduce the "solar noise" like $p$-modes superimposing the convective blueshift. Such comprehensive studies imply unrestricted repeatability of the measurements. A valid and careful wavelength calibration needs to be performed to attain an outstanding spectral accuracy of the measurement. The determination of the variation of the solar spectrum from disk center to the limb involves a reliable pointing of the telescope. An adequate statistical sampling has to be obtained to account for other systematical solar effects interfering with the convective motion. In this work, all of these requirements are met at an unprecedented scale. In this section, we will describe the scientific instrument, the systematic observations performed during two observational campaigns in 2016, and the absolute calibration of the solar spectra enabled by a laser frequency comb. 

\subsection{LARS instrument}\label{sec2_lars}
The observations were performed with the Laser Absolute Reference Spectrograph. LARS is a scientific instrument developed to perform solar observations with the high-resolution echelle spectrograph of the German Vacuum Tower Telescope (VTT, Observatorio del Teide). A laser frequency comb generates an emission spectrum of comb modes which works as an absolute wavelength ruler for the solar spectrum and enables the calibration of the spectrograph. In a pioneering work, \citet{Steinmetz+etal2008} demonstrated the calibration of the echelle spectrograph with a laser frequency comb. The prototype instrument \citep{Doerr+etal2012,Doerr+etal2012a,Doerr2015,Probst+etal2015,ediss18701} was then developed in a cooperation between the Kiepenheuer Institute for Solar Physics, Freiburg, the Max Planck Institute of Quantum Optics, Garching, and Menlo Systems GmbH, Martinsried. In 2016, a substantial upgrade of the laser frequency comb was performed which enabled the stable operation of LARS as a turn-key instrument. The current state of the instrument in its science-ready configuration was described by \citet{2017A&A...607A..12L}. For the sake of clarity, we briefly explain the setup of LARS and highlight its main subsystems.

The sunlight is collected with VTT's 70\,cm primary mirror with a focal length of 45\,m. By means of a 45$^\circ$ fold mirror in front of the spectrograph entrance and two collimating lenses, the beam focus is transferred to the optical laboratory of LARS. The optical setup at the time of the observation campaigns is shown in Fig.\,\ref{fig_sec2_lars}. After passing the field-stop wheel and a fold mirror, a cubic beamsplitter divides the incoming beam. 10\% of the light is reflected toward the Context Imager. The camera records two-dimensional images ($100\arcsec\times75\arcsec$) of the targeted solar region. An interference filter limits the transmitted light to a narrow wavelength band. In this work, we used a G-band filter with a central wavelength of 430.8\,nm, a full width at half maximum of 0.39\,nm, and a transmission of 18\%. The G-band images provided us with reference information on, both, the convective and magnetic context of the solar region. Passing the beamsplitter, the other 90\% of the incoming light directs toward an optical fiber-coupling unit. The light from the central 10\arcsec\ (in diameter) wide region of the field-of-view is fed to a single-mode fiber. The integrated fiber-coupled light is guided to a fiber switch device and, later on, to the spectrograph for spectroscopic observations. Other artificial light sources like a tungsten lamp for spectral flatfielding or the laser frequency comb for absolute wavelength calibration are coupled in the same manner. One single output fiber guides the light of the selected input channel to the coupling unit at the spectrograph entrance. This guarantees an identical illumination of the spectrograph by all light sources and enables a valid calibration. The spectrograph consists of a pre-disperser and a high-dispersion echelle grating. We achieve a spectral sampling of ${\rm2.73\,m\AA\,pixel^{-1}}$ at a wavelength of ${\rm \lambda=6302\,\AA}$ which leads to a spectral range of 5.6\,\AA recorded with the 2048\,$\times$\,512 pixel camera. The spectral resolution R=$\lambda/\Delta\lambda$ exceeds 700\,000 at ${\rm \lambda=6302\,\AA}$, with $\Delta\lambda$ as the width of the instrumental point spread function. 

\begin{figure}[htpb]
\includegraphics[trim=0.4cm 0cm 0.35cm 0cm,clip,width=\columnwidth]{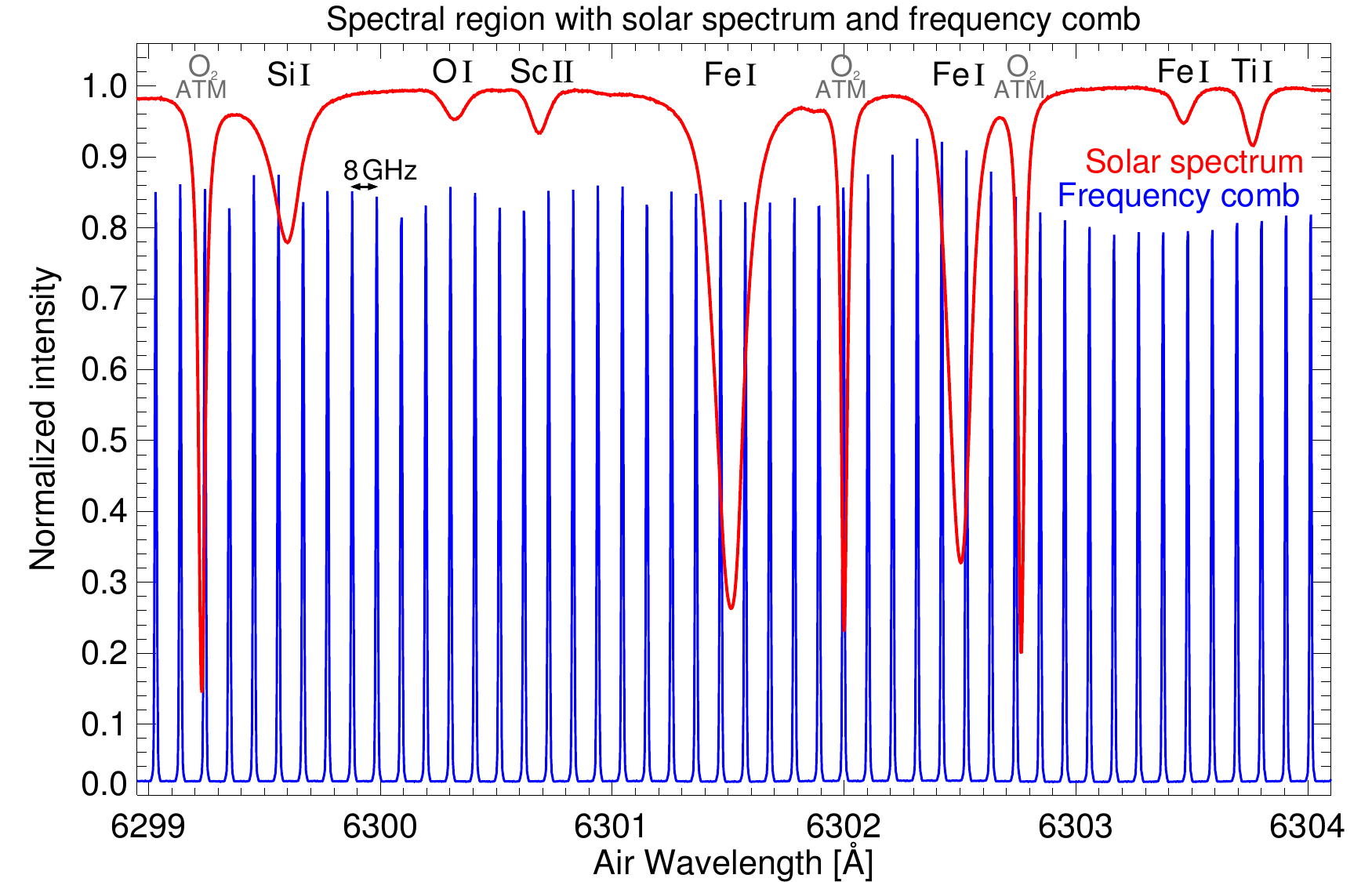}
\caption{Observation cycle of LARS consisting of a solar spectrum (red) and a frequency comb spectrum (blue). The intensities were normalized to the solar continuum level and displayed against the air wavelength in \AA. The solar lines are labeled in black, the telluric lines in gray. The solar spectrum was observed in the Quiet Sun at disk center. The emission modes of the comb spectrum have an equidistant spacing of 8.0\,GHz.}
\label{fig_sec2_solar_comb}
\end{figure}

The frequency comb serves as an absolute wavelength calibrator for the spectrograph. A pulsed femtosecond-laser operates at a wavelength of 1060\,nm. It generates a comb-shaped emission spectrum with an equidistant mode spacing of 250\,MHz. To suit the frequency comb spectrum to the spectrograph capabilities, a pair of Fabry-P\'erot cavities filters the spectrum to transmit only every 32nd mode. Thus, the final repetition frequency is 8.0\,GHz (or ${\rm 106\,m\AA}$\ at ${\rm 6302\,\AA}$) . Since the offset frequency is known, the unambiguous determination of one reference mode by its proximity to a well-known solar line is sufficient to obtain an absolute wavelength calibration. To operate LARS in the spectral range from 480\,nm to 700\,nm, the initial frequency comb signal is boosted by a high-power amplifier and spectrally broadened by a photonic crystal fiber. Fig.\,\ref{fig_sec2_solar_comb} displays the analyzed 5\,\AA\ wide part of the final comb spectrum (blue curve), overplotted to the solar spectrum at the 6302\,\AA\ range recorded in the Quiet Sun. 

\subsection{Systematic observations}\label{sec2_observation}
LARS is a unique instrument in solar science. It provides the opportunity to observe the spectrum of a limited spatial region of the Sun at an absolute wavelength scale. With its high spectral and temporal resolution it is ideally suited to perform most accurate investigations of the convective blueshift. To improve the significance of our study, we created the systematic observation procedure which we describe in the following.

\subsubsection{Spectral setup}
We centered the 5.6\,\AA\ wide wavelength range at 6301.5\,\AA\ (see Fig.\,\ref{fig_sec2_solar_comb}, red curve). The spectrum includes seven identifiable solar lines (\ion{Si}{I}\,6299.6\,\AA, \ion{O}{I}\,6300.3\,\AA, \ion{Sc}{II}\,6300.7\,\AA, \ion{Fe}{I}\,6301.5\,\AA, \ion{Fe}{I}\,6302.5\,\AA, \ion{Fe}{I}\,6303.5\,\AA, \ion{Ti}{I}\,6303.8\,\AA) and three telluric oxygen lines (at 6299.2\,\AA, 6302.0\,\AA, 6302.8\,\AA). The solar lines are formed in the photosphere, almost entirely in the lower layers, and provide comprehensive information about the convective processes there. On the spectrograph camera, the spectrum illuminates only two to three pixel rows over the full chip size of 2048\,pixel. Since the spectrum can slowly drift perpendicular to the dispersion axis, a band of 20\,pixels (centered on the illuminated pixels) was binned vertically. To decrease the detection noise to a negligible level, the spectrograph camera was cooled to a temperature of $-40^{\circ}$C. One observation cycle consisted of one frequency comb spectrum followed by one solar spectrum. An exemplary cycle is displayed in Fig.\,\ref{fig_sec2_solar_comb}. The fiber switch changes between the selected channels within milliseconds. The exposure time of the camera was set to 0.5\,s. In total, this resulted in a temporal cadence of 1.5\,s. In a sequence of cycles, the comb and solar spectrum were always observed in an alternating order. This enables the individual calibration of each solar spectrum with the previous and subsequent comb spectra.

\begin{figure}[htpb]
\includegraphics[width=\columnwidth]{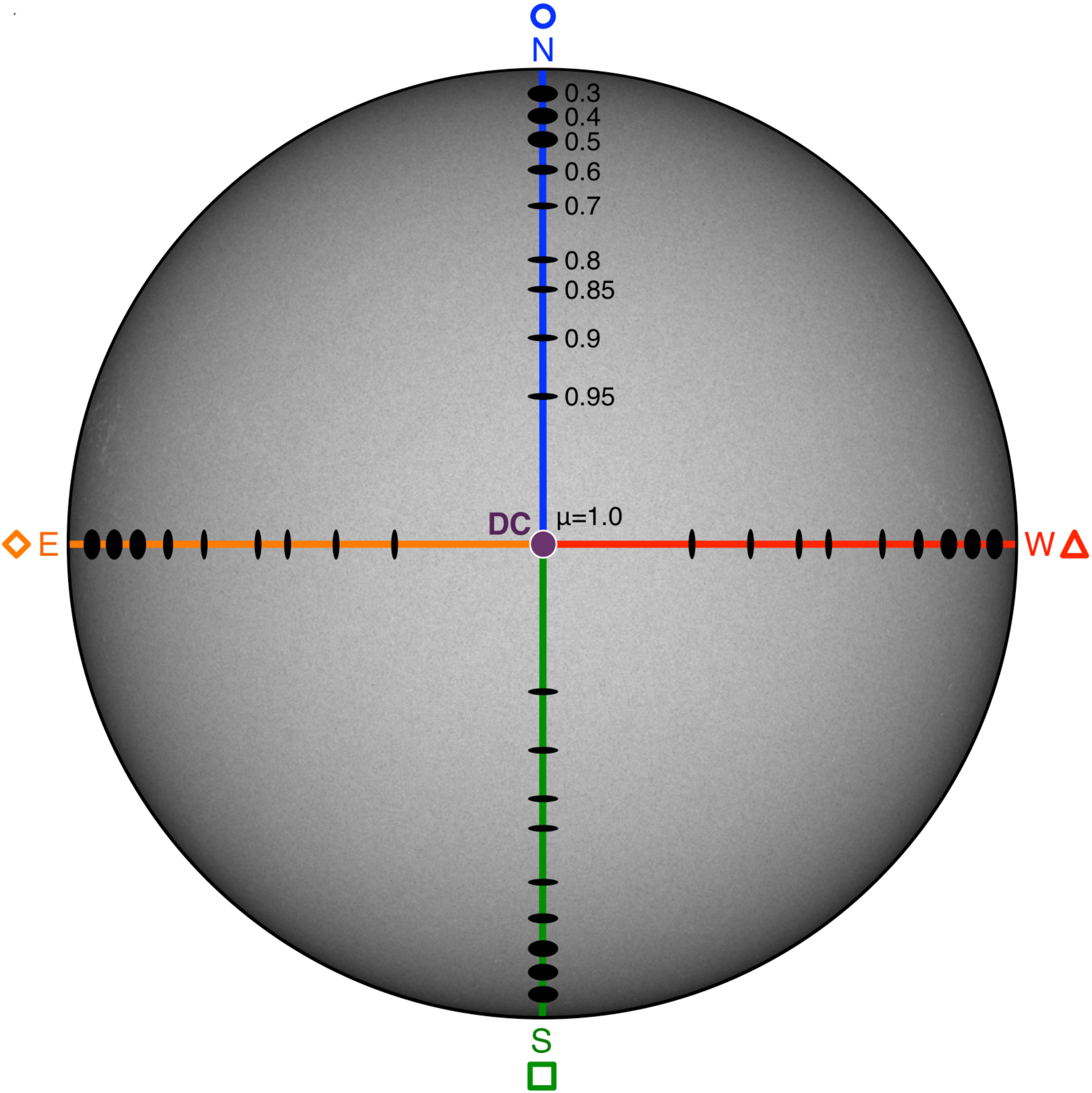}
\caption{Observation scheme. The measurements were performed at ten heliocentric positions ($\mu=\cos\theta$) from disk center (DC, $\mu=1.0$) toward the solar limb ($\mu=0.3$) along four radial directions: northern (blue, circle), southern (green, rectangle), eastern (orange, diamond), western (red, triangle) axis. The positions are drawn as black ellipses indicating the spatially covered area.}
\label{fig_sec2_observations_scheme}
\end{figure}

\subsubsection{Spatial scanning} Each observation is limited to a 10\arcsec\ wide region by the fiber-coupling unit. In a Quiet Sun region, the averaged area typically covers around 50 granules and its intergranulum. To investigate large-scale systematics of the convective blueshift, we had to develop a systematic observation scheme which allowed us to increase the spatial scanning, optimally exploit the allocated telescope operation time and increase the statistical significance of our measurements. To capture the center-to-limb variation of the convective shift, we observed the Sun at well-defined positions on the solar disk. According to the heliocentric angle $\theta$ which constitutes the angle between the line-of-sight and the solar surface vertical, we chose ten positions $\mu=\cos\theta$, from $\mu=1.0$ (at disk center) to $\mu=0.3$ (near the solar limb). The step size was set to $\Delta\mu=0.05$ from disk center to $\mu=0.8$, and $\Delta\mu=0.1$ toward the solar limb. The selection allows a direct comparison with recent theoretical investigations of the center-to-limb variation, for example with \citet{2011A&A...528A.113D} who synthesized line profiles at the stated heliocentric positions. Like indicated in the observation scheme shown in Fig.\,\ref{fig_sec2_observations_scheme}, we measured along the equator (eastern and western axes) and the meridian (northern and southern axes). Throughout this work, we maintain a convention of standardized colors and symbols for the respective axes, i.e., north (blue circles), south (green rectangles), east (orange diamonds), and west (red triangles). 

The telescope has a pointing accuracy of around 1\arcsec\ on the coordinate grid. At disk center, the resultant heliocentric error $\delta\mu$ is negligible. Toward the solar limb, it increases to $\delta\mu=\pm0.005$. We neglect the error in our statistical analysis. To guarantee the consistency of our study, only Quiet Sun regions were observed. Since magnetic fields lead to a modification of the convective energy transport and consequently to a reduced vertical velocity of granulation and less pronounced convective blueshift, we had to omit magnetic regions in our analysis. This criterion was verified by the G-band image of the LARS context camera and the most recent magnetogram of the Helioseismic Magnetic Imager (HMI). 

\begin{table}[htbp]
\caption{Summary of observation details. The total amount of observation sequences, cycles, and time is given for each heliocentric position $\mu$ on the solar disk. The size of the elliptical telescope movement is listed.}
\label{table_sec2_observations}
\centering
\begin{tabular}{ccccccccc}
\hline\hline
Pos.&\multicolumn{5}{c}{Sequences\,/\,axis}&Cycles&Time&Ellipse\\ 
$\mu$&N&S&E&W&total&total&(min)&a,b (\arcsec)\\ 
\hline
1.0&\multicolumn{4}{c}{--}&13&8562&214&10, 10\\
0.95&3&1&1&1&6&3980&100&15, 0\\
0.9&1&2&3&3&9&5495&137&15, 0\\
0.85&1&2&2&2&7&4785&120&15, 0\\
0.8&2&2&4&3&11&6784&170&15, 0\\
0.7&2&2&4&3&11&6683&167&15, 0\\
0.6&2&2&3&3&10&6047&151&15, 5\\
0.5&2&2&3&3&10&6181&155&15, 10\\
0.4&2&3&3&3&11&6877&172&15, 10\\
0.3&3&2&3&3&11&6594&165&15, 10\\
\hline
Total&18&18&26&24&99&61988&1550\\
\hline
\end{tabular}
\tablefoot{Each sequence consists of up to 800 cycles, but at least 400 cycles. The exposure time was set to ${\rm 0.5\,s}$ giving a cycle cadence of ${\rm 1.5\,s}$. The orientation of the elliptical oscillation of the telescope is sketched in Fig.\,\ref{fig_sec2_observations_scheme} for the different axes.}
\end{table}

A major hindrance to precise measurements of the convective blueshifts comes from the Sun itself. Large-scale convective motions, especially  supergranulation, and the superpositions of acoustic \textit{p}-modes provide a strong background of "solar noise". This contribution can only be reduced by taking long and repeated measurement sequences, as well as spatial averaging over an increased solar region. To cover a bigger spatial area, we performed a fast (5\,Hz) elliptical oscillation around the nominal heliocentric positions. The covered areas are illustrated in Fig.\,\ref{fig_sec2_observations_scheme} by the black ellipses. The amplitudes of the two-axis-oscillation of the telescope's scanner mirror were defined so that the main axis of the ellipses is parallel to the respective solar axis limb. Consequently, along the solar east-west axis the projected north-south oscillation dominates, and vice versa. In Table \ref{table_sec2_observations}, the amplitudes of the oscillation are listed for each heliocentric position. At disk center ($\mu=1.0$), we performed a circulation motion arising from an $90^{\circ}$ phase-shifted oscillation with an amplitude of 10\arcsec\ for both axes. The covered area has a diameter of 30\arcsec\ in total. At disk center, the larger averaged area limits the effect of $p$-mode oscillations with amplitudes of several $100\,{\rm m\,s^{-1}}$ which superimpose the convective shift in the data. At heliocentric positions from $\mu=0.95$ to $0.7$, the telescope oscillation is performed only along one axis (parallel to the solar limb) with an amplitude of 15\arcsec. Thus, the covered area has a size of around $40\arcsec\times10\arcsec$. Closer to the solar limb, the line-of-sight effect of $p$-modes decreases slowly while horizontal supergranular flows with up to $300-500\,{\rm m\,s^{-1}}$ \citep{1989A&A...221..137R} become the dominating error component. Supergranules have a typical diameter of around 40\arcsec\ and rotate very slowly at the solar limb. Consequently, measuring a time sequence of less than an hour with an integrated region of only 10\arcsec\ involves the danger of capturing only one side of the supergranular flow. At the solar limb, this can easily lead to a statistical scatter for the convective shift of several hundred meters per second. At heliocentric positions from $\mu=0.5$ to $0.3$, the amplitude of the elliptical oscillations was set to 15\arcsec\ along the axis parallel to the solar limb and 10\arcsec\ perpendicular to it. The total elliptical area covers $40\arcsec\times30\arcsec$, approximately the size of a supergranule. Obviously, we gain a far better accuracy of the average convective shift in exchange for a larger  uncertainty of the observed heliocentric position $\mu$. 

\subsubsection{Sequential observation} Due to the presence of solar $p$-mode oscillations, specific analyses of the convective blueshift require time series which are long enough to eliminate the wave component by temporal averaging. Since $p$-modes are characterized by wave periods of predominantly 5\,min, we set the regular time series to 20\,min. By this even multiple, we capture around four 5-min modes within one sequence. Averaging the sequence should balance the oscillations to yield the mean convective shift. The temporal sampling was defined by 800 cycles per sequence with a cadence of 1.5\,s. We aimed to observe each heliocentric position along each axis at least twice. The numbers of performed sequences at each position are listed in Table \ref{table_sec2_observations}, for each axis and in total. At disk center, we observed 13 sequences, with 8562 cycles or 214\,min in total. The other positions with equal heliocentric value were observed six to eleven times, summing up to 100\,--\,172\,min. In total, 99 sequences were observed. The number of cycles added up to 61988 with a total observation time of 1550\,min, or almost 26\,h. Over the entire time, the frequency comb guaranteed identical conditions for direct spectral comparisons. We rate the performed systematic observations as a good statistical sample to yield conclusive results on the convective shift and its center-to-limb variation.

\subsection{Data calibration}\label{sec2_reduction}
For long or repeated sequences, proper wavelength calibration is mandatory. Otherwise, systematic effects would make the results unreliable or even wrong. These systematic effects include the time-dependent relative motions between the telescope and the observed area on the Sun (rotation of the Sun, rotation of the Earth, and all orbital motions), as well as the influence of the ambient temperature, air pressure and humidity on the measured wavelengths. 

With LARS, we were able to perform a careful instrumental calibration and to determine absolute wavelengths for each observation with an accuracy of around ${\rm 1\,m\,s^{-1}}$. The emission spectrum of the laser frequency comb served as a ruler for the absolute wavelength scale. The alternate measurement of the comb and solar spectrum (see Fig.\,\ref{fig_sec2_solar_comb}) allowed for an individual calibration of the solar spectrum at each instant in time, given by the temporal interpolation of the adjacent comb spectra. We used the data pipeline developed by \citet{Doerr2015} to process the spectroscopic LARS data. An overview of the current software and included calibration steps therein was given by \citet{2017A&A...607A..12L}. In the following, we briefly describe the steps which have been executed to obtain the final comb-calibrated data.

Using a tungsten lamp, we generated white-light spectra for spectral flatfield calibration. We used the flatfield spectrum closest in time (<\,30\,min) to calibrate the observation sequences. The data reduction included the elimination of background noise, intensity gradients and camera defects. In the next step, we determined the absolute wavelength scale given by the comb spectra. Each comb emission mode was fitted with a Gaussian profile to define its central pixel position on the camera. The pixel positions of all modes were fitted with a polynomial function. To translate the preliminary spectral calibration to an absolute wavelength grid, each pixel had to be assigned to its frequency. Since the comb modes have a constant frequency spacing of 8.0\,GHz and an offset frequency of $-100$\,MHz, it was sufficient to identify only one of the comb numbers to unambiguously obtain the full frequency spectrum. 
{Since the wavelengths of most lines of the solar spectrum are well known, we defined one of the observed spectral lines as the required unambiguous reference. In doing so, we performed a parabolic fit to the core of the \ion{Fe}{I}\,6302.5\,\AA\ line and ascribed the nominal air wavelength to its center position.}
We emphasize that throughout this work, we adopt common practice in astrophysics and use air wavelengths when referring to spectral lines or observed wavelengths. The difference to vacuum wavelength is about 1.7\,\AA\ in the 6302\,\AA\ range \citep{Edlen1953, Edlen1966}.
After the determination of the most proximate comb mode, this reference mode was then ascribed to its definite frequency. It has to be noted that each solar spectrum was calibrated with the interpolation between both temporally adjacent comb spectra. Since each solar spectrum was calibrated individually at a cadence of 1.5\,s, we accounted for eventual drifts of the spectrograph. A spectral accuracy of a few femtometer was reached, which corresponds to an uncertainty in Doppler velocities of the order of ${\rm 1\,m\,s^{-1}}$.

To focus the spectral analysis on the mere solar activity, we removed all inherent systematic effects from the observed data. We used the ephemerides code developed by \citet{Doerr2015}, which in turn is based on NASA's Navigation and Ancillary Information Facility Spacecraft Planet Instrument C-matrix Events (SPICE) toolkit \citep{Acton1996}. For each time step, the software computed all systematic relative motions between the telescope and the Sun with a precision of ${\rm  0.1\,m\,s^{-1}}$. It included the line-of-sight components of the orbital motions of the Earth and Sun, as well as the terrestrial rotation at the location of the telescope. The differential rotation of the Sun at the observed heliographic position was computed according to a model by \citet{1990ApJ...351..309S}. In addition, a constant gravitational redshift of ${\rm 633\,m\,s^{-1}}$, caused by the Sun and Earth according to the principle of equivalence and the General Theory of Relativity, was considered. Taken together, all of these effects can sum up to systematic Doppler shifts of the spectral lines of a few ${\rm km\,s^{-1}}$. After removal, the spectral lines contain only wavelength shifts caused by local motions and activity in the solar atmosphere.

To account for residual gradients of the spectral continuum over the full wavelength region, we compared the LARS spectra with the {profile from the} most widely used Fourier Transform Spectrometer \citep[FTS,][]{1999SoPh..184..421N} {disk center atlas. The FTS atlas was made up by seven separate FTS scans and provides a spectral resolution of the order of 350\,000, barely half the resolution of LARS. With a size of around $100\,\arcsec$, the resultant area of the observed region at disk center was larger by one magnitude than the region covered with LARS. Based on an adequate comparability of the data,} a polynomial correction function was applied to the {observed LARS spectra in order to rectify the continuum level with respect to the FTS spectrum.} 

For a qualitative global analysis of the convective blueshift (from Section \ref{sec3_bisectors} on), we averaged each calibrated observation sequence in time. Exemplarily, a calibrated solar spectrum is shown in Fig.\,\ref{fig_sec2_solar_comb} (red curve).

\section{Results and discussion}\label{sec3_results}
In preparation for the presentation and discussion of the convective blueshift results, we start this section with the evaluation of the temporally variant components which superimpose the convective shift measurements. In Section \ref{sec3_pmodes}, we analyze the inherent $p$-mode oscillations and their spatial evolution across the solar disk in order to validate the subsequent elimination of the component by temporal averaging. The discussion of the center-to-limb variation of the convective blueshift in Section \ref{sec3_clv} is divided into the analysis of the 
line profiles (Section \ref{sec3_clv_profiles}), their bisectors (Section \ref{sec3_bisectors}), and the derived Doppler velocities (Section \ref{sec3_clv_Doppler}). In this context, we compare the profiles and obtained convective shifts of the different spectral lines within the observed wavelength region. To ensure the applicability of our study to other observations, we performed an additional conversion of our results to the spectral resolution of other spectroscopic instruments (Section \ref{sec3_resolution_comparison}). 

Throughout this work, we determined all spectral line positions in units of absolute wavelength $\lambda$. For the conversion of wavelength shifts into Doppler velocities, an undisturbed wavelength reference $\lambda_0$ is required. As reference, we used the observed air wavelength provided by the National Institute of Standards and Technology Atomic Spectra Database \citep[NIST ASD,][]{NIST_ASD}. The statistical error of the reference wavelengths is about 0.2\,m\AA\ ($\pm10\,{\rm m\,s^{-1}}$) for the strong iron lines at 6301.5\,\AA\ and 6302.5\,\AA\ \citep{1994ApJS...94..221N} and at the order of 2\,m\AA\ for the weaker lines. All Doppler velocities $v_{\rm LOS}$ were calculated from the wavelength shift ($\Delta\lambda=\lambda-\lambda_0$) by
\begin{equation}
v_{\rm LOS}=\frac{\Delta\lambda}{\lambda_0}\cdot c
\end{equation}
 with $c$ being the speed of light. 
 
\begin{figure*}[htpb]
\includegraphics[trim=0cm 0cm 2.45cm 0cm,clip,height=5.25cm]{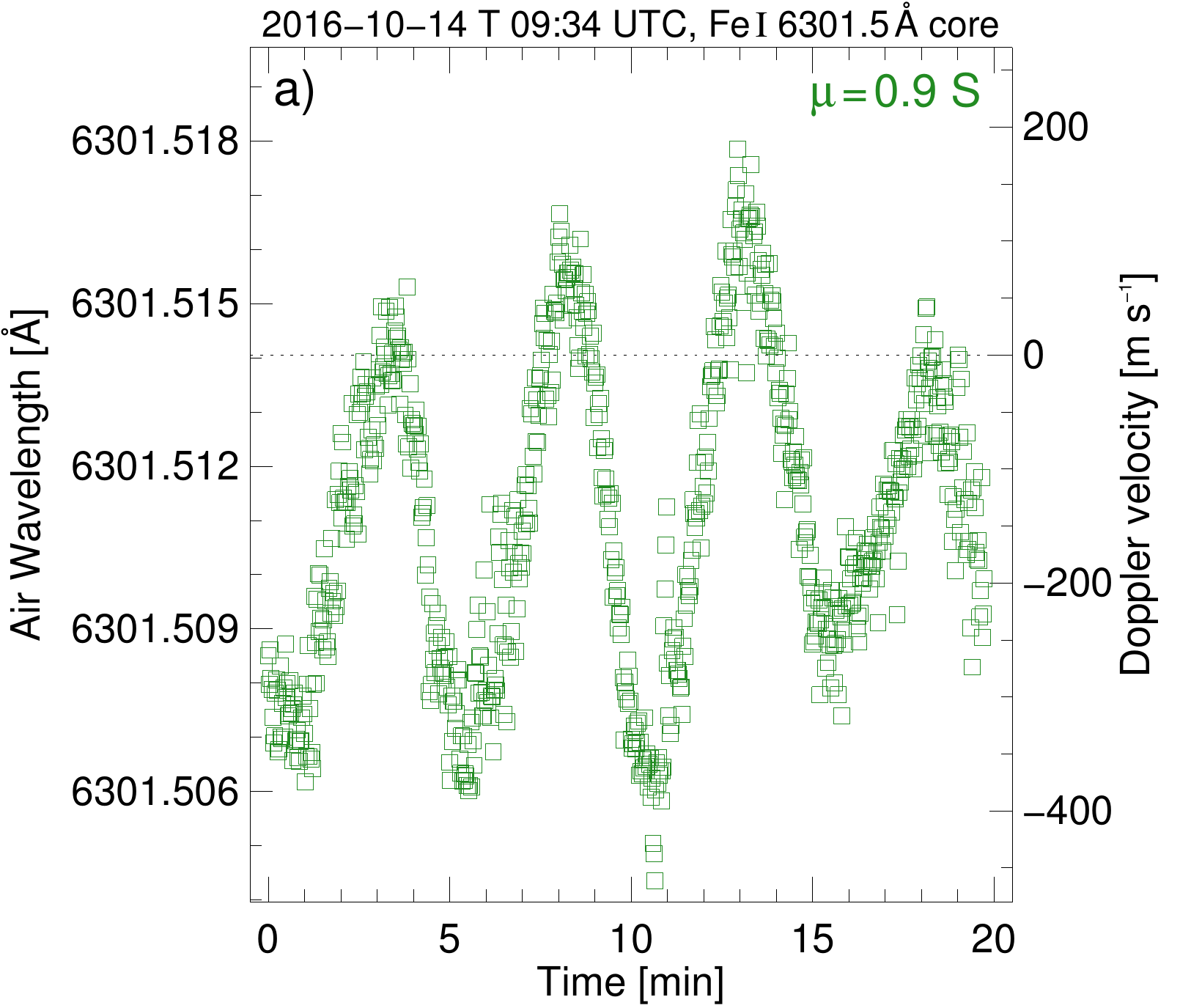}
\includegraphics[trim=3.45cm 0cm 2.45cm 0cm,clip,height=5.25cm]{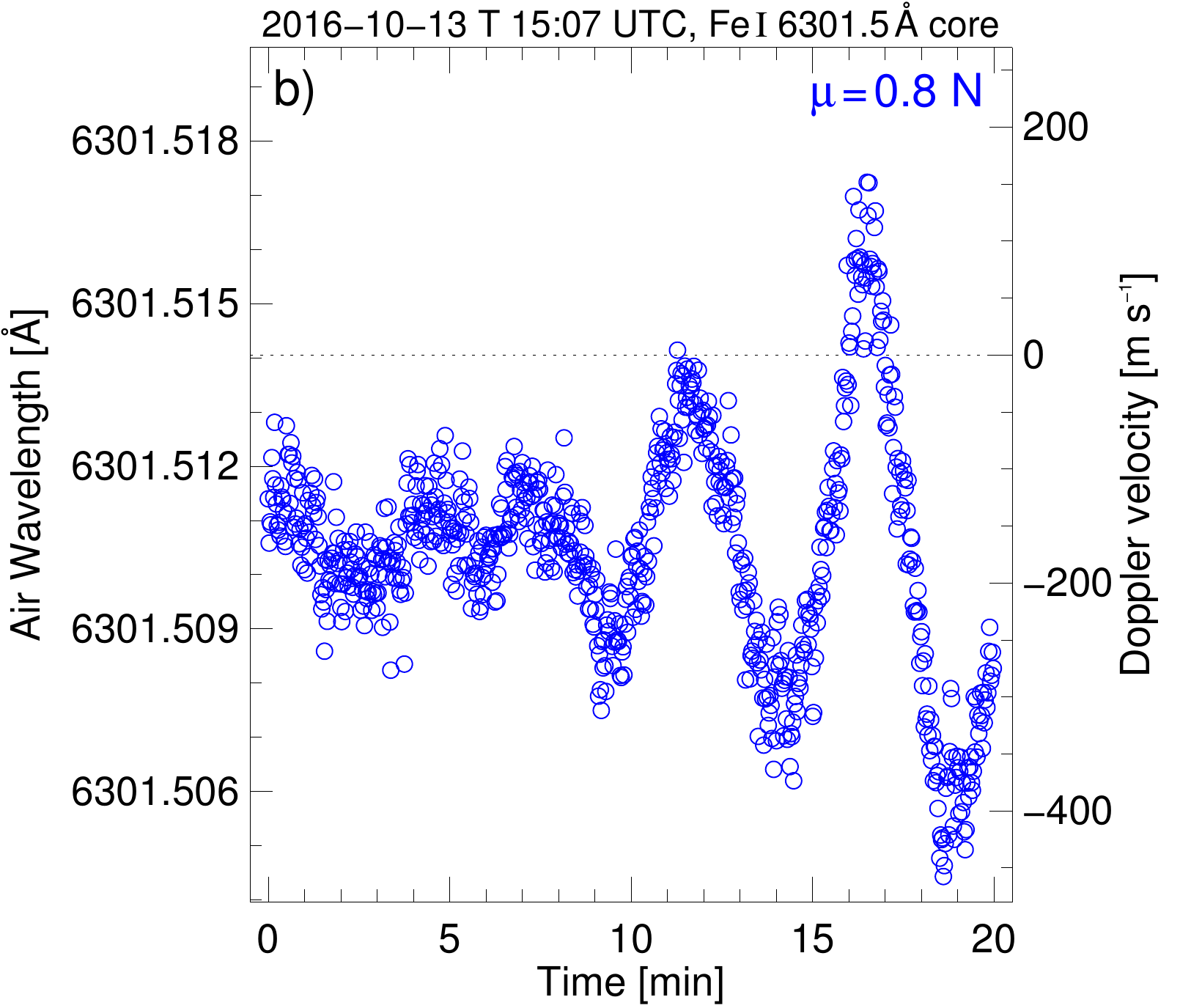}
\includegraphics[trim=3.45cm 0cm 2.45cm 0cm,clip,height=5.25cm]{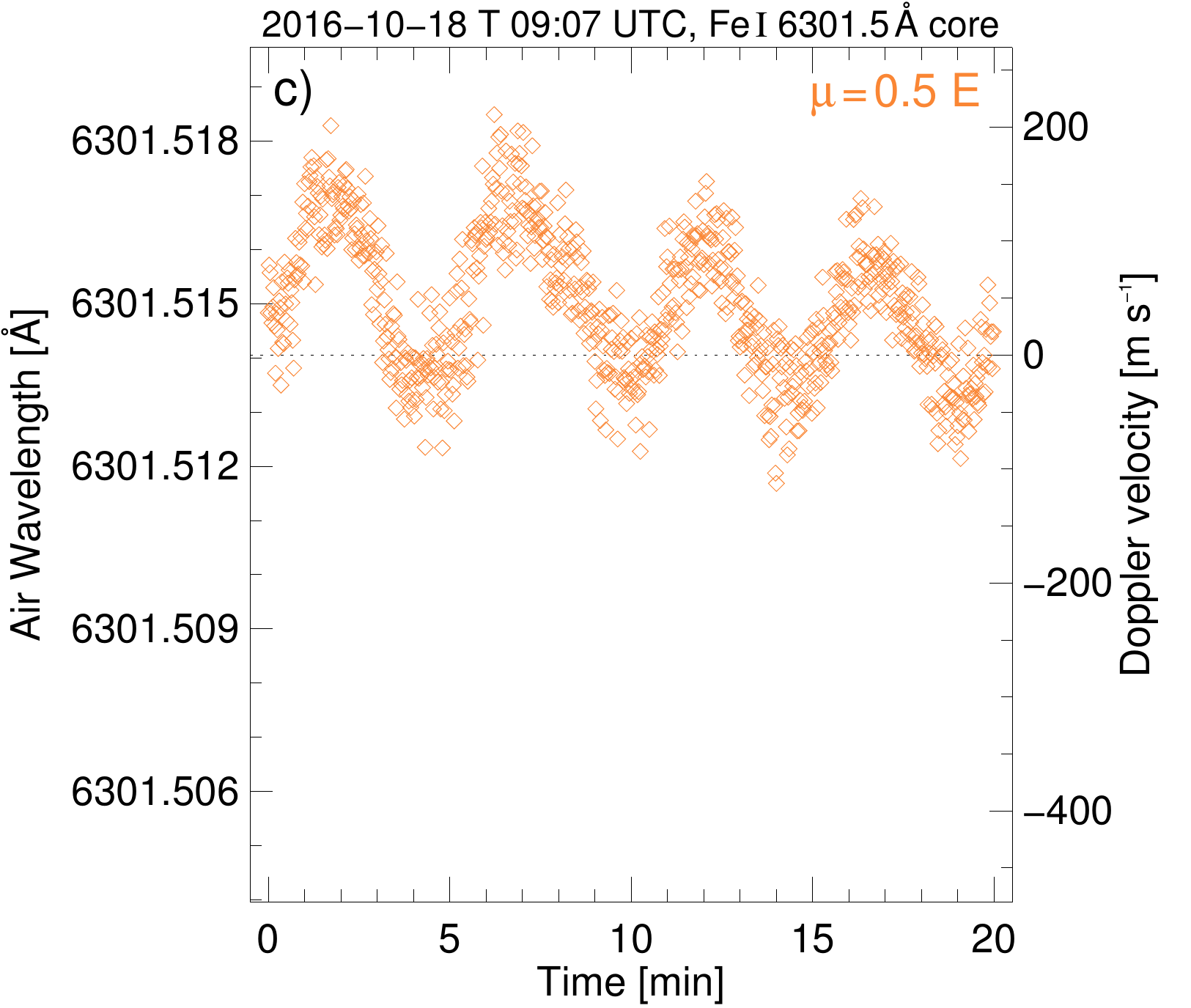}
\includegraphics[trim=3.45cm 0cm 0cm 0cm,clip,height=5.25cm]{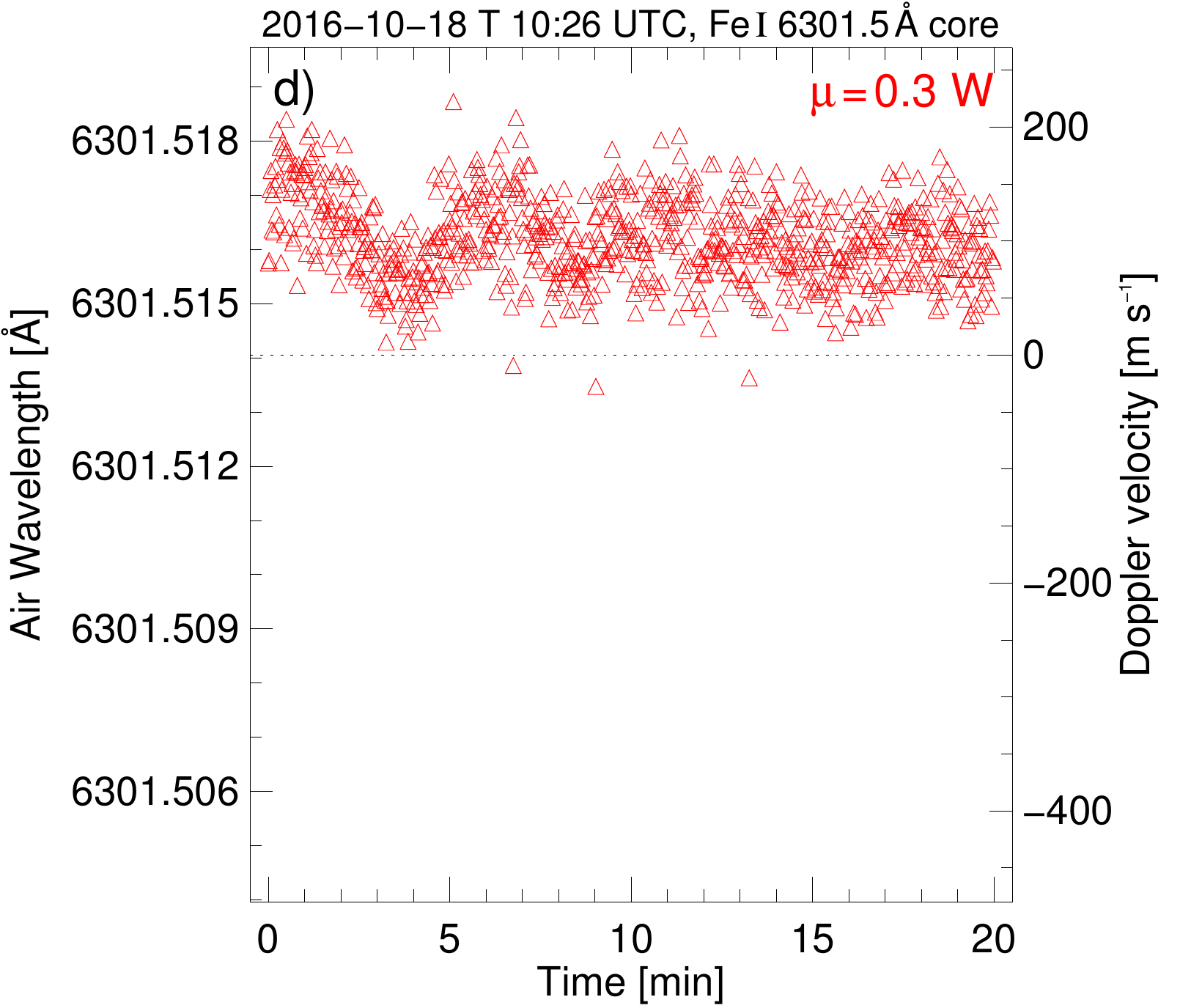}
\caption{Time series of the observed line minimum position of \ion{Fe}{I}\,6301.5\,\AA. The characteristic evolution is shown for four sequences with different heliocentric position on the solar disk: {a)} $\mu=0.9$ on the south axis (green rectangles), {b)} $\mu=0.8$ on the north axis (blue circles), {c)} $\mu=0.5$ on the east axis (orange diamonds), {d)} $\mu=0.3$ on the west axis (red triangles). The wavelength scale (left y-axes in \AA) is identical for all panels. The corresponding Doppler velocities (in $\rm{ m\,s^{-1}}$) can be identified by the right y-axes. The zero velocity is indicated by the dotted gray line. 
}
\label{fig_sec3_pmodes}
\end{figure*}
 
\subsection{Temporal and spatial evolution of $p$-modes}\label{sec3_pmodes}
All observations were carried out as time sequences of 20\,min with a cadence of 1.5\,s for the single measurement. Thus, the solar $p$-modes were clearly identifiable as oscillations in the sequence. We present some typical examples in Fig.\,\ref{fig_sec3_pmodes}. 

Close to the solar disk center, the oscillatory component is most pronounced. Fig.\,\ref{fig_sec3_pmodes}a shows the temporal evolution of the \ion{Fe}{I}\,6301.5\,\AA\ line core position observed at a heliocentric position $\mu=0.9$ on the southern axis. The 20\,min series displays four distinct 5-min oscillations with amplitudes of up to ${\rm 300\,m\,s^{-1}}$ in Doppler velocity, or 6\,m\AA\ in wavelength shift. The temporal average of the sequence amounts to an absolute wavelength of 6301.511\,\AA, or a Doppler velocity of ${\rm -145\,m\,s^{-1}}$. We interpret this average blueshift to have a purely convective origin.

The second example shown in Fig.\,\ref{fig_sec3_pmodes}b stresses that solar $p$-modes can affect each other causing positive or destructive interference. The observation was performed at a position $\mu=0.8$ on the northern axis, so still in the range of small heliocentric angles. In the second half of the time series, two distinct 5-min oscillations are recognizable. However, the first half suffers from destructive interference. The $p$-modes cancel each other and we see no clear oscillatory trend. We point out that this does not affect our measurement of the convective blueshift component. The mean blueshift of the entire sequence amounts to ${\rm -170\,m\,s^{-1}}$, which is close to the value of the fairly constant velocity during the destructive interference. 

Toward the solar limb, the scenery changes. Due to the increasing heliocentric angle, the amplitude of the $p$-mode oscillations get less pronounced \citep[in line with][]{1999A&A...346..633S}. In Fig.\,\ref{fig_sec3_pmodes}c, the temporal evolution at the solar disk position $\mu=0.5$ on the eastern axis still shows distinct 5-min oscillations, but the amplitude has decreased to below ${\rm 150\,m\,s^{-1}}$ in Doppler velocity, or 3\,m\AA\ in wavelength shift. Close to the limb, the amplitude further decreases. At $\mu=0.3$ (Fig.\,\ref{fig_sec3_pmodes}d), the oscillatory imprint has almost vanished. The mean Doppler velocity near the limb got redshifted to positive velocities of ${\rm 50\,m\,s^{-1}}$ and ${\rm 100\,m\,s^{-1}}$.

In summary, we consider the temporal average of each 20\,min sequence as a good estimator for the mere convective blueshift. We therefore limit our further investigations to the temporal averages. The statistical error will be represented by the standard deviation of each time series, making the $p$-mode oscillation to the main error component. With regard to accurate $p$-mode observations, we conclude that LARS provides best conditions for the measurement of absolute oscillatory velocities which are crucial for Helioseismology \citep[e.g.,][]{2011Sci...332..213C}.

\begin{figure}[htpb]
\includegraphics[width=\columnwidth]{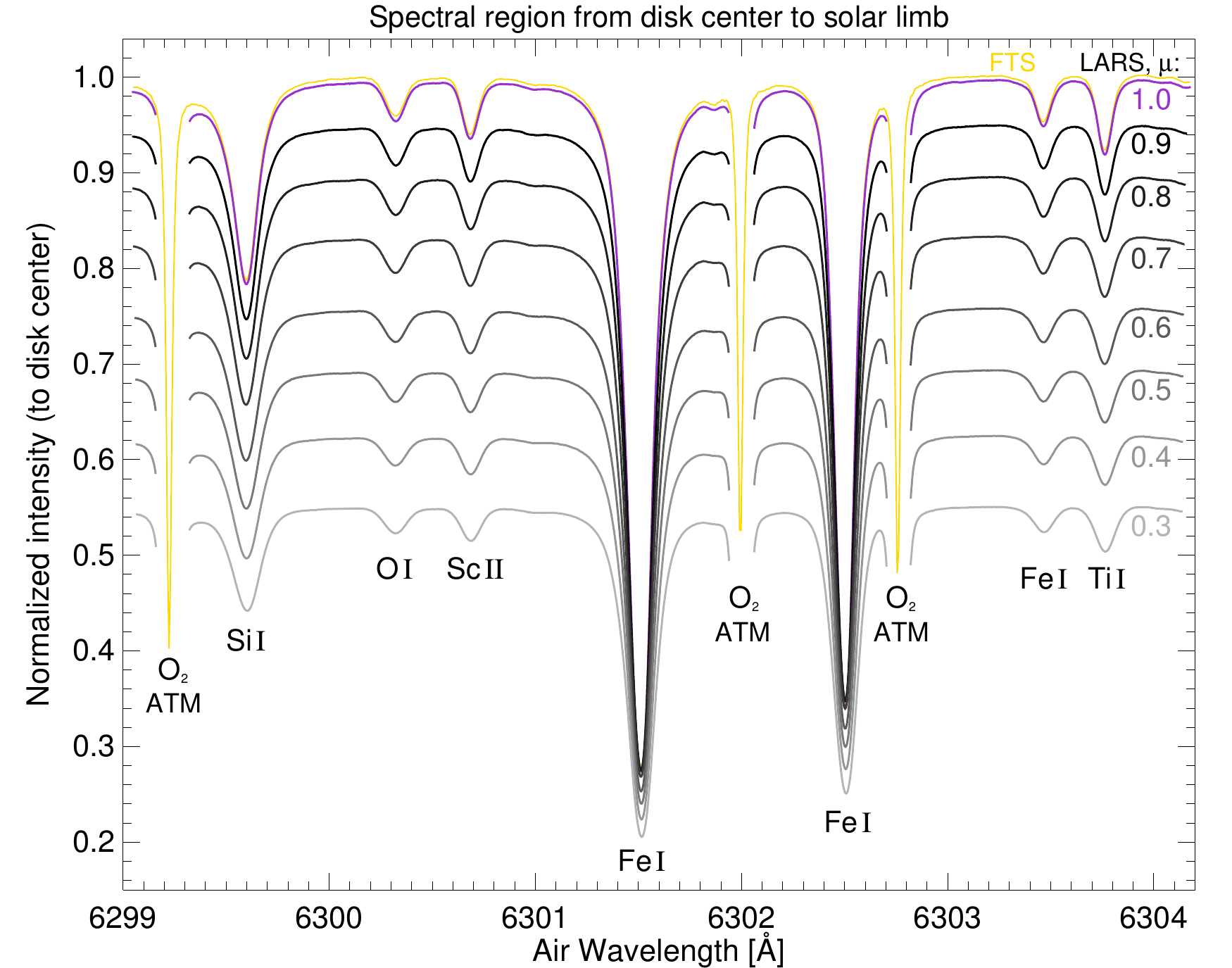}
\caption{Solar spectrum observed with LARS in the 6302\,\AA\ region from disk center ($\mu=1.0$, purple line) toward the solar limb ($\mu=0.3$, with decreasing gray scale). The spectra were normalized to the continuum intensity at disk center and are displayed against the air wavelength in \AA. The FTS atlas profile is added in yellow (slight y-shifted by $+0.5\%$ for better visualization).}
\label{fig_sec3_profileCLV}
\end{figure}

\subsection{Center-to-limb variation of the convective blueshift}\label{sec3_clv}
In the following subsections, we focus our analysis on the convective blueshift of the sequentially averaged spectral profiles in the 6302\,\AA\ region. Therein, we comparatively study the center-to-limb variation of the spectral lines shift and shape.

\begin{figure}[htpb]
\includegraphics[width=\columnwidth]{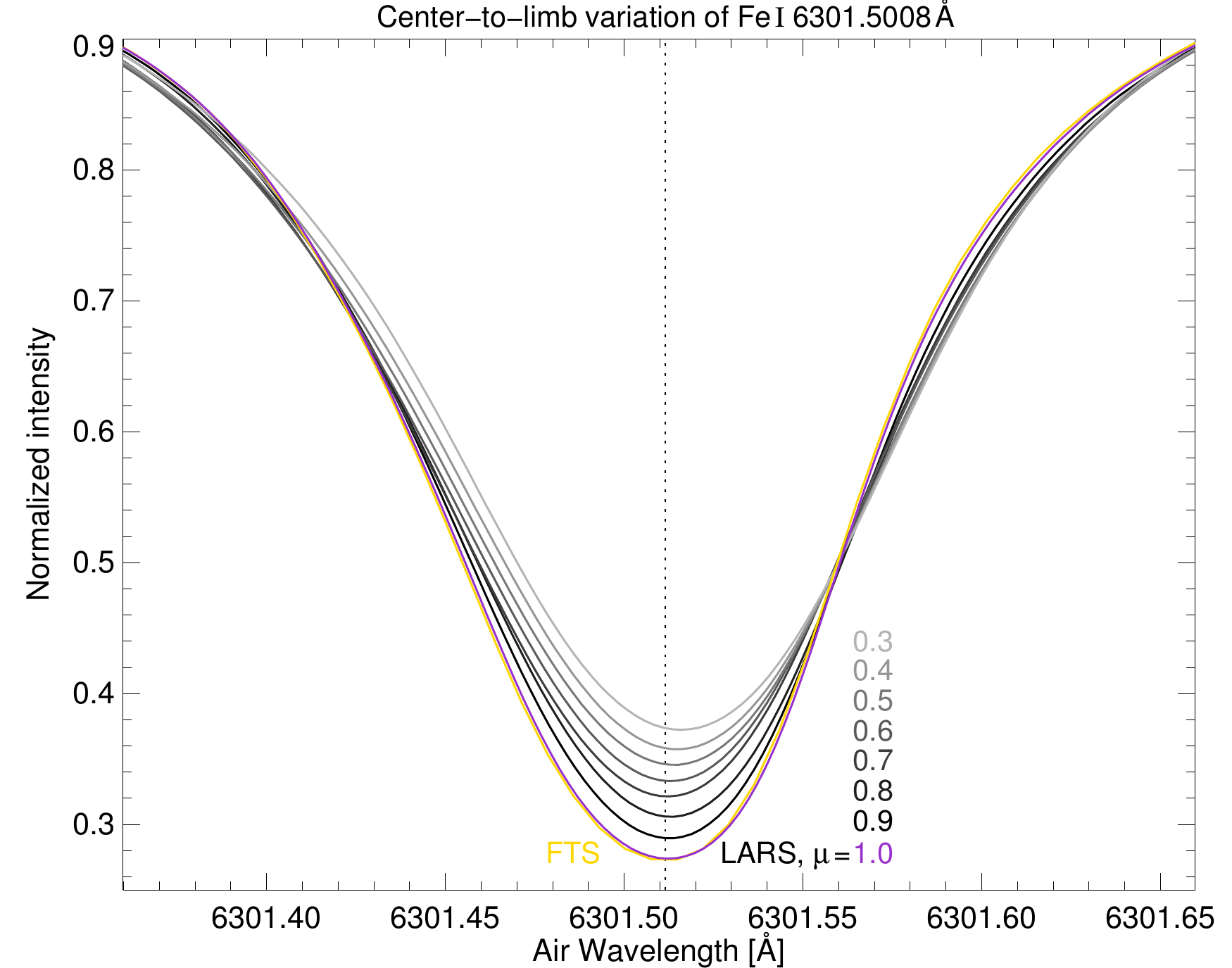}\\[0.1cm]
\includegraphics[width=\columnwidth]{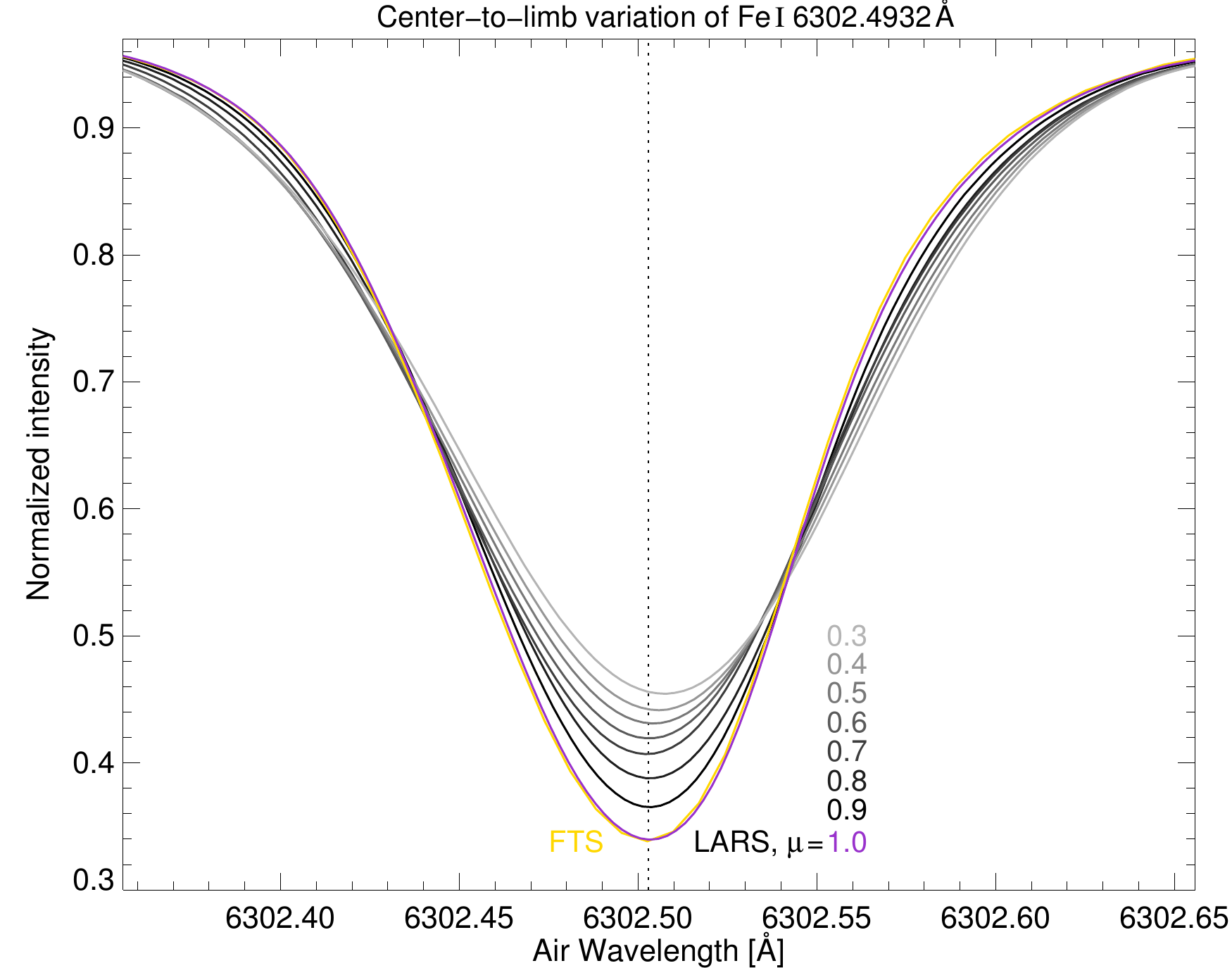}
\caption{Variation of the \ion{Fe}{I}\,6301.5\,\AA\ (upper panel) and \ion{Fe}{I}\,6302.5\,\AA\ (lower panel) line shapes from disk center ($\mu=1.0$, purple line) toward the solar limb ($\mu=0.3$, with decreasing gray scale). The intensities of the profiles were normalized to the spectral continuum and are plotted against the air wavelength (in \AA). The FTS atlas profile is added in yellow for direct comparison. The line minimum at $\mu=1.0$ is marked by the dotted line.}
\label{fig_sec3_profiles}
\end{figure}

\subsubsection{Center-to-limb variation of the spectral profiles}\label{sec3_clv_profiles}
The evolution of the solar spectrum from disk center to the limb is displayed in Fig.\,\ref{fig_sec3_profileCLV}. The spectra describe the average at each heliocentric position $\mu$, omitting the measurement at $\mu=0.95$ and $\mu=0.85$ for better visualization. On that account, the FTS atlas spectrum \citep[yellow curve,][]{1999SoPh..184..421N} at disk center was shifted by $+0.5\%$ along the intensity axis with respect to the LARS spectrum. Otherwise, both spectra would be indistinguishable. The LARS observations were normalized to the spectral continuum intensity measured at disk center. From $\mu=1.0$ (purple curve) to $\mu=0.3$ (bottom grey curve), the intensity decreased by around 45\% due to the limb darkening. Caused by the temperature gradient in the photosphere and the shallower depths along the line-of-sight when approaching the solar limb, we observe higher and therefore cooler layers with increasing distance from the disk center \citep{2002tsai.book.....S}.

For the further analysis, we normalized all spectra to their specific continuum intensity to improve the comparability of the profiles from disk center to the solar limb. The center-to-limb variation of the normalized line profiles of \ion{Fe}{I}\,6301.5\,\AA\ (upper panel) and \ion{Fe}{I}\,6302.5\,\AA\ (lower panel) are shown in Fig.\,\ref{fig_sec3_profiles}. The line profiles of the weaker lines (\ion{Si}{I}\,6299.6\,\AA, \ion{Sc}{II}\,6300.7\,\AA, \ion{Fe}{I}\,6303.5\,\AA, \ion{Ti}{I}\,6303.8\,\AA) are displayed in Fig.\,\ref{fig_secA_analysis_SiI62996}--\ref{fig_secA_analysis_TiI63038} in the appendix. We note that all line profiles still include the gravitational redshift of 13.3\,m\AA.

In Fig.\,\ref{fig_sec3_profiles}, both iron lines feature a similar center-to-limb variation of the line profile. With respect to the disk center, the line depths decrease by around 10\% when measured close to the solar limb at $\mu=0.3$. In addition, the line width increases steadily. But even more conspicuous is the line shift to longer wavelengths. In the case of \ion{Fe}{I}\,6301.5\,\AA, the line minimum at $\mu=1.0$ has a wavelength of 6301.512\,\AA, marked by the dotted line. In comparison, the line minimum at $\mu=0.3$ has shifted by almost 4\,m\AA\ to 6301.516\,\AA. In the case of \ion{Fe}{I}\,6302.5\,\AA, the line center shifts from 6302.503\,\AA\ (at $\mu=1.0$) to 6302.507\,\AA\ (at $\mu=0.3$). In units of Doppler velocities, this shift amounts to almost ${\rm 200\,m\,s^{-1}}$.

In the case of the weaker lines, the line shift to longer wavelength is identifiable too (see Fig.\,\ref{fig_secA_analysis_SiI62996}--\ref{fig_secA_analysis_TiI63038}). But contrary to the decreasing line depths of the stronger lines, the line depths maintains (Fig.\,\ref{fig_secA_analysis_SiI62996} and Fig.\,\ref{fig_secA_analysis_FeI63035}) or increases by a few percent (Fig.\,\ref{fig_secA_analysis_ScII63007} and Fig.\,\ref{fig_secA_analysis_TiI63038}) when approaching the solar limb. 

\subsubsection{Center-to-limb variation of the line bisectors}\label{sec3_bisectors}

\begin{figure}[htpb]
\includegraphics[width=\columnwidth]{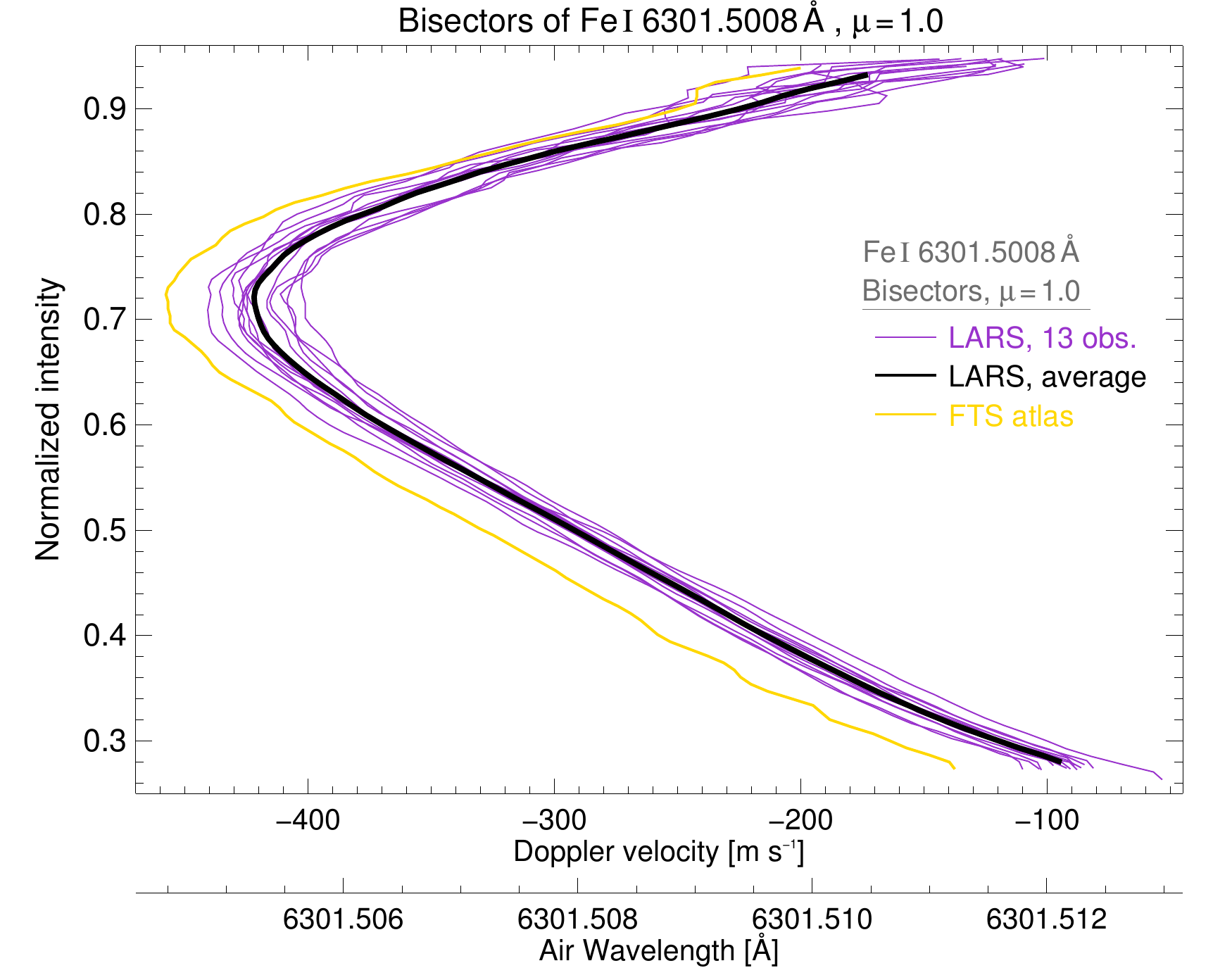}\\[0.1cm]
\includegraphics[width=\columnwidth]{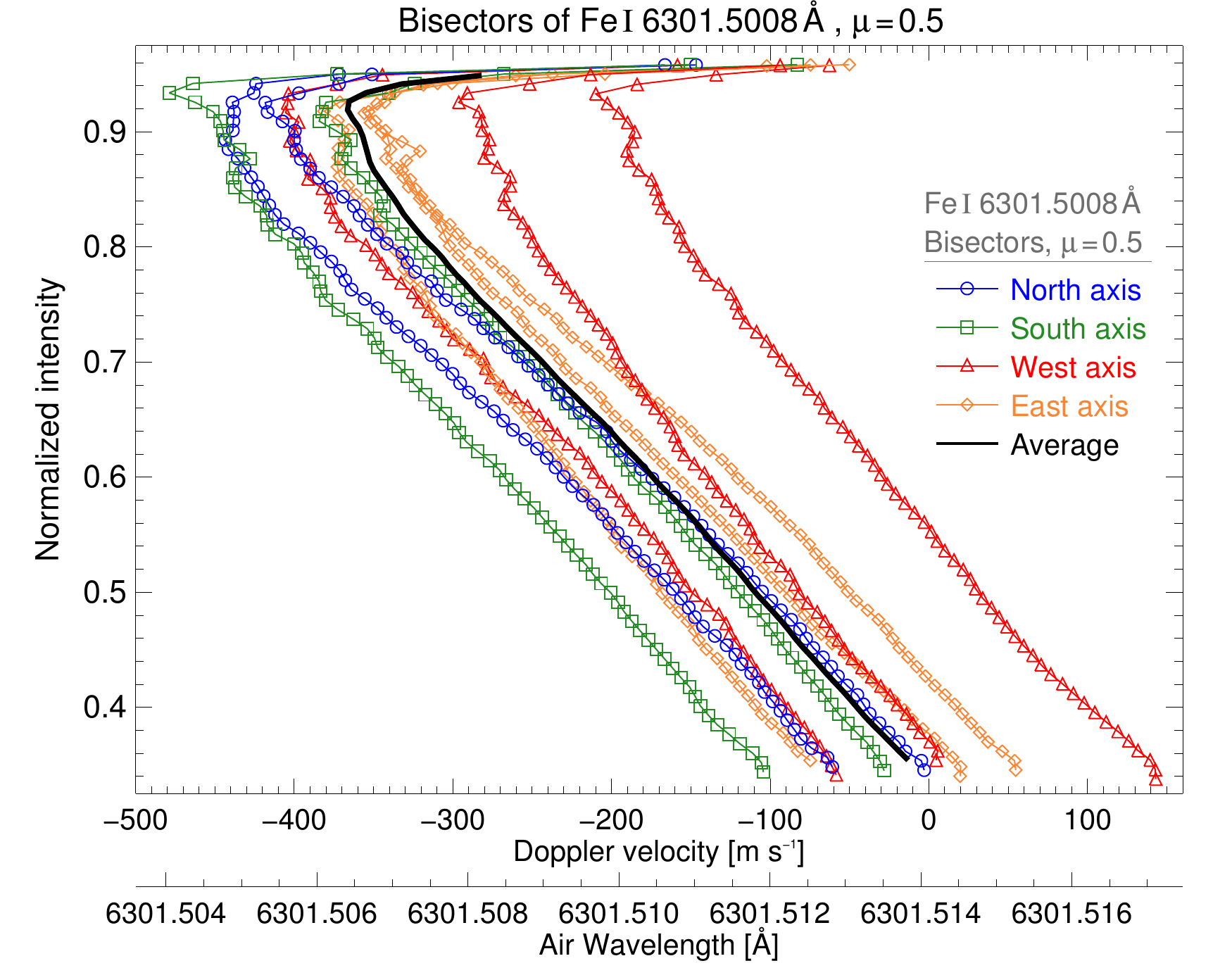}\\[0.1cm]
\includegraphics[width=\columnwidth]{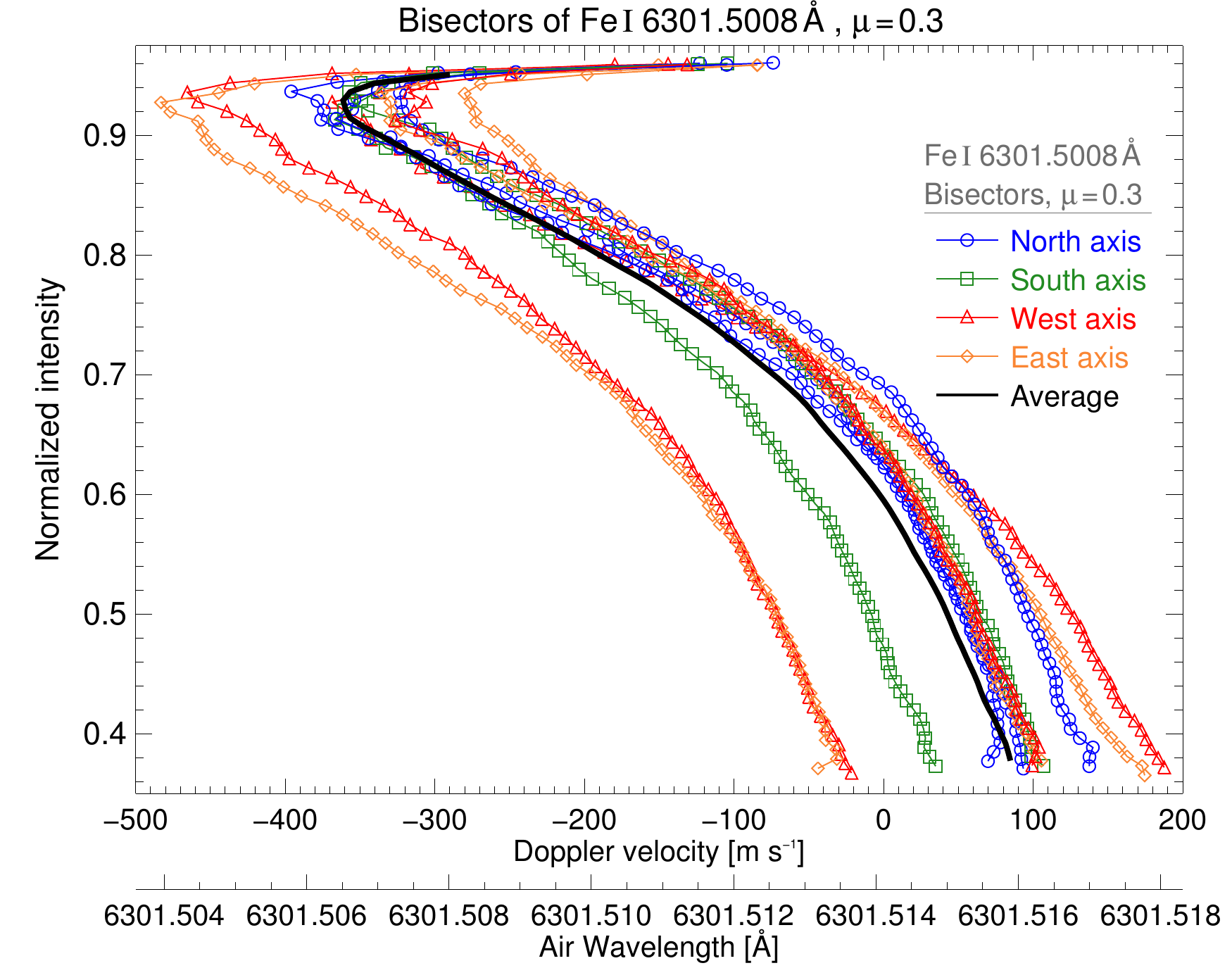}
\caption{Bisectors of \ion{Fe}{I}\,6301.5\,\AA\ at three different heliocentric positions on the solar disk: $\mu=1.0$ (top panel), $\mu=0.5$ (center panel), and $\mu=0.3$ (bottom panel). The intensity was normalized to the spectral continuum, the convective shifts are given in air wavelength (in \AA) and derived Doppler velocities (in ${\rm m\,s^{-1}}$). Each curve represents the average bisectors of a sequence. The colors and symbols correspond to the respective solar axis (see Fig.\,\ref{fig_sec2_observations_scheme}). The black solid lines are the average bisectors of all measurements at the respective heliocentric position. The FTS line bisector is added as yellow curve in the top panel.}
\label{fig_sec3_bisectors}
\end{figure}

For a closer inspection of the line profiles, we performed a detailed bisector analysis. Eminently, this is a conventional tool to analyze the asymmetry of a spectral line. The bisector curve is defined as the connection of the midpoints of horizontal line segments at distinct line depths. Bisectors were calculated in the range from the line minimum to a line-dependent upper threshold intensity. Due to the proximity of telluric lines, we set the threshold level for \ion{Fe}{I} 6301.5\,\AA, 6302.5\,\AA, and \ion{Si}{I} 6299.6\,\AA\ to 95\% of the continuum intensity. For the remaining weaker lines, the level was set to 98\%.
We used an equidistant sampling of bisector positions and defined a total number of positions. Depending on the line depth, we adapted the total number to gain an equidistant step size between 0.5\% and 1\% to avoid oversampling.

The resulting bisector curves for the \ion{Fe}{I} 6301.5\,\AA\ line at disk center are shown in the top panel of Fig.\,\ref{fig_sec3_bisectors}. Each of the curves (purple lines) represents the average bisector for one of the 13 observation sequences at $\mu=1.0$. All bisectors show the same C-shaped profile and lie close to each other. We thus limit our description to the average bisector (black curve) over all measurements at disk center. The bisector curve first describes an increasing blueshift from the continuum level down to a normalized intensity of 0.72 with a maximum Doppler shift of ${\rm -420\,m\,s^{-1}}$. Subsequently, the blueshift decreases monotonically to a Doppler velocity of ${\rm -95\,m\,s^{-1}}$ at the line minimum at an intensity of 0.28. In comparison, the bisector from the FTS atlas profile (yellow curve) shows an identical trend but with an velocity offset of ${\rm -40\,m\,s^{-1}}$. As an outlook to the further analysis, we will see that this offset is apparent for all lines. 

\begin{figure}[htpb]
\includegraphics[width=\columnwidth]{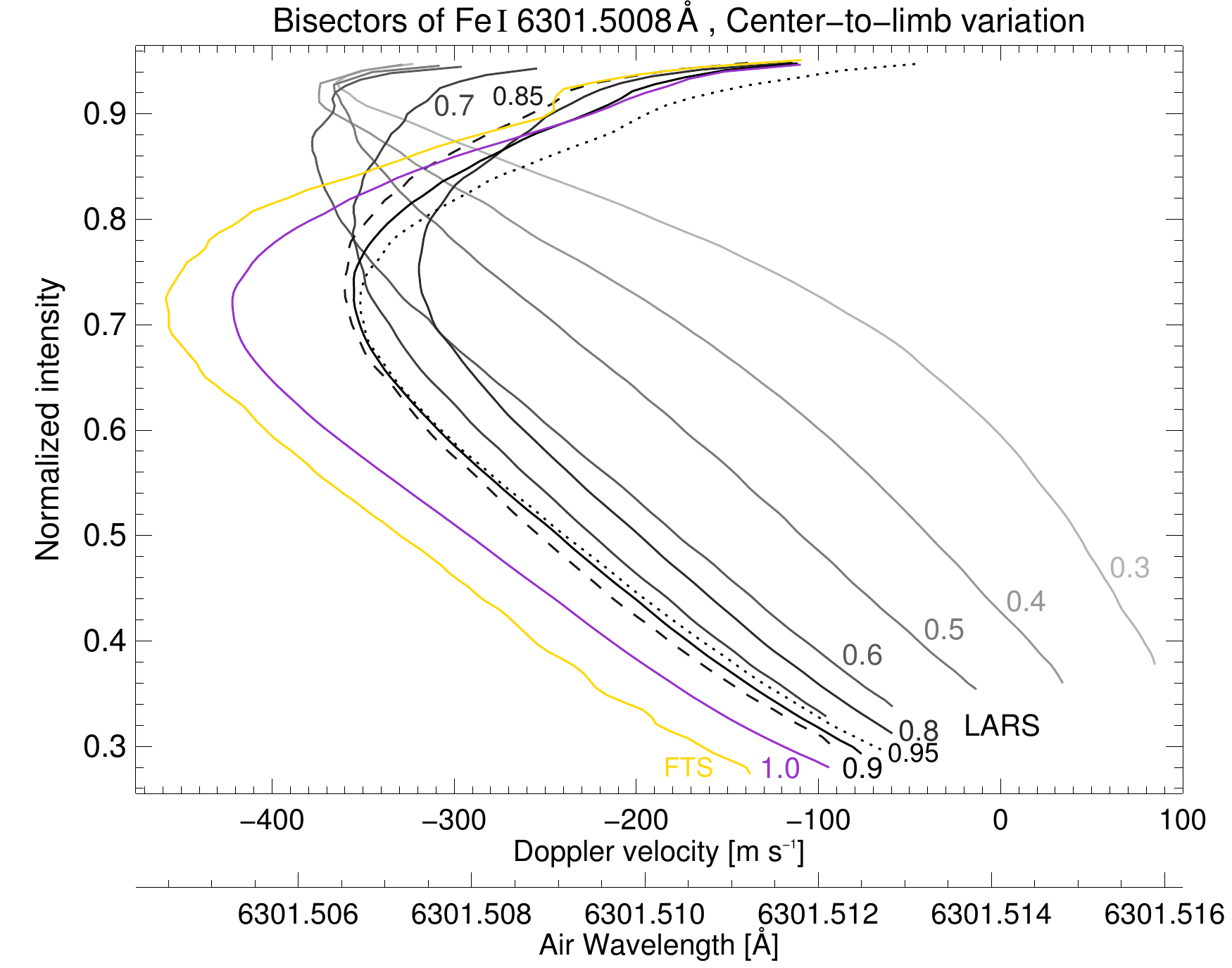}\\[0.1cm]
\includegraphics[width=\columnwidth]{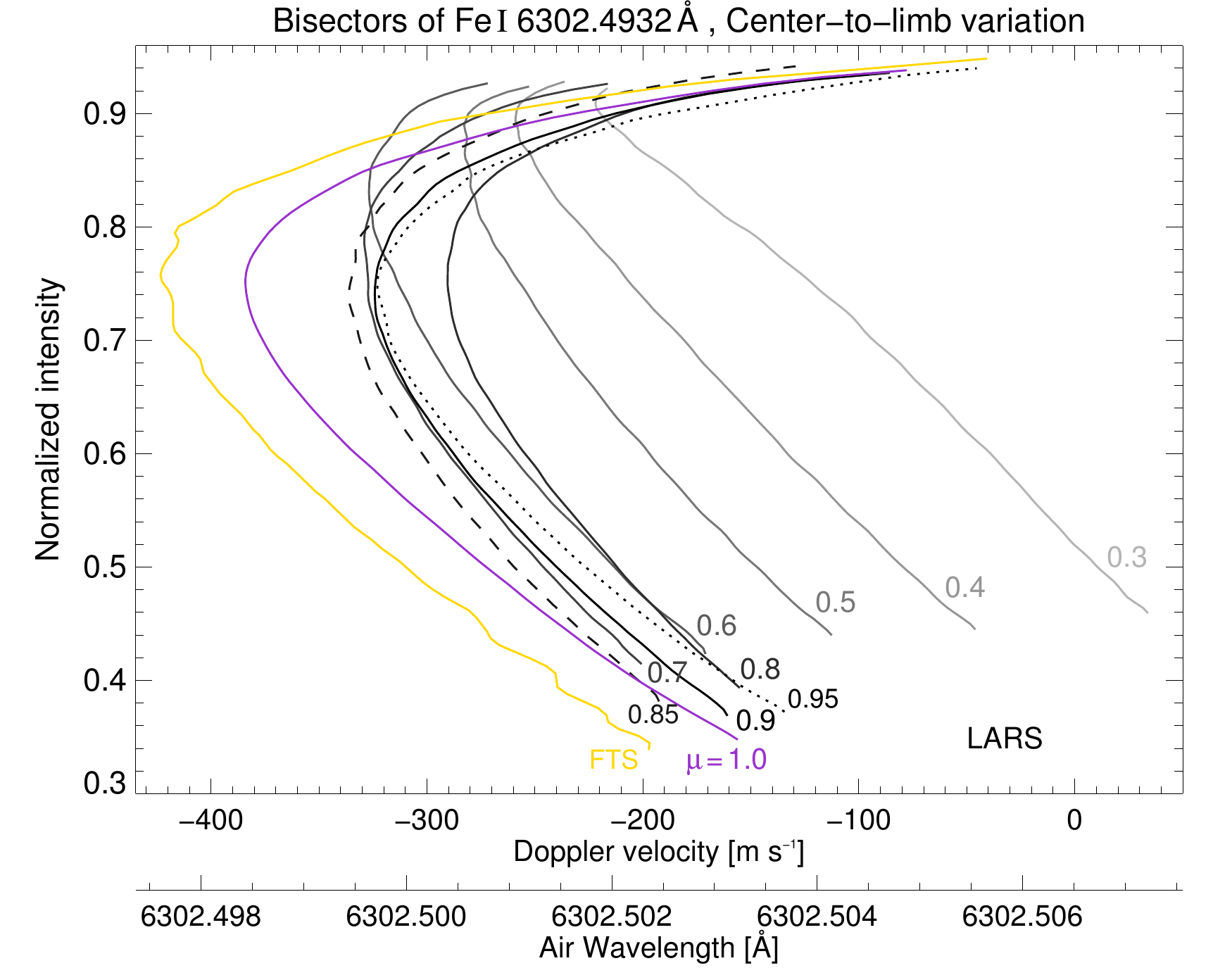}
\caption{Center-to-limb variation of the line bisectors of \ion{Fe}{I}\,6301.5\,\AA\ (upper panel) and \ion{Fe}{I}\,6302.5\,\AA\ (lower panel) from disk center (\mbox{$\mu=1.0$}, purple line) toward the solar limb ($\mu=0.3$, with decreasing gray scale). The intensities were normalized to the spectral continuum and are plotted against the air wavelength (in \AA) and Doppler velocity (in ${\rm m\,s^{-1}}$). The FTS atlas bisector is added in yellow for direct comparison.}
\label{fig_sec3_bisectorsFe}
\end{figure}

In the next step, we analyzed the bisector profiles observed at larger heliocentric angles. For each position $\mu$, we compared the bisectors measured along the four axes (highlighted in Fig.\,\ref{fig_sec2_observations_scheme}). Below, we discuss the results for the positions $\mu=0.5$ and $\mu=0.3$ close to the solar limb.

The bisectors of \ion{Fe}{I} 6301.5\,\AA\ at $\mu=0.5$ are displayed in the middle panel of Fig.\,\ref{fig_sec3_bisectors}. In contrast to the C-shape curve at disk center, the bisectors have turned into "\textbackslash"-shaped curves. The average bisector (black curve) first described a steep increase in blueshift from normalized intensities of 0.95 to an average maximum Doppler shift of ${\rm -367\,m\,s^{-1}}$ at intensities around 0.92. The bisector then turns into an almost curvature-free decrease in blueshift and finally arrives at a Doppler velocity of ${\rm -13\,m\,s^{-1}}$ at an intensity of little less than 0.36. We find that the mere shape of all individual bisectors invariably resemble each other. This is not only the case at $\mu=0.5$, but at all other heliocentric positions as well. However, the distribution of the bisectors shows a scatter in wavelength. {We} can exclude systematic errors from the telescope pointing and the solar rotation model as the reason for this spreading. {Consequently, the scatter is caused by the Sun itself, for example by supergranular or large scale flows, $p$-modes, and the impact of the distribution of the large-scale magnetic field on the convection. }

Approaching the solar limb, the line asymmetry undergoes a further change. The bisectors of \ion{Fe}{I} 6301.5\,\AA\ at $\mu=0.3$ are displayed in the bottom panel of Fig.\,\ref{fig_sec3_bisectors}. Alike the bisector at $\mu=0.5$, the average bisector (black curve) at $\mu=0.3$ performs the same steep increase in blueshift to ${\rm -361\,m\,s^{-1}}$ at intensities between 0.95 and 0.93. Thereafter, the bisector exhibits a straight decrease in blueshift by around ${\rm 250\,m\,s^{-1}}$ within the next 20\% in intensity. Unlike to other heliocentric position, the curve at the limb seems to saturate and bends concave downwards in the lower half of the bisector. The line minimum at an intensity of 0.38 features a redshift of ${\rm +85\,m\,s^{-1}}$. 

The center-to-limb variation of the line bisectors of \ion{Fe}{I} 6301.5\,\AA\ is synoptically plotted in the upper panel of Fig.\,\ref{fig_sec3_bisectorsFe}. Each bisector represents the average curve for all measurements at the respective heliocentric position $\mu$. As indicated in the last paragraph, the bisector curves depend strongly on the position on the solar disk. At disk center (purple curve), the bisector exhibits the most pronounced C-shape with the maximum blueshift of ${\rm -420\,m\,s^{-1}}$ at a normalized intensity of 0.72. With decreasing $\mu$ value, or increasing distance from disk center, the bisector shape transforms its downward curvature from convex to concave \citep[in line with][]{1976MNRAS.177..687A}. In addition to this variation, the line depth decreases continuously while the blueshift in the line core drops and even turns into a redshift at the solar limb. The intensity level of the maximum blueshift of around ${\rm -350\,m\,s^{-1}}$ in the upper half of the bisectors rises steadily from 0.72 ($\mu=1.0$) to 0.93 ($\mu=0.3$). By setting the observed bisector variation in relation to the bisector shape classification of spectral lines \citep{1984ssdp.conf..317B}, we confirm that the \ion{Fe}{I} 6301.5\,\AA\ line belongs to class II \citep[in line with][]{1984SoPh...93..219B}.

The second strong line in the selected wavelength region is \ion{Fe}{I} 6302.5\,\AA. The center-to-limb variation of its normalized profile was shown in the lower panel of Fig.\,\ref{fig_sec3_profiles}. The corresponding bisector curves are plotted in the lower panel of Fig.\,\ref{fig_sec3_bisectorsFe}. The bisector at disk center (purple curve) describes a distinct C-shape. From the continuum to the line minimum, the curve describes an increase in blueshift to ${\rm -385\,m\,s^{-1}}$ at an intensity of 0.75 followed by a decrease in Doppler velocity to around ${\rm -160\,m\,s^{-1}}$ at an intensity of 0.35. In comparison, the FTS atlas line bisector (yellow curve) indicates a similar shape but again with a wavelength shift of ${\rm -0.8\,m\AA}$, or ${\rm -40\,m\,s^{-1}}$ in Doppler velocity. When departing the disk center toward the solar limb, the C-shape transforms into a \textbackslash-shape. In connection with a decrease of line depth by some 10\%, and a decline of convective blueshift, the bisector at $\mu=0.3$ describes an almost straight line (with a slight concave curvature) from a maximum blueshift of  ${\rm -220\,m\,s^{-1}}$ at an intensity of 0.91 ending in a redshift of ${\rm +35\,m\,s^{-1}}$ at the line minimum intensity of 0.46. Contrary to \citet{1984SoPh...93..219B} who grouped \ion{Fe}{I} 6302.5\,\AA\ into class I characterized by solely blueshifts, we proved that the line core shows a supergravitational redshift at the solar limb. We therefore suggest to sort \ion{Fe}{I} 6302.5\,\AA\ into class II (or an intermediate class I--II). 

The bisector analysis for the weaker absorption lines in the spectral range is presented in the middle panels of Fig.\,\ref{fig_secA_analysis_SiI62996}--\ref{fig_secA_analysis_TiI63038} in the  appendix. In case of \ion{Si}{I}\,6299.6\,\AA\ (Fig.\,\ref{fig_secA_analysis_SiI62996}), the line asymmetry is strongly influenced by the telluric line in the close proximity in the blue. We therefore confine ourself to the fact that the bisector shape describes a downward concave curvature which remains largely the same for all heliocentric positions. Only the wavelength shifts feature a distinct center-to-limb variation with decreasing blueshifts. In case of the very weak lines, \ion{Sc}{II}\,6300.7\,\AA\ (Fig.\,\ref{fig_secA_analysis_ScII63007}) and \ion{Fe}{I}\,6303.5\,\AA\ (Fig.\,\ref{fig_secA_analysis_FeI63035}) with depth below 0.05, the bisectors at disk center do not reach a reversal in blueshift for decreasing intensities. No C-shape is identifiable. For both lines, the bisectors from $\mu=1.0$ to $\mu=0.6$ resemble their shape. For \ion{Sc}{II}\,6300.7\,\AA, the blueshift increases monotonically toward the line core to some ${\rm -350\,m\,s^{-1}}$. For \ion{Fe}{I}\,6303.5\,\AA, the blueshift of around ${\rm -420\,m\,s^{-1}}$ is fairly constant for intensities from 0.98 to 0.95. Only when approaching the solar limb ($\mu=0.5-0.3$), the bisector shape changes to a decline in blueshift from the continuum to line minimum. The bisectors of \ion{Ti}{I}\,6303.8\,\AA\ (Fig.\,\ref{fig_secA_analysis_TiI63038}) for $\mu=1.0-0.6$ exhibit a C-shape from the continuum to the line minimum intensity at 0.92. The maximum blueshift ranges around  ${\rm -350\,m\,s^{-1}}$. Toward $\mu=0.3$, the asymmetry turns into a \textbackslash-shaped bisector with a blueshift of ${\rm -230\,m\,s^{-1}}$ at the line minimum. According to the bisector shape classification, the \ion{Ti}{I}\,6303.8\,\AA\ is a class I line.

\begin{figure}[htpb]
\includegraphics[width=\columnwidth]{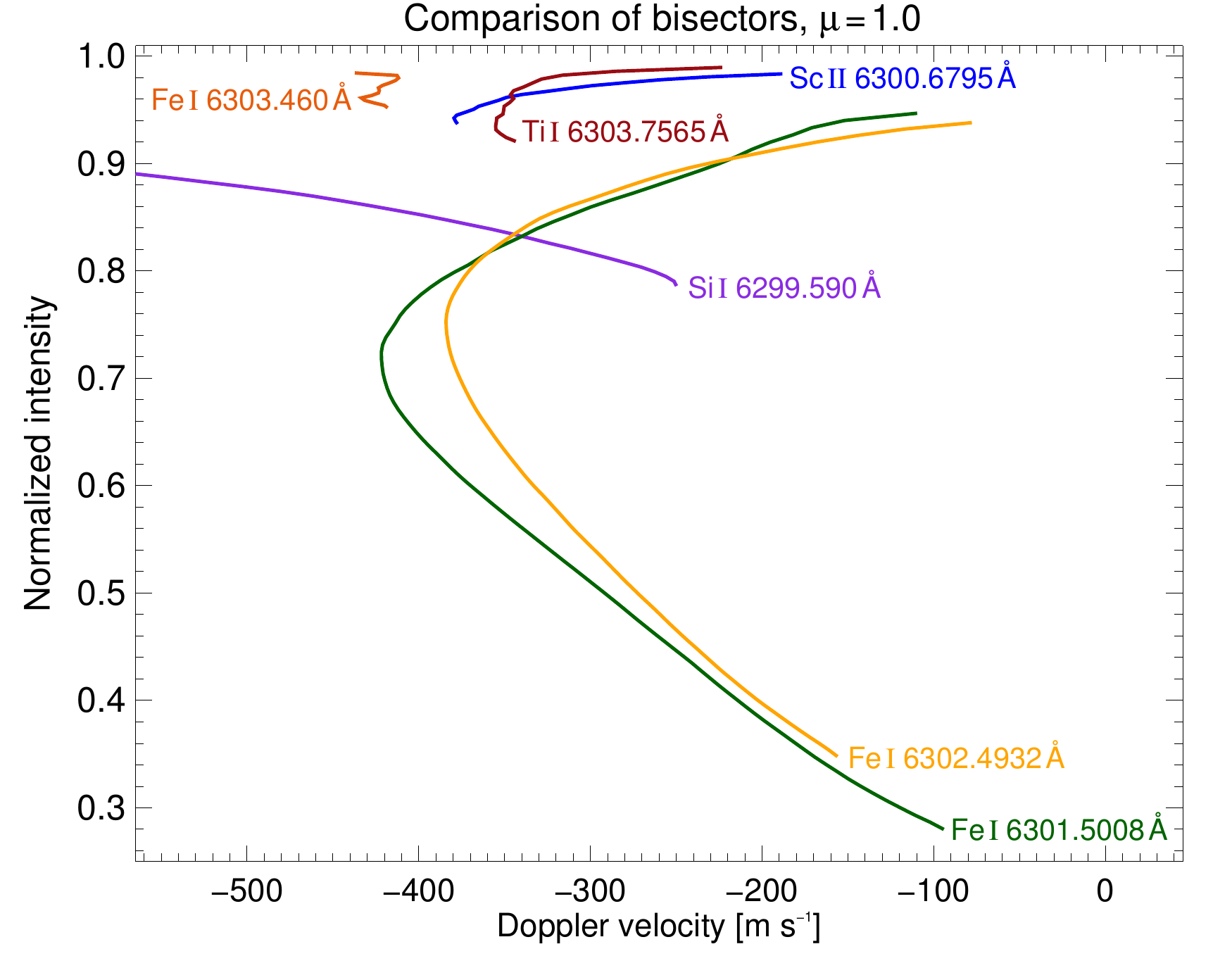}\\[0.1cm]
\includegraphics[width=\columnwidth]{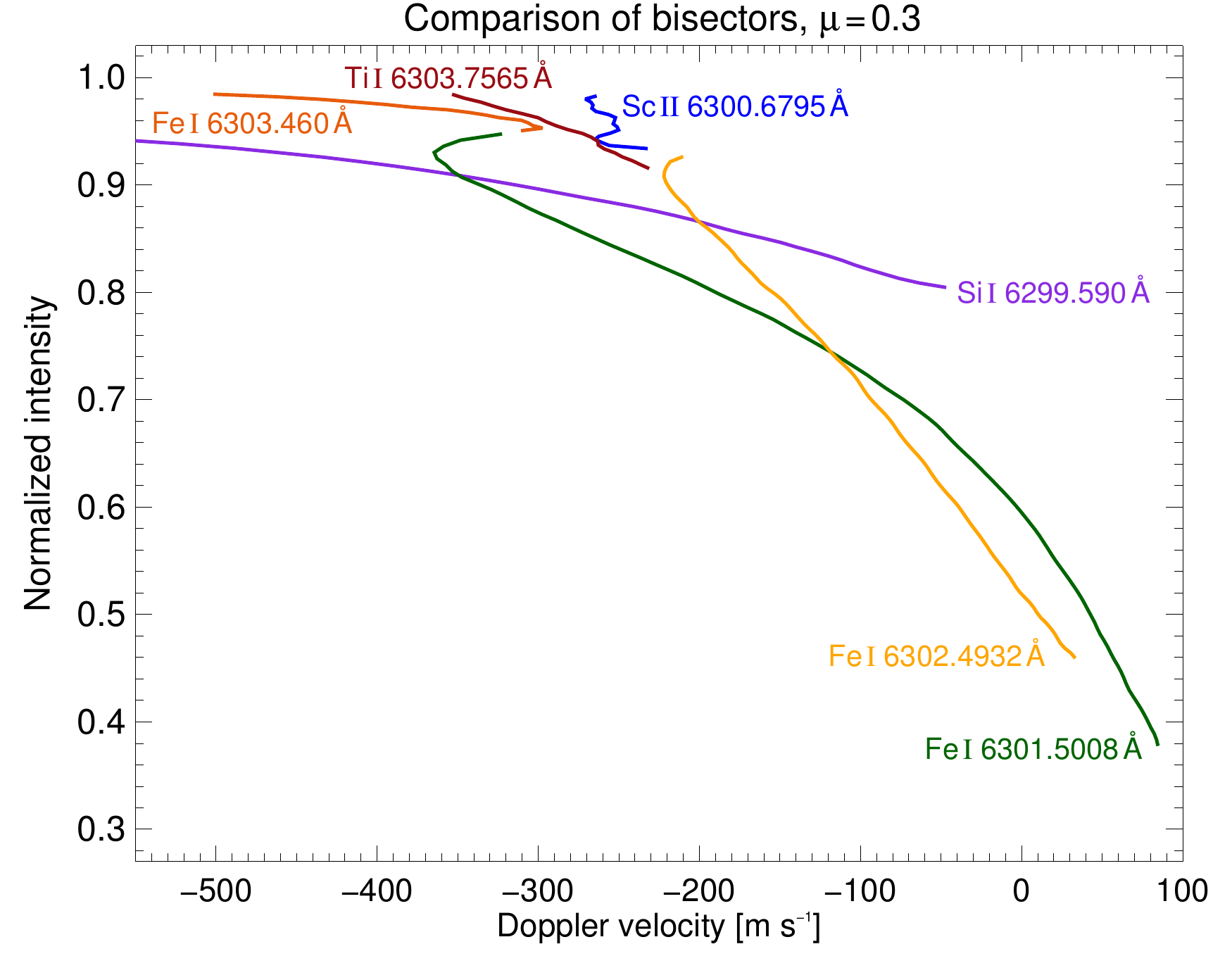}
\caption{Bisectors of the different spectral lines within the observed wavelength range. The upper panel shows the Doppler velocities (in ${\rm m\,s^{-1}}$) of the spectral profiles measured at disk center ($\mu=1.0$). The lower panel displays the results near the solar limb ($\mu=0.3$).}
\label{fig_sec3_bisectors_all10}
\end{figure}

We note that in all cases, the FTS atlas profiles reveal a systematic wavelength shift of ${\rm -0.8\,m\AA}$, or ${\rm -40\,m\,s^{-1}}$ in Doppler velocity against the LARS data. Since the accuracy of our measurements is of the order of ${\rm 1\,m\,s^{-1}}$, we suppose that the calibration of the FTS atlas \citep{1999SoPh..184..421N} suffers from a systematic {wavelength offset. This offset is of the order of the stated relative wavelength error of $10^{-7}$ in the visible range. Most likely, the error is linked to uncertainty of the laboratory wavelength of the telluric line used for the wavelength calibration of the entire FTS spectrum in the visible. Similar findings about systematic offsets have} been reported by \citet{Reiners+etal2016}. 

In Fig.\,\ref{fig_sec3_bisectors_all10}, we give a comparative view of the bisectors of the different spectral lines. Exemplarily, the bisectors at disk center ($\mu=1.0$) are shown in the upper panel. The rigorous change toward the solar limb is demonstrated by the bisectors at $\mu=0.3$ displayed in the lower panel. 

\subsubsection{Center-to-limb variation of the Doppler velocities}\label{sec3_clv_Doppler}
In the previous sections, the center-to-limb variation of the convective blueshift did become apparent in the line profiles and bisectors.
To raise a more quantifiable and resilient measuring parameter for the changing line shift, we ascertain the Doppler shift of restricted parts of the spectral lines
and describe their evolution from disk center toward the solar limb. After the examination of different line segments, we confine the presentation of the results to the Doppler velocities computed for two regimes. These are, firstly, the line core velocity determined by a parabolic fit to the line minimum ($\pm20\,{\rm m}\AA$) and, secondly, the average line shift of all bisector positions from the line minimum up to a defined upper threshold ($95\%$, or $98\%$ for the weakest lines) close to the continuum level. The summarized results for all lines are displayed in Table \ref{table_sec3_line_comparison}. 

\begin{figure}[htpb]
\includegraphics[width=\columnwidth]{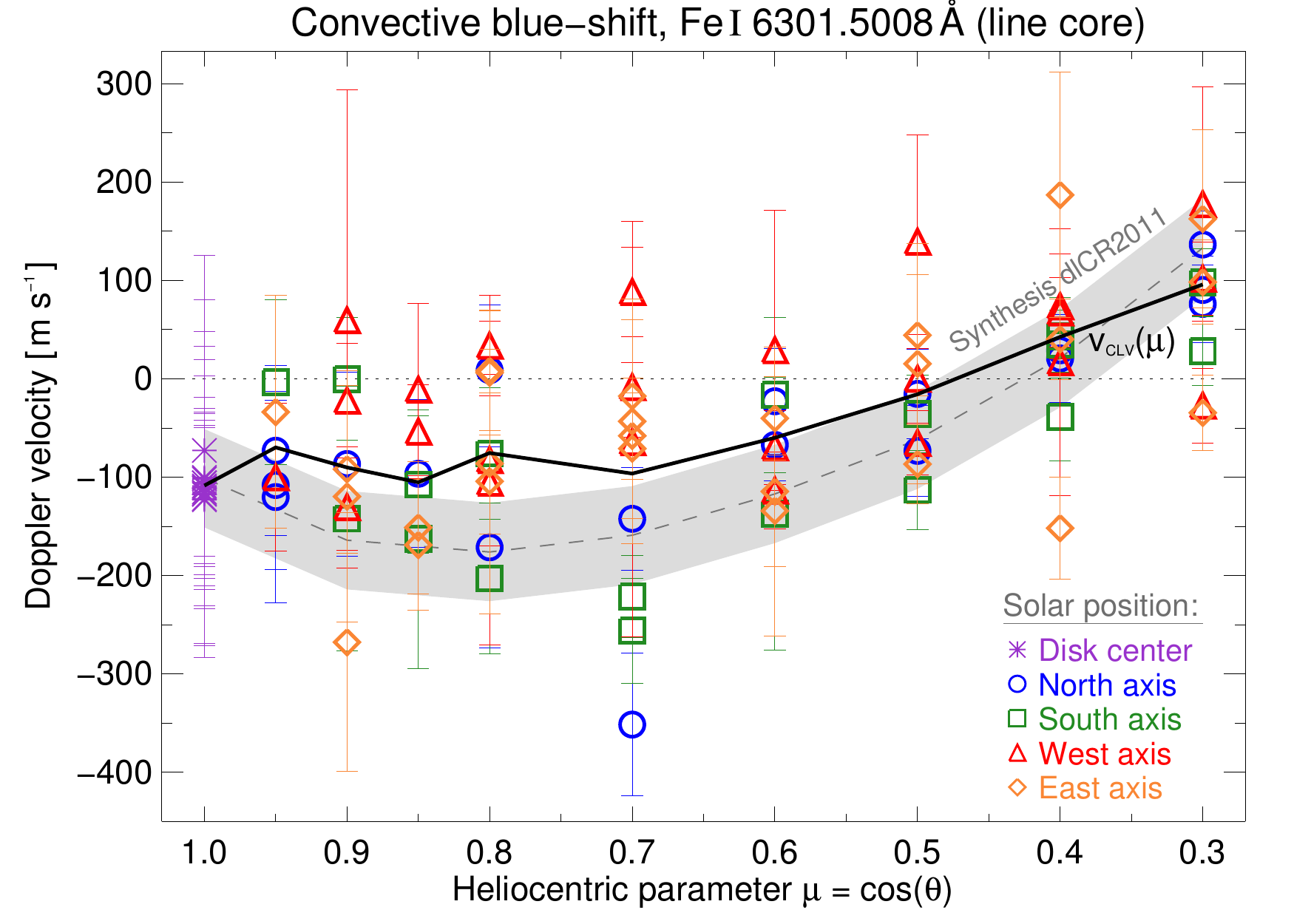}\\[0.1cm]
\includegraphics[width=\columnwidth]{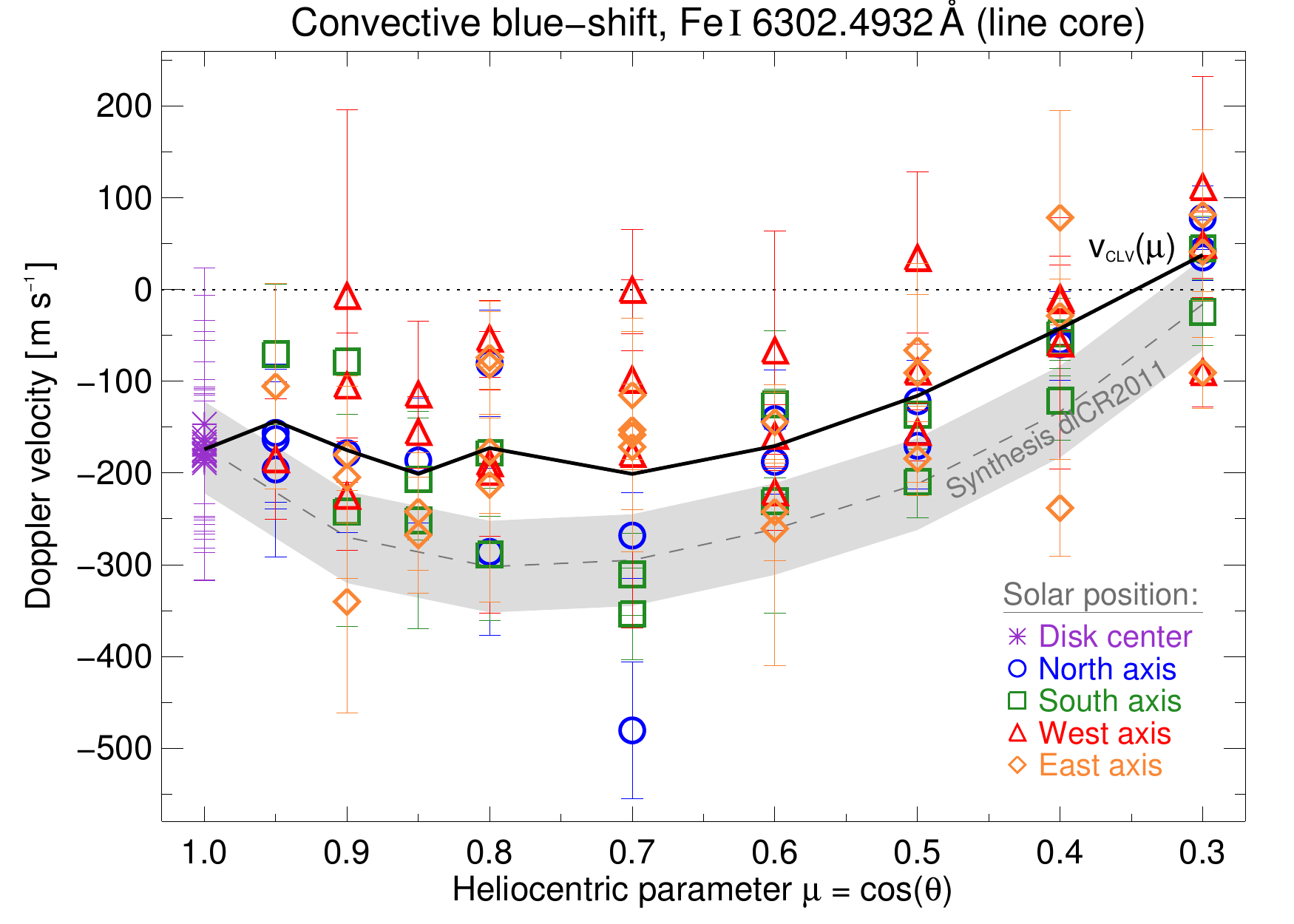}
\caption{Center-to-limb variation of the convective shift. The Doppler velocities (in ${\rm m\,s^{-1}}$) from the line core of \ion{Fe}{I}\,6301.5\,\AA\ (upper panel) and \ion{Fe}{I}\,6302.5\,\AA\ (lower panel) are plotted against the heliocentric parameter from disk center ($\mu=1.0$) toward the solar limb ($\mu=0.3$). The data points represent the average for each sequence. The colored symbols indicate the solar axis. The error bars show the standard deviation of the respective sequence. The black solid line displays the average curve $v_{\rm CLV}(\mu)$. The gray dashed line and shaded area depict the synthesis of \citet{2011A&A...528A.113D} and its assumed error.}
\label{fig_sec3_clv_linecore}
\end{figure}

In the following, we focus our attention to the center-to-limb variation of the line core velocities of \ion{Fe}{I}\,6301.5\,\AA\ and \ion{Fe}{I}\,6302.5\,\AA. The results are illustrated in Fig.\,\ref{fig_sec3_clv_linecore}. Each of the 99 data points represents the average Doppler shift over all time steps of the respective 20\,min-sequence. The error bars indicate the standard deviation within the sequence. We point out that $p$-mode oscillations with amplitudes of some ${\rm 100\,m\,s^{-1}}$ constitute the major error component. The intrinsic instrumental error of ${\rm 1\,m\,s^{-1}}$ is negligible. We stick to the convention of symbols and color code for observations along the four axes indicated in Fig.\,\ref{fig_sec2_observations_scheme}. For each heliocentric position $\mu$, the measurements were averaged yielding the mean curve ($v_{\rm CLV}$, black solid line) as the overall trend. In addition, we compare our observations with the synthetic center-to-limb variation (grey dashed line) of the convective blueshift from \citet{2011A&A...528A.113D}.

\begin{table*}[htbp]
\caption{Center-to-limb variation of the convective shifts. The Doppler velocities (in ${\rm m\,s^{-1}}$) were obtained either from the line core position or the mean bisector. The heliocentric parameter $\mu$ is the cosine of the heliocentric angle $\theta$, ordered from disk center ($\mu=1.0$) to the solar limb ($\mu=0.3$).}
\label{table_sec3_line_comparison}
\centering
\begin{tabular}{c c c | c c c c c c c c c c}
\hline\hline
\multicolumn{3}{c|}{Spectral line}& \multicolumn{10}{c}{Heliocentric parameter $\mu = \cos\theta$}\\ 
Element & $\lambda_0\ (\AA)$&Part&1.0&0.95&0.9&0.85&0.8&0.7&0.6&0.5&0.4&0.3\\ 
\hline
\ion{Fe}{I} & 6301.5008&Line core&$-107$&$-74$&$-81$&$-99$&$-82$&$-101$&$-64$&$-18$&$+33$&$+85$\\
& & Mean bis. &$-295$&$-244$&$-251$&$-269$&$-246$&$-271$&$-255$&$-206$&$-156$&$-87$\\
& & Synthesis &$-307$&&$-349$&&$-358$&$-351$&$-325$&$-279$&$-211$&$-101$\\[0.1cm]
\ion{Fe}{I} & 6302.4932&Line core&$-164$&$-139$&$-163$&$-195$&$-178$&$-204$&$-173$&$-115$&$-47$&$+33$\\
& & Mean bis. &$-295$&$-249$&$-262$&$-281$&$-260$&$-286$&$-274$&$-221$&$-170$&$-94$\\
& & Synthesis &$-335$&&$-387$&&$-404$&$-403$&$-381$&$-342$&$-280$&$-176$\\[0.1cm]
\ion{Fe}{I} & 6303.460&Line core&$-420$&$-398$&$-394$&$-423$&$-392$&$-434$&$-420$&$-385$&$-359$&$-277$\\
& & Mean bis. &$-422$&$-388$&$-390$&$-419$&$-381$&$-444$&$-455$&$-417$&$-402$&$-347$\\[0.1cm]
\ion{O}{I} & 6300.304&Line core&+272&+327&+306&+265&+263&+224&+232&+221&+201&+235\\
& & Mean bis. &$+214$&$+262$&$+241$&$+201$&$+232$&$+169$&$+142$&$+161$&$+163$&$+191$\\[0.1cm]
\ion{Sc}{II} & 6300.6795&Line core&$-387$&$-316$&$-351$&$-372$&$-338$&$-382$&$-387$&$-355$&$-294$&$-241$\\
& & Mean bis. &$-344$&$-288$&$-312$&$-339$&$-298$&$-357$&$-366$&$-332$&$-304$&$-257$\\[0.1cm]
\ion{Si}{I} & 6299.590&Line core&$-253$&$-223$&$-253$&$-296$&$-277$&$-334$&$-328$&$-269$&$-185$&$-48$\\
& & Mean bis. &$-484$&$-460$&$-485$&$-515$&$-482$&$-534$&$-530$&$-467$&$-385$&$-259$\\[0.1cm]
\ion{Ti}{I} & 6303.7565&Line core&$-353$&$-317$&$-331$&$-366$&$-335$&$-380$&$-379$&$-342$&$-296$&$-225$\\
& & Mean bis. &$-347$&$-304$&$-325$&$-354$&$-320$&$-379$&$-395$&$-353$&$-324$&$-277$\\
\hline
\end{tabular}
\tablefoot{The line core velocity was determined by a parabolic fit to the intensities around the line minimum. The mean bisector is the average of all bisector centers from the line core to the spectral continuum. For the deeper lines \ion{Fe}{I} 6301.5\,\AA, 6302.5\,\AA, and \ion{Si}{I} 6299.6\,\AA\ the upper threshold was set to 95\% of the continuum intensity. The values of the synthesis (same threshold region) were taken from Table 2 of \citet{2011A&A...528A.113D}. For the weaker lines \ion{Fe}{I} 6303.5\,\AA, \ion{O}{I} 6300.3\,\AA, \ion{Sc}{II} 6300.7\,\AA, and \ion{Ti}{I} 6303.8\,\AA, the upper intensity level was defined as 98\%.}
\end{table*}

In case of \ion{Fe}{I}\,6301.5\,\AA\ (upper panel of Fig.\,\ref{fig_sec3_clv_linecore}), we find that all observed Doppler velocities range between ${\rm -350\,m\,s^{-1}}$ and ${\rm +200\,m\,s^{-1}}$. At disk center ($\mu=1.0$), the measurements match each other very well. The scatter is below ${\rm 50\,m\,s^{-1}}$ with a mean Doppler velocity of ${\rm -107\,m\,s^{-1}}$. The very small scatter can be explained by the line-of-sight for the disk center observation. Since mesogranular, supergranular and large-scale flows describe predominantly horizontal flow fields, their velocity component is absent for observations at small heliocentric angles for which the current and the line of sight become perpendicular. The remaining vertical $p$-mode oscillation then makes the main error component at $\mu=1.0$. At positions between $\mu=0.95$ and $\mu=0.7$, we register a considerably larger scatter of the Doppler velocities. Especially at $\mu=0.7$, the spreading has reached around ${\rm 400\,m\,s^{-1}}$. In this region of the solar disk, the sum of line-of-sight effects ($p$-modes, supergranulation, large-scale flows) superimposing the mere convective blueshift becomes largest. The mean values range between ${\rm -101\,m\,s^{-1}}$ and ${\rm -74\,m\,s^{-1}}$. The trend of the average curve suggests a rather constant blueshift. From $\mu=0.6$ to $\mu=0.3$, the scenery changes when observations approach the solar limb. The scatter decreases to around ${\rm 200\,m\,s^{-1}}$ due to the declining imprint of $p$-mode oscillations and foreshortening effects of supergranules allowing for a more effective spatial averaging over the flow field. The mean curve shows a distinct monotonic decrease in blueshift from ${\rm -64\,m\,s^{-1}}$ at $\mu=0.6$ to a supergravational redshift of ${\rm +85\,m\,s^{-1}}$ at $\mu=0.3$. 

To gain theoretical context information for our measurements, we draw a comparison with the numerical synthesis of the \ion{Fe}{I}\,6301.5\,\AA\ line and its center-to-limb convective blueshift \citep{2011A&A...528A.113D}. The Doppler shifts of the synthesized line core are overplotted as gray dashed line in Fig.\,\ref{fig_sec3_clv_linecore}. The error range is estimated to ${\rm \pm50\,m\,s^{-1}}$ (gray colored area). At disk center ($\mu=1.0$), the observed velocities of around ${\rm -100\,m\,s^{-1}}$ are perfectly in line with the theoretical ones. Near the solar limb ($\mu=0.6-0.3$), the measured and synthetic values also agree within one standard deviation, with slightly differing slopes of the curves. 

{The distribution of convective blueshifts at positions between $\mu=0.9$ and $\mu=0.7$ merits special attention. Especially at $\mu=0.7$ in Fig.\,\ref{fig_sec3_clv_linecore}, we notice that the observations along the meridian yield a stronger convective blueshift than the measurements along the equator. At the extreme, the Doppler blueshift measured at the northern and southern axis is about ${\rm 200\,m\,s^{-1}}$ stronger compared to the eastern and western axis. Consequentially, the often reported initial increase in convective blueshift from $\mu=1.0$ to $\mu=0.7$ is only measured for the meridional axes. In comparison with the numerical simulation, we find that the measurements are well distributed around the predicted blueshifts with an increase in velocity by around ${\rm 60\,m\,s^{-1}}$ from disk center to $\mu=0.7--0.8$. As observed in other observational studies \citep[e.g.,][]{1984SoPh...93..219B} and discussed in \citet{1985SoPh...99...31B}, this initial drop of convective blueshift is caused by the impact of horizontal granular motions.
We can only report on such a strong drop for the observation along the meridional axis. In contrast, the center-to-limb variation along the equatorial axis clearly deviates while describing a rather flat slope of the convective blueshift. To our best knowledge, this difference of the center-to-limb variation between the meridian and the equator has not been reported so far. An explanation by the meridional flow toward the poles does not hold, since the flow speeds are assumed to be around ${\rm 10-20\,m\,s^{-1}}$ which is one magnitude smaller than the observed effect. Other causes like supergranular and large-scale flows can explain the scatter around the nominal convective Doppler shift, but not the systematic difference between the meridional and equatorial axes. Additional measurements would possibly reduce the strong scatter in our data, but we do not expect a significant change of the finding. The most likely reason for the difference between the north-south and east-west distribution is the solar magnetic field. As observed by \citet{1985A&A...143..116C}, the convective blueshift increases when moving away from magnetically active regions. The same has been seen in three-dimensional hydrodynamical simulations when adding magnetic fields to the calculations. Since the solar activity belts assimilate a slightly stronger background magnetic field around the equator, one may expect a more disturbed granulation and consequently less pronounced convective blueshift for the measurements at the eastern and western axes. To verify our measurements of the convective blueshift, simulations have to adapt the atmospheric parameters and include a different treatment of the magnetic field.}

The center-to-limb variation of the convective blueshift of the \ion{Fe}{I} 6302.5\,\AA\ line core is displayed in the lower panel of Fig.\,\ref{fig_sec3_clv_linecore}. At disk center, all measurement lie close together with a little scatter around the mean Doppler shift of ${\rm -164\,m\,s^{-1}}$. The synthesized Doppler shift of the line core is in perfect agreement with the observations. Departing from disk center toward the solar limb, the variance of the individual observation data is larger, resembling the case of  \ion{Fe}{I} 6301.5\,\AA\ which was discussed in the paragraph above. Thus, we focus on the average trend (black curve) of the center-to-limb variation. Contrary to \ion{Fe}{I} 6301.5\,\AA, the line core of \ion{Fe}{I} 6302.5\,\AA\ performs a slight increase in blueshift by ${\rm 40\,m\,s^{-1}}$ from $\mu=1.0$ to $\mu=0.7$. Toward the solar limb ($\mu=0.7-0.3$), the blueshift decreases monotonically and turns into a redshift of ${\rm +33\,m\,s^{-1}}$ at $\mu=0.3$. In comparison to the theoretical evolution \citep{2011A&A...528A.113D}, the slopes of the observed and synthesized curve are quite similar. However, we report an offset of ${\rm 50-100\,m\,s^{-1}}$ whereas the theoretical evolution predicts a more pronounced increase in blueshift between disk center and $\mu=0.8$. 

\begin{figure}[htpb]
\includegraphics[width=\columnwidth]{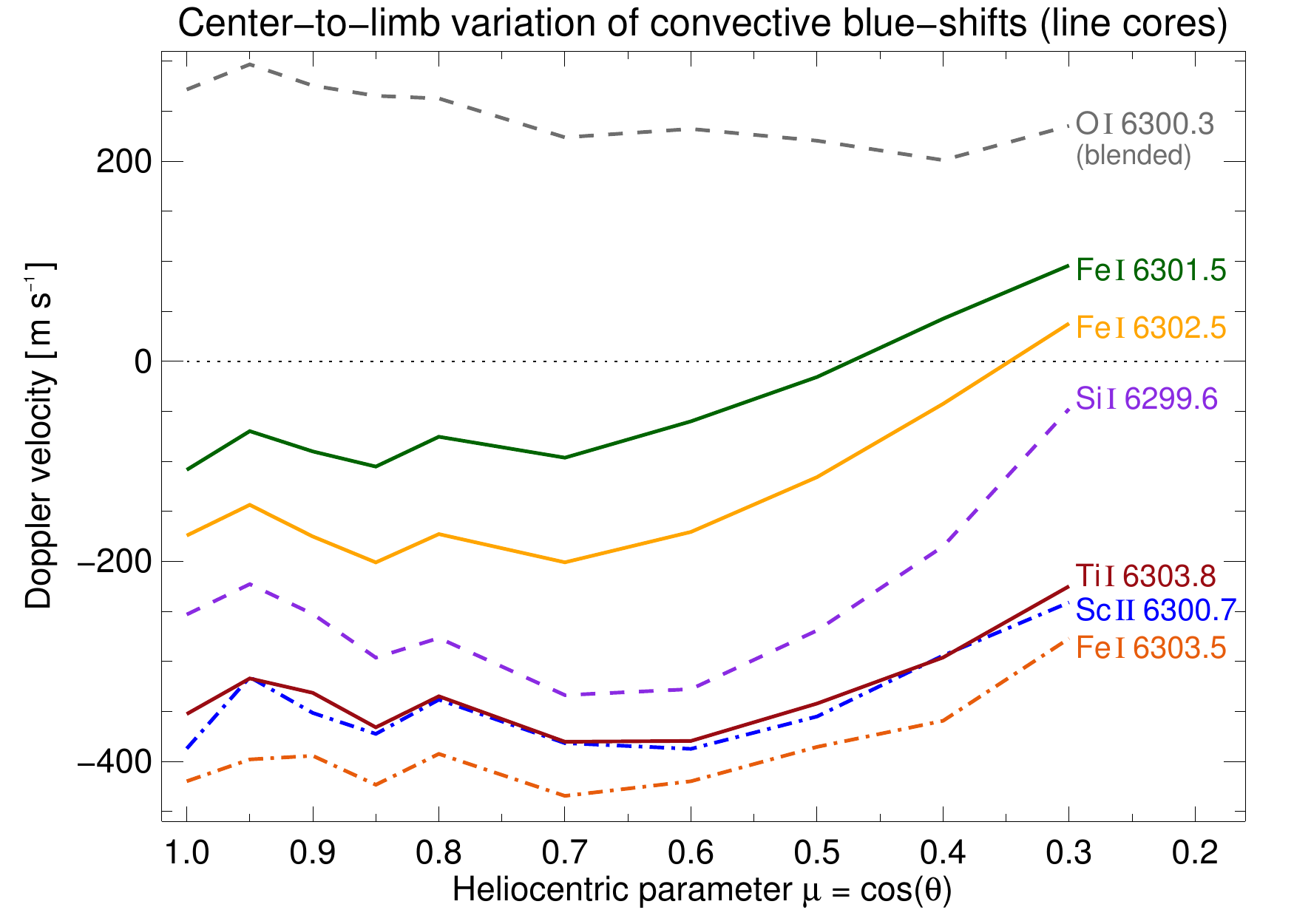}
\caption{Center-to-limb variation of the convective shift of the spectral lines within the observed wavelength range. The Doppler velocities (in ${\rm m\,s^{-1}}$) determined from the line core are plotted against the heliocentric parameter from disk center ($\mu=1.0$) toward the solar limb ($\mu=0.3$).}
\label{fig_sec3_bisectors_all03}
\end{figure}

For the weaker lines within the spectral range, the center-to-limb variation of the line core blueshift shows a similar trend and data scatter (displayed in the respective bottom panel of Fig.\,\ref{fig_secA_analysis_SiI62996}--\ref{fig_secA_analysis_TiI63038} in the  appendix). For \ion{Si}{I} 6299.6\,\AA, the initial increase in blueshift from ${\rm -253\,m\,s^{-1}}$ at $\mu=1.0$ to ${\rm -334\,m\,s^{-1}}$ at $\mu=0.7$ is most pronounced. Moreover, the line also features the steepest decrease in blueshift toward the limb (to ${\rm -48\,m\,s^{-1}}$ at $\mu=0.3$). For \ion{Fe}{I} 6303.5\,\AA, \ion{Sc}{II} 6300.7\,\AA, and \ion{Ti}{I} 6303.8\,\AA, the data distribution and mean curves closely resemble each other. Disk center velocities range around ${\rm -420\,m\,s^{-1}}$ and ${\rm -350\,m\,s^{-1}}$. Toward $\mu=0.7$, the blueshift increases only slightly by a few ${\rm 10\,m\,s^{-1}}$ or remains constant. Toward the solar limb, the blueshift decreases by around ${\rm 150\,m\,s^{-1}}$. Since the \ion{O}{I} 6300.3\,\AA\ line is affected by a line blend, we could not determine its convective blueshift or observe a distinct center-to-limb variation. The direct comparison of the convective blueshift of all seven spectral line cores is displayed in Fig.\,\ref{fig_sec3_bisectors_all03}. In line with \citet{1984SoPh...93..219B}, the convective blueshift of the spectral lines describes a similar center-to-limb variation with a maximum blueshift at heliocentric positions around $\mu=0.7$.

The average convective shifts we measured with LARS are listed in Table \ref{table_sec3_line_comparison}. For each spectral line, the first row specifies the Doppler shifts of the line core in ${\rm m\,s^{-1}}$ for ten heliocentric positions. The respective second row lists the average shift of the mean bisector. For lines with strong asymmetries these two values can differ significantly. Thus, especially for lines with a pronounced C-shape profile like in case of \ion{Fe}{I} 6301.5\,\AA, the velocities can differ by up to ${\rm 190\,m\,s^{-1}}$. In addition to our observations, the table states the synthesized mean line shifts for \ion{Fe}{I} 6301.5\,\AA\ and \ion{Fe}{I} 6302.5\,\AA\ \citep[taken from][]{2011A&A...528A.113D}. 

Our measurements at $\mu=1.0$ are in very good agreement with recent observational and theoretical studies. \citet{Reiners+etal2016} developed a solar FTS atlas in which they report on a synoptic analysis of absolute convective blueshift. Enabled by the large number of spectral lines, they investigated the convective blueshift as a function of line depth. For a normalized line depth of 0.72 like in case of \ion{Fe}{I} 6301.5\,\AA, they yielded an average Doppler shift of ${\rm -300\,m\,s^{-1}}$. Our observed Doppler shift for the mean bisector at disk center amounts to ${\rm -295\,m\,s^{-1}}$. The synthesis obtained a Doppler velocity of ${\rm -307\,m\,s^{-1}}$. In case of \ion{Fe}{I} 6302.5\,\AA, the normalized line depth would suggest an average Doppler shift of around ${\rm -350\,m\,s^{-1}}$. While the synthesis gives a similar value of ${\rm -335\,m\,s^{-1}}$, we measure an average Doppler velocity of ${\rm -295\,m\,s^{-1}}$. For normalized line depth below 0.3, \citet{Reiners+etal2016} find a saturating mean convective blueshift which amounts to around ${\rm -500\,m\,s^{-1}}$. Likewise, for \ion{Si}{I} 6299.6\,\AA\ with a line depth of 0.21, we yield an average blueshift of ${\rm -484\,m\,s^{-1}}$. For very weak lines with a line depth of less than 0.05, the compared atlas lines yield a large scatter around a blueshift of ${\rm -450\,m\,s^{-1}}$.

Further empirical analysis of the convective blueshift have been performed by \citet{1981A&A....96..345D} and \citet{1984SoPh...93..219B}. Investigating a large number of \ion{Fe}{I} lines, both studies brought the convective shift in context with the line depth, excitation potential, and line formation region. For \ion{Fe}{I}\,6301.5\,\AA\ and \ion{Fe}{I}\,6302.5\,\AA, the lower excitation potentials are ${\rm \chi=3.65\,eV}$ and ${\rm \chi=3.69\,eV}$, respectively. But since the distribution of convective lineshifts with respect to the excitation potentials is large, we struggle to find a reference value for comparison. As a rough estimate we can say that our mean bisector shifts of both lines are well within the given range of ${\rm -300\,(\pm200)\,m\,s^{-1}}$ for their excitation potentials.

The formation height of the line minimum at solar disk center was denoted by \citet{1984SoPh...93..219B} to $\log\tau=-2.9$ for \ion{Fe}{I}\,6301.5\,\AA\ and $\log\tau=-2.0$ for \ion{Fe}{I}\,6302.5\,\AA\ {under the assumption of a  one-dimensional atmospheric model}. According to \citet{1985SoPh...99...31B}, the convective shift follows an exponential regression depending on the formation height of the line core. Starting from a vertical Doppler velocity of around ${\rm -500\,m\,s^{-1}}$ at $\log\tau=-1.0$ (or 130\,km above the surface), the blueshift decreases to ${\rm -100\,m\,s^{-1}}$ at $\log\tau=-2.0$ (or 280\,km), crosses the ${\rm 0\,m\,s^{-1}}$ level at $\log\tau=-2.9$ (or 420\,km), and saturates at a supergravitational redshift of around ${\rm +50\,m\,s^{-1}}$ for $\log\tau<-4.0$ (or >\,550\,km). In comparison, our line core measurements show a somewhat larger blueshift by ${\rm 50-100\,m\,s^{-1}}$ \citep[in line with][]{1984SoPh...93..219B}. On the one hand, the bigger blueshift can be caused partially by the parabolic fit of the line core to obtain the shift of the line center. On the other hand, the line formation height could have been overestimated and the line cores are actually formed at slightly larger $\log\tau$ values. {In general, it must be noted that values of formation height steeming from simple one-dimensional models can only give a rough estimate for an idealized average atmosphere but can not account for horizontal distributions of inhomogeneities across the observed spatially averaged region.}

The center-to-limb variation of the convective blueshift has its origin in the vertical and horizontal granular motion and its change with height in the solar atmosphere \citep{1978SoPh...58..243B}. The observational origin was discussed by \citet{1985SoPh...99...31B} in great detail. From disk center toward the solar limb, the formation height increases monotonically. While at disk center the line center of \ion{Fe}{I}\,6301.5\,\AA\ is formed at $\log\tau=-2.9$, the formation height changes to $\log\tau=-3.6$ at $\mu=0.3$. Since the vertical Doppler velocity decreases with atmospheric height, this instantaneously leads to a decrease in blueshift from disk center to the limb. The characteristic trend of the center-to-limb blueshift is explained by the line-of-sight angle onto the height-dependent granular convection. The initial increase in blueshift from $\mu=1.0$ to $\mu=0.7$ is facilitated by the growing impact of the granular horizontal flow toward the observer. The supplementary redshift at the solar limb is achieved by the reversed granulation with increasing atmospheric heights. A reverse in convective motion and temperature stratification of granules and intergranular lanes starts at around 150\,km above the solar surface \citep{1996ASPC..109..155R,2007A&A...461.1163C} or $\log\tau=-2.6$ \citep{1985SoPh...99...31B}. Observations near the solar limb thus capture diverging horizontal flows to a larger extent.

\begin{figure}[htpb]
\includegraphics[width=\columnwidth]{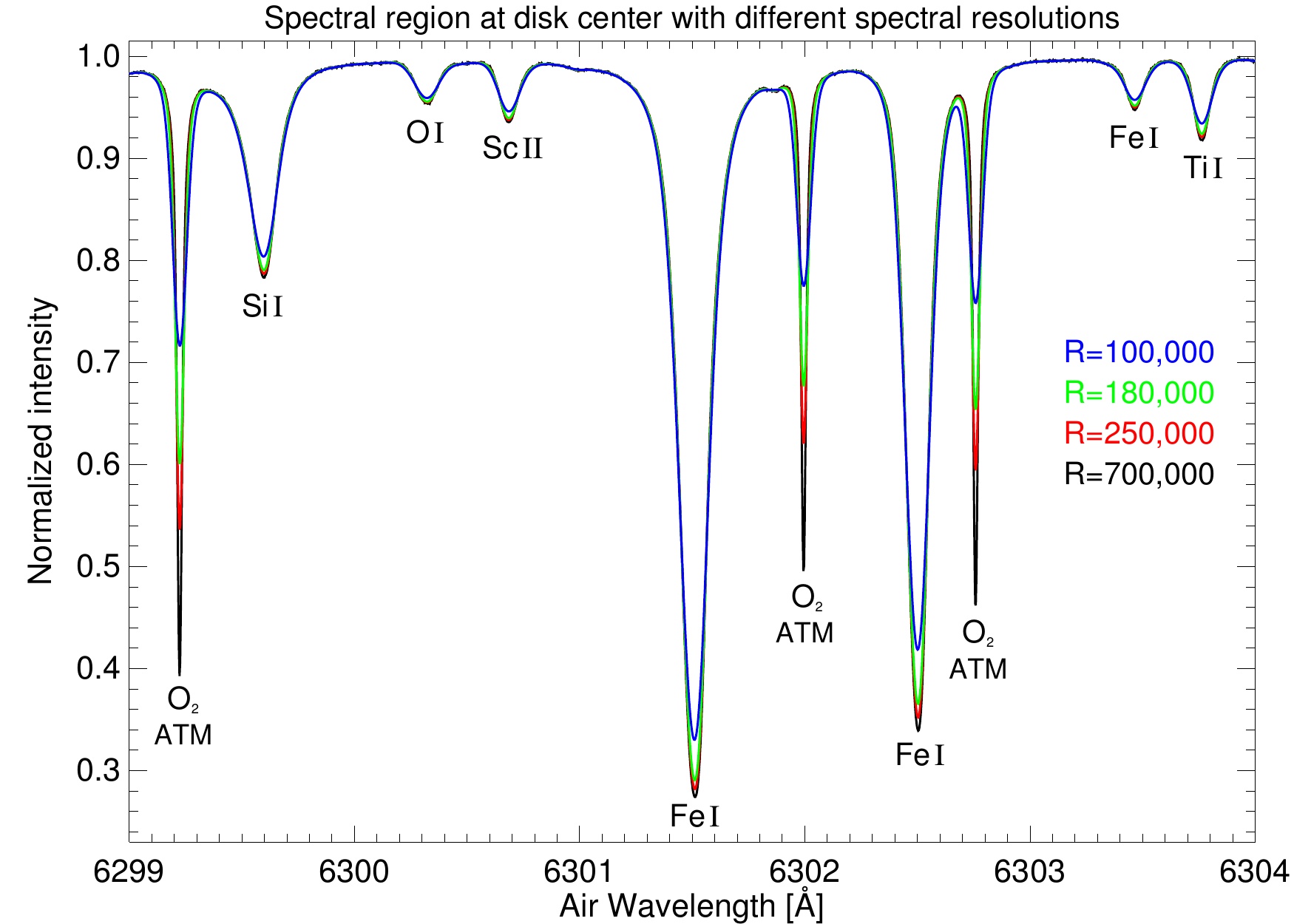}\\[0.25cm]
\includegraphics[width=\columnwidth]{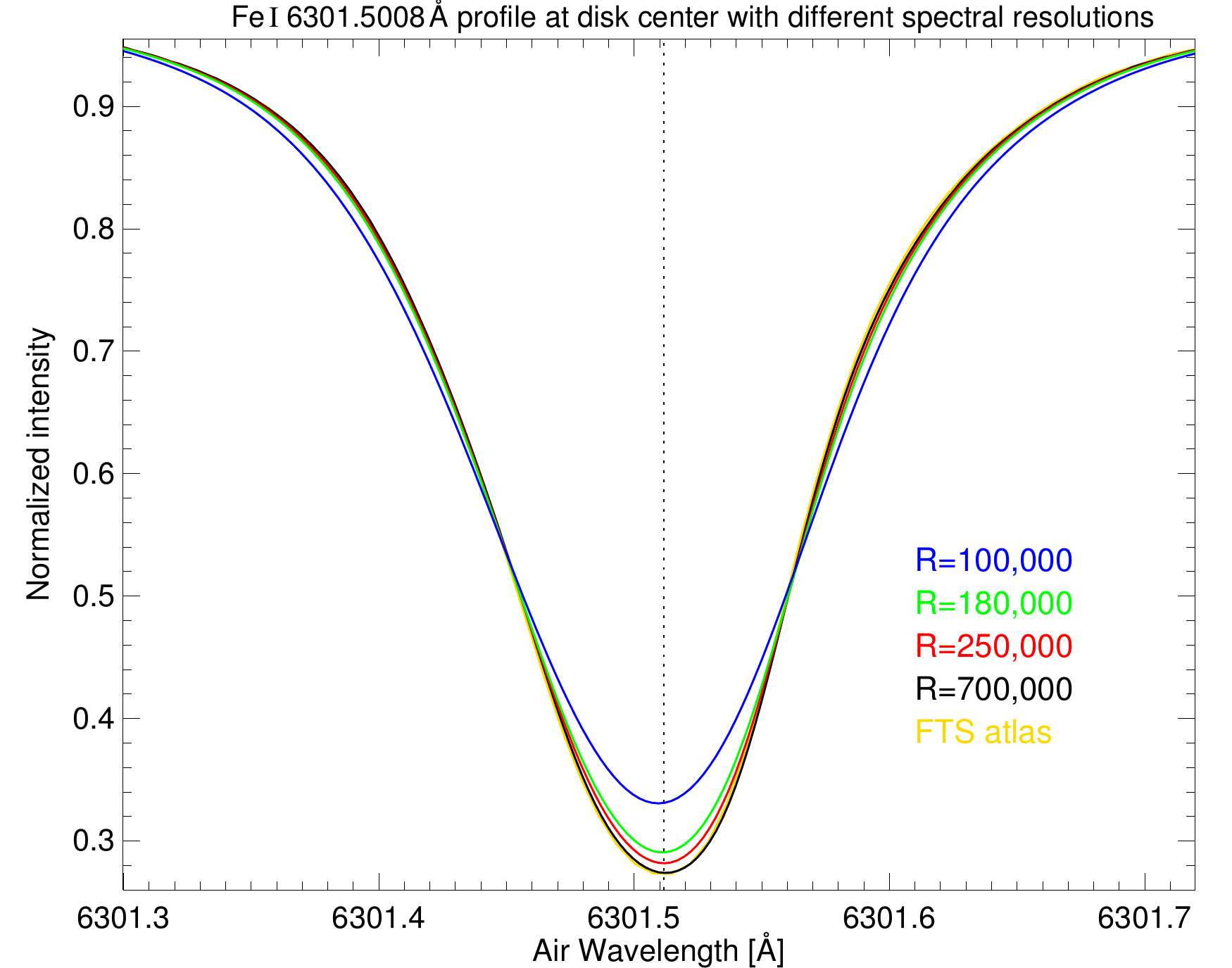}\\[0.25cm]
\includegraphics[width=\columnwidth]{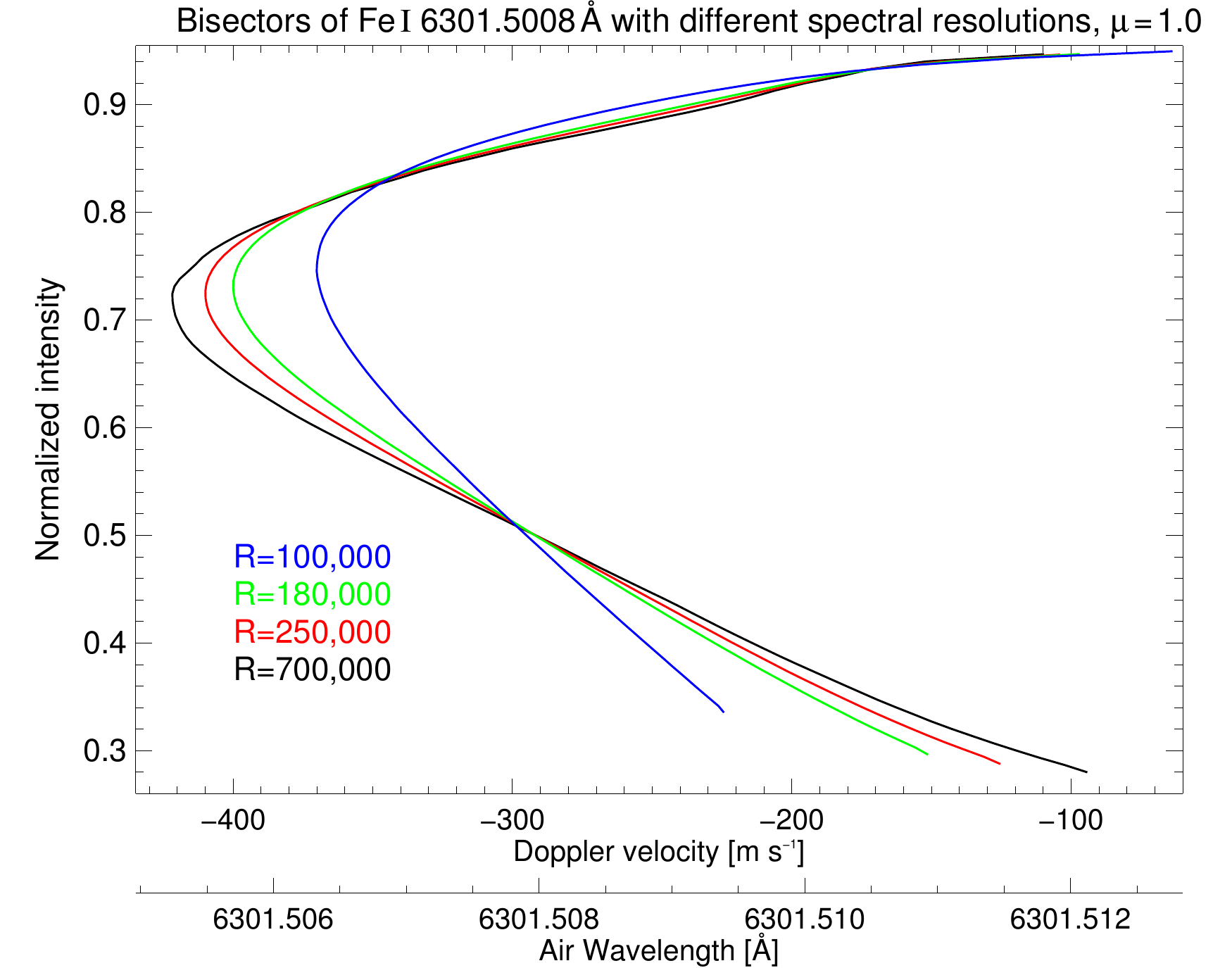}
\caption{Variation of the solar spectrum under different spectral resolutions from 700\,000 (black), 250\,000 (red), 180\,000 (green), to 100\,000 (blue). The three panels show the disk center profile of the full spectral region (top panel), the \ion{Fe}{I}\,6301.5\,\AA\ line profile (center panel), and the \ion{Fe}{I}\,6301.5\,\AA\ line bisector (bottom panel). The FTS atlas profile of \ion{Fe}{I}\,6301.5\,\AA\ is added in yellow. The profiles were normalized to the continuum intensity and plotted against the air wavelength in \AA. Line shifts were translated into Doppler velocities (in ${\rm m\,s^{-1}}$).}
\label{fig_sec3_resolution_profiles}
\end{figure}

\subsection{Convolution to lower spectral resolution}\label{sec3_resolution_comparison}
The observations presented in this work have been performed with LARS and the echelle spectrograph of the VTT. Since the laser frequency comb generates emission modes with a theoretical delta-peak, we can directly monitor the impact of the instrument on the spectral measurement. The observed output profile is the spectral point spread function (PSF) or instrumental profile. LARS yields an approximate Gaussian PSF with a full-width at half-maximum of around ${\rm \Delta\lambda=9\,m\AA}$ at ${\rm \lambda=6302\,\AA}$. This leads to a spectral resolution ${\rm R=\lambda/\Delta\lambda}$ of more than 700\,000. In general, most other spectrographs and spectrometers do not operate at this enormous spectral resolution. To permit a comparison of our observations with data and studies from other instruments, we degraded our results to the spectral properties of several other solar instruments.

To compute spectra with a lower spectral resolution, we had to convolve the original spectrum with the respective degraded instrumental profile.  
Based on the known spectral point spread function of LARS, we calculated normalized Gaussian functions with larger full-width at half-maximum which correspond to the spectral resolutions of 250\,000, 180\,000, and 100\,000. The LARS spectra were then convolved with these Gaussians. We note that the spectral sampling was retained.
The results of the degraded solar spectra are shown in the top panel of Fig.\,\ref{fig_sec3_resolution_profiles}. The black curve displays the original spectrum which was measured with LARS at disk center and normalized to the continuum intensity. By decreasing the spectral resolution to 250\,000 (red curve), 180\,000 (green curve), and 100\,000 (blue curve), the decline in line depth becomes apparent. In case of the narrow telluric lines, the smoothing impact of the convolution is most conspicuous. The depth of the telluric lines halves from the highest to lowest resolution.

The change of the \ion{Fe}{I}\,6301.5\,\AA\ line shape can be seen in the middle panel of Fig.\,\ref{fig_sec3_resolution_profiles}. The line depth decreases from a normalized line minimum intensity of 0.27 at a resolution of ${\rm R=700\,000}$, to 0.33 at ${\rm R=100\,000}$. The full-width at half-maximum of the line increases from 144\,m\AA\ to 161\,m\AA. But most importantly, we notice a shift of the line core to shorter wavelength when degrading the profile to smaller resolutions. Overplotted in the figure, the shape of the FTS atlas profile (yellow curve) is in agreement with the original LARS profile (black curve). The respective change of the \ion{Fe}{I}\,6302.5\,\AA\ line profile is displayed in the top panel of Fig.\,\ref{fig_secA_resolution} in the  appendix.

For a closer inspection of the profile variation, we performed a bisector analysis on the degraded \ion{Fe}{I}\,6301.5\,\AA\ line profiles. The results are shown in the bottom panel of Fig.\,\ref{fig_sec3_resolution_profiles}. With decreasing spectral resolution, the line shape suffers from a significant change in asymmetry. The C-shaped bisector gets less pronounced. Especially the velocity gradient from the maximum blueshift position down to the line minimum gets far less steep. Starting from a spectral resolution of ${\rm R=700\,000}$ (black curve), the maximum blueshift of around ${\rm -420\,m\,s^{-1}}$ is reached at a normalized intensities of 0.72 which then declines to a Doppler shift of around ${\rm -100\,m\,s^{-1}}$ in the line core. At a spectral resolution of ${\rm R=250\,000}$ (red curve), the maximum blueshift of the line wing has decreased to ${\rm -410\,m\,s^{-1}}$ while the line core velocity has grown to around ${\rm -130\,m\,s^{-1}}$. At ${\rm R=180\,000}$ (green curve), the maximum blueshift amounts to ${\rm -400\,m\,s^{-1}}$ at an intensity slightly above 0.73. The Doppler velocity of the line core has changed to around ${\rm -160\,m\,s^{-1}}$. The most noticeable modification of the line bisector happens for the transition to the lowest spectral resolution of ${\rm R=100\,000}$ (blue curve). The maximum blueshift is reached at an intensity level of 0.75 and has a Doppler velocity of ${\rm -370\,m\,s^{-1}}$. The Doppler velocity of the line core has grown to around ${\rm -230\,m\,s^{-1}}$. In comparison to the original line core position, this makes an additional blueshift of ${\rm 130\,m\,s^{-1}}$.

\begin{table*}[htbp]
\caption{Comparison of the center-to-limb variation of the convective shifts for different solar instruments. The Doppler velocities (in ${\rm m\,s^{-1}}$) from the line core and the mean bisector are given for heliocentric parameters $\mu$ from disk center ($\mu=1.0$) to the solar limb ($\mu=0.3$).}
\label{table_sec3_resolution_comparison}
\centering
\begin{tabular}{c | c c | c c c c c c c c c c}
\hline\hline
Spectral line&\multicolumn{2}{c|}{Instrument}& \multicolumn{10}{c}{Heliocentric parameter $\mu = \cos\theta$}\\ 
and part & Name&Resolution&1.0&0.95&0.9&0.85&0.8&0.7&0.6&0.5&0.4&0.3\\ 
\hline
\ion{Fe}{I} 6301.5008\,\AA&LARS&700\,000&$-107$&$-74$&$-81$&$-99$&$-82$&$-101$&$-64$&$-18$&$+33$&$+85$\\
line core &ViSP,\,GFPI&250\,000&$-135$&$-98$&$-104$&$-123$&$-103$&$-119$&$-81$&$-32$&$+23$&$+79$\\
 &--&180\,000&$-159$&$-119$&$-125$&$-144$&$-121$&$-136$&$-97$&$-45$&$+12$&$+72$\\
 &VTF&100\,000&$-228$&$-182$&$-186$&$-202$&$-175$&$-186$&$-146$&$-87$&$-25$&$+46$\\
\hline
\ion{Fe}{I} 6301.5008\,\AA&LARS&700\,000&$-295$&$-244$&$-251$&$-269$&$-246$&$-271$&$-255$&$-206$&$-156$&$-87$\\
mean bisector &ViSP,\,GFPI&250\,000&$-297$&$-245$&$-253$&$-271$&$-248$&$-273$&$-258$&$-208$&$-158$&$-89$\\
 &--&180\,000&$-299$&$-246$&$-254$&$-273$&$-249$&$-275$&$-260$&$-210$&$-160$&$-91$\\
 &VTF&100\,000&$-304$&$-253$&$-261$&$-280$&$-256$&$-280$&$-267$&$-217$&$-168$&$-99$\\
\hline
 \ion{Fe}{I} 6302.4932\,\AA&LARS&700\,000&$-164$&$-139$&$-163$&$-195$&$-178$&$-204$&$-173$&$-115$&$-47$&$+33$\\
line core &ViSP,\,GFPI&250\,000&$-195$&$-165$&$-179$&$-214$&$-193$&$-218$&$-188$&$-130$&$-62$&$+20$\\
 &--&180\,000&$-218$&$-185$&$-197$&$-230$&$-206$&$-231$&$-200$&$-141$&$-74$&$+10$\\
 &VTF&100\,000&$-276$&$-233$&$-242$&$-268$&$-242$&$-263$&$-236$&$-175$&$-111$&$-25$\\
\hline
\ion{Fe}{I} 6302.4932\,\AA&LARS&700\,000&$-295$&$-249$&$-261$&$-281$&$-260$&$-286$&$-274$&$-221$&$-170$&$-94$\\
mean bisector &ViSP,\,GFPI&250\,000&$-297$&$-250$&$-261$&$-282$&$-260$&$-287$&$-275$&$-222$&$-171$&$-95$\\
 &--&180\,000&$-298$&$-250$&$-261$&$-284$&$-260$&$-288$&$-276$&$-223$&$-172$&$-95$\\
 &VTF&100\,000&$-301$&$-252$&$-264$&$-286$&$-262$&$-289$&$-278$&$-224$&$-173$&$-94$\\
\hline
\end{tabular}
\tablefoot{To reduce the spectral resolution ($\lambda_0/\Delta\lambda$), the spectral profiles from LARS were convolved with gaussian point spread functions corresponding to the given spectral resolution. The line core velocity and mean bisector were then obtained like explained in Table \ref{table_sec3_line_comparison}.}
\end{table*}

We have emphasized the importance and effect of the spectral resolution of the spectroscopic instrument on the obtained Doppler velocity. To allow other studies a direct comparison with our results, we have selected spectral resolutions (at ${\rm \lambda=6302\,\AA}$) which are representative for most of the current and future solar spectrometers and spectrographs. Of special interest will be the Visible Tunable Filter \citep[VTF,][]{2012SPIE.8446E..77K,2014SPIE.9147E..0ES} with a spectral resolution of ${\rm R=100\,000}$ and the Visible Spectro-Polarimeter \citep[ViSP,][]{2012SPIE.8446E..6XD} with ${\rm R=250\,000}$. Both will be first-light instruments of the 4\,m-class Daniel K. Inouye Solar Telescope \citep[DKIST,][]{2010AN....331..609K,2012ASPC..463..377R}, that is currently under construction at the Haleakala Observatory on Maui. Moreover, we aim to encourage observational comparisons with the current prime spectrometers which operate in the visible range. These are 
the GREGOR Fabry-P{\'e}rot Interferometer \citep[GFPI,][]{2013OptEn..52h1606P} with ${\rm R=250\,000}$,
the Interferometric BIdimensional Spectropolarimeter \citep[IBIS,][]{2006SoPh..236..415C,2008A&A...481..897R} with ${\rm R=280\,000}$,
the CRisp Imaging SpectroPolarimeter \citep[CRISP,][]{2006A&A...447.1111S} with ${\rm R=120\,000}$ \citep{2015A&A...573A..40D},
the Triple Etalon SOlar Spectrometer \citep[TESOS,][]{1998A&A...340..569K,2002SoPh..211...17T} with ${\rm R=300\,000}$,
and the HELioseismological Large Regions Interferometric DEvice \citep[HELLRIDE,][]{2011A&A...535A..83S} with ${\rm R=35\,000}$.

\begin{figure}[htpb]
\includegraphics[width=\columnwidth]{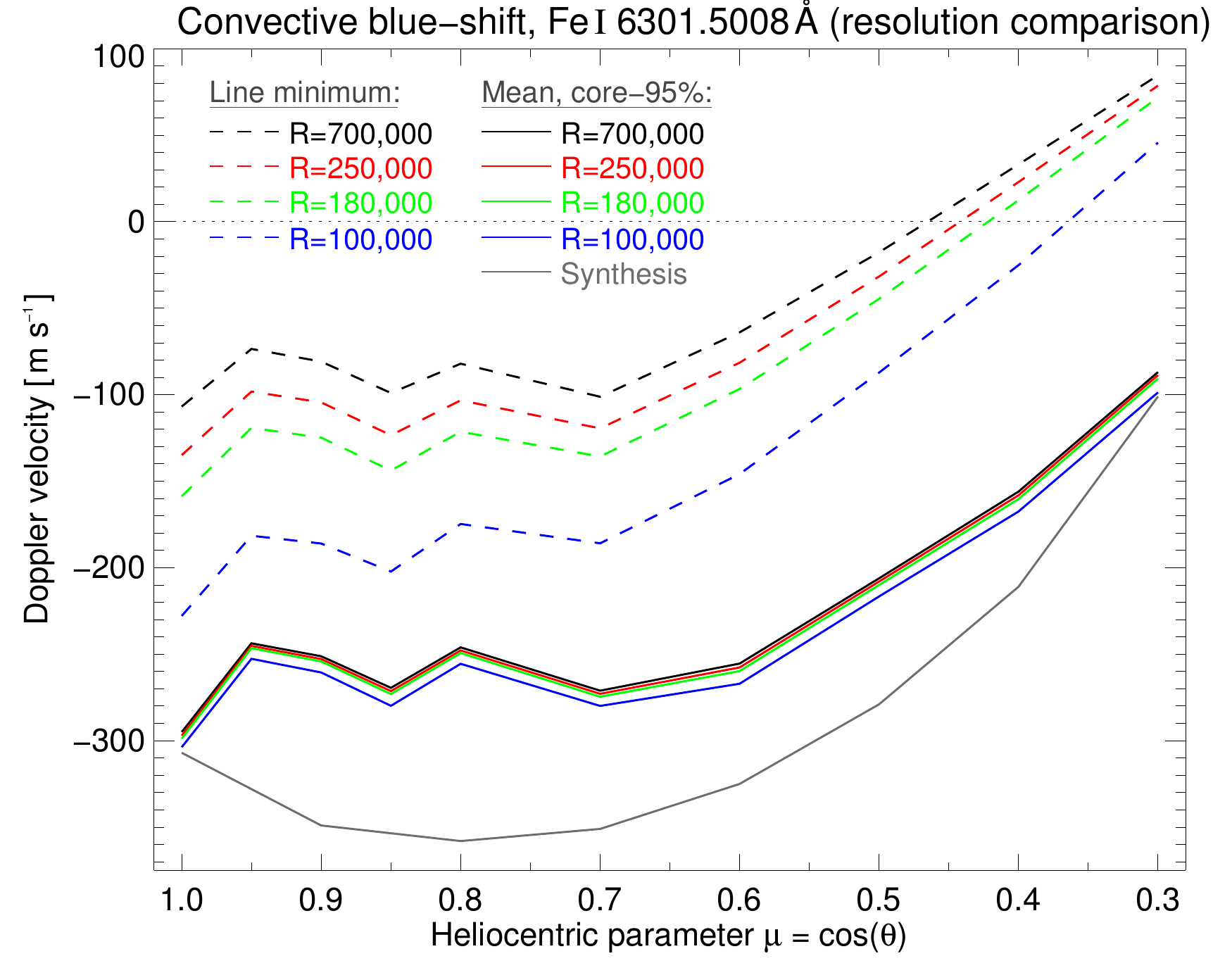}
\caption{Center-to-limb variation of the convective shift from \ion{Fe}{I}\,6301.5\,\AA\ under different spectral resolutions. The Doppler velocities (in ${\rm m\,s^{-1}}$) are plotted against the heliocentric parameter from disk center ($\mu=1.0$) toward the solar limb ($\mu=0.3$). The dashed curves were determined for the line core. The solid lines were computed for the average bisector from line core to an upper threshold of 95\% of the continuum intensity. The respective synthesis from \citet{2011A&A...528A.113D} is depicted in gray.}
\label{fig_sec3_resolution_clv}
\end{figure}

To complete this study, we applied the convolution to lower spectral resolution on the entire analysis of the center-to-limb convective blueshift (Section \ref{sec3_clv}). All original spectra were degraded to a resolution of 250\,000, 180\,000, and 100\,000. For all spectral lines, the Doppler shifts were calculated for the entire line (mean bisector) or the line core (parabolic fit to the line center). In Table \ref{table_sec3_resolution_comparison}, we list the results for \ion{Fe}{I}\,6301.5\,\AA\ and  \ion{Fe}{I}\,6302.5\,\AA. All Doppler velocities are given in ${\rm m\,s^{-1}}$ with negative values indicating blueshifts. For a better illustration of the effect, we plot the values and resultant center-to-limb curves for \ion{Fe}{I}\,6301.5\,\AA\ in Fig.\,\ref{fig_sec3_resolution_clv} (and for \ion{Fe}{I}\,6302.5\,\AA\ in Fig.\,\ref{fig_secA_resolution} in the  appendix). The original center-to-limb variation of the convective blueshift at a resolution of 700\,000 is shown by the black curves. The solid curve represents the average line shift, the dashed curve displays the line core shift. As indicated by the dashed curves for 250\,000 (red), 180\,000 (green), and 100\,000 (blue), the decrease in spectral resolution strongly affects the convective blueshift obtained from the line core. At the extreme, the Doppler velocities close to the solar disk center can change by up to ${\rm 130\,m\,s^{-1}}$. For heliocentric positions approaching the solar limb, the maximum difference declines to around ${\rm 40\,m\,s^{-1}}$. This variation is due to the effective change of the line asymmetry (compare Fig.\,\ref{fig_sec3_bisectorsFe}). 
{Contrary to the severe shift of the line core position, the average value for the entire line bisector remains almost unaffected by a change in spectral resolution.} As indicated by the solid curves in Fig.\,\ref{fig_sec3_resolution_clv}, the convective blueshift increases by only up to ${\rm 10\,m\,s^{-1}}$ from the maximum (black) to the minimum (blue) resolution, independently from the observed position on the solar disk. 

\section{Conclusions}\label{sec4_conclusions}
We performed systematic observations of the convective blueshift of the Sun with the highest spectral resolution that is currently attainable in solar physics. Our measurements sampled specific heliocentric positions from disk center to near the solar limb which permitted us to investigate the center-to-limb variation of the spectral line profiles in the 6302\,\AA\ region. The accurate bisector analysis yielded comprehensive information about the line asymmetries and convective shifts caused by the solar convection and atmospheric granular motion. We presented exact wavelength positions and convective Doppler shifts at an instrumental accuracy of ${\rm 1\,m\,s^{-1}}$. On this basis, we revealed a systematic wavelength offset of ${\rm -0.8\,m\AA}$ at 6302\,\AA\ (about ${\rm -40\,m\,s^{-1}}$) of the FTS atlas \citep{1999SoPh..184..421N} against LARS data.

Major deviations from numerically simulated values like indicated in Fig.\,\ref{fig_sec3_clv_linecore} have to originate from the superimposing solar activity. Whereas theoretical syntheses \citep[e.g.,][]{2011A&A...528A.113D} base on an undisturbed invariable atmosphere, observations capture all solar dynamics at all time scales. To a large degree, we overcame the issue of $p$-mode oscillation and other short-term variations by averaging the observed 20\,min time series and observing several sequences per disk position. However, due to the plethora of acoustic mode periods and their interference, a thorough suppression of all temporal variations would require longer time series and a significant increase of the statistical sample. The existence of larger convective cells, like supergranulation (30\,Mm) or mesogranulation (5\,Mm), with lifespans of several hours and horizontal flow speeds of a few $100\,{\rm m\,s^{-1}}$ can lead to substantial Doppler shifts which increase toward the solar limb. Temporal averaging would work only at the middle-term scale. Assuming a stable stationary supergranule, the solar rotation velocity of $2\,{\rm km\,s^{-1}}$ would change the longitudinal position of the supergranule by $10\,{\rm \arcsec\,h^{-1}}$ at disk center, and way less at the solar limb. It would thus take several hours to scan and average over one standard supergranule. We mastered this issue by performing an elliptically spatial scanning around the nominal heliocentric position. At the cost of positional accuracy, this enabled the instantaneous averaging over the spatial area and supergranular dynamics.

On the long-term scale, it has been argued that the line profiles and convective blueshift vary with the solar activity cycle \citep{1982Natur.297..208L,1986A&A...158..275C}. This seems convincing since the magnetic field activity changes with the cycle, which in turn influences the convection. Due to the suppression of convective motion, magnetic field regions in general exhibit a weaker convective blueshift. Though, the cyclical variation of the convective blueshift is expected to be of the order of a few ${\rm \,m\,s^{-1}}$. It requires the optimum repetition accuracy of ${\rm 1\,m\,s^{-1}}$ over a time scale of many years. Only consequent repetitive observations with LARS will yield the long-term variation of the absolute convective blueshift of the Sun.

In this work, we used reference wavelengths from NIST to compute the Doppler shifts of spectral lines observed with LARS. To further improve the validity of the obtained Doppler velocities in future studies with LARS, we aim to include own high-precision measurements of reference wavelengths using hollow cathode lamps.

In a broader sense, our observational insights based on measurements with unprecedented quality will foster the improvement of physical models and "realistic" numerical simulations \citep{Pereira+etal2013,Scott+etal2014b} of the quiet solar atmosphere. Beyond, the Sun serves as the main benchmark for other stars. Models of the solar atmosphere are fundamental for the interpretation of stellar spectra and lead to a better understanding of their atmospheres. By implication, such models of stellar atmospheres must be able to reproduce the observed spectral line profiles of the Sun and their center-to-limb variation.

\begin{acknowledgements} 
We thank all colleagues at the Kiepenheuer Institute for Solar Physics, at Menlo Systems GmbH, and at the Max Planck Institute of Quantum Optics who worked on the development of the instrument. We are deeply grateful to Hans-Peter Doerr for the preparation of the LARS prototype and his guidance, support, and helpfulness for the instrumental practice. We thank Rolf Schlichenmaier and Nazaret Bello Gonz\'alez as co-investigators, and Thomas Keller, Karin Gerber, Oliver Wiloth, Roberto Simoes, and Thomas Sonner for their assistance during the observation campaigns. The development of the LARS instrument and the operation of the Vacuum Tower Telescope at the Observatorio del Teide on Tenerife were performed by the Kiepenheuer Institute for Solar Physics Freiburg, which is a public law foundation of the State of Baden-W\"urttemberg. This work is part of a Post-Doc project funded by the Deutsche Forschungsgemeinschaft (DFG, Ref.-No. Schm-1168/10). The initial astro-comb project for the VTT had been funded by the Leibniz-Gemeinschaft through the "Pakt f\"ur Forschung und Innovation". Finally, we are obliged to Catherine Fischer for her fruitful comments on the manuscript.
\end{acknowledgements}
\bibliographystyle{aa} 
\bibliography{LARS.bib}

\clearpage
\begin{appendix}
\section{Additional figures}

\begin{figure}[htpb]
\includegraphics[width=0.96\columnwidth]{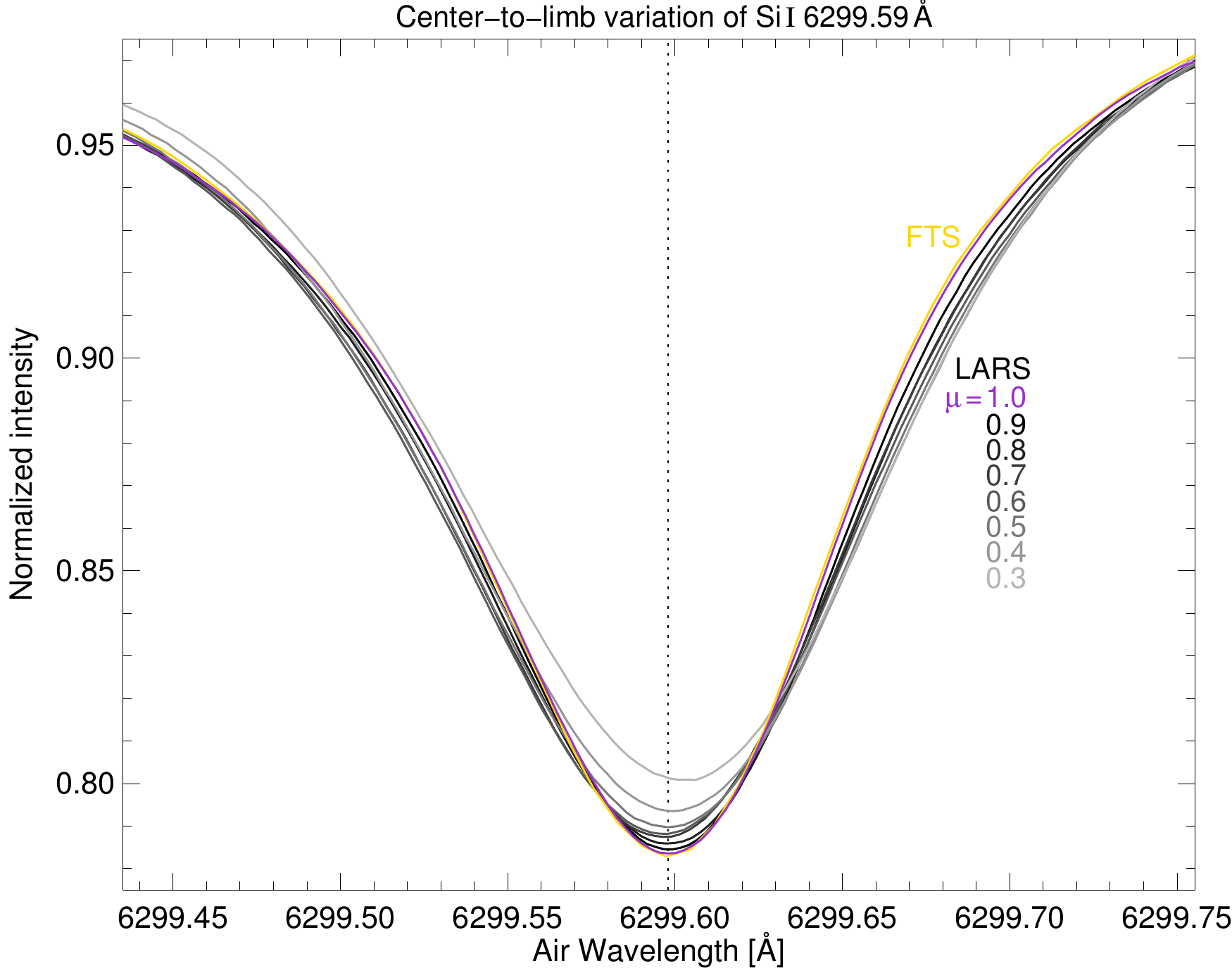}\\[0.1cm]
\includegraphics[width=0.96\columnwidth]{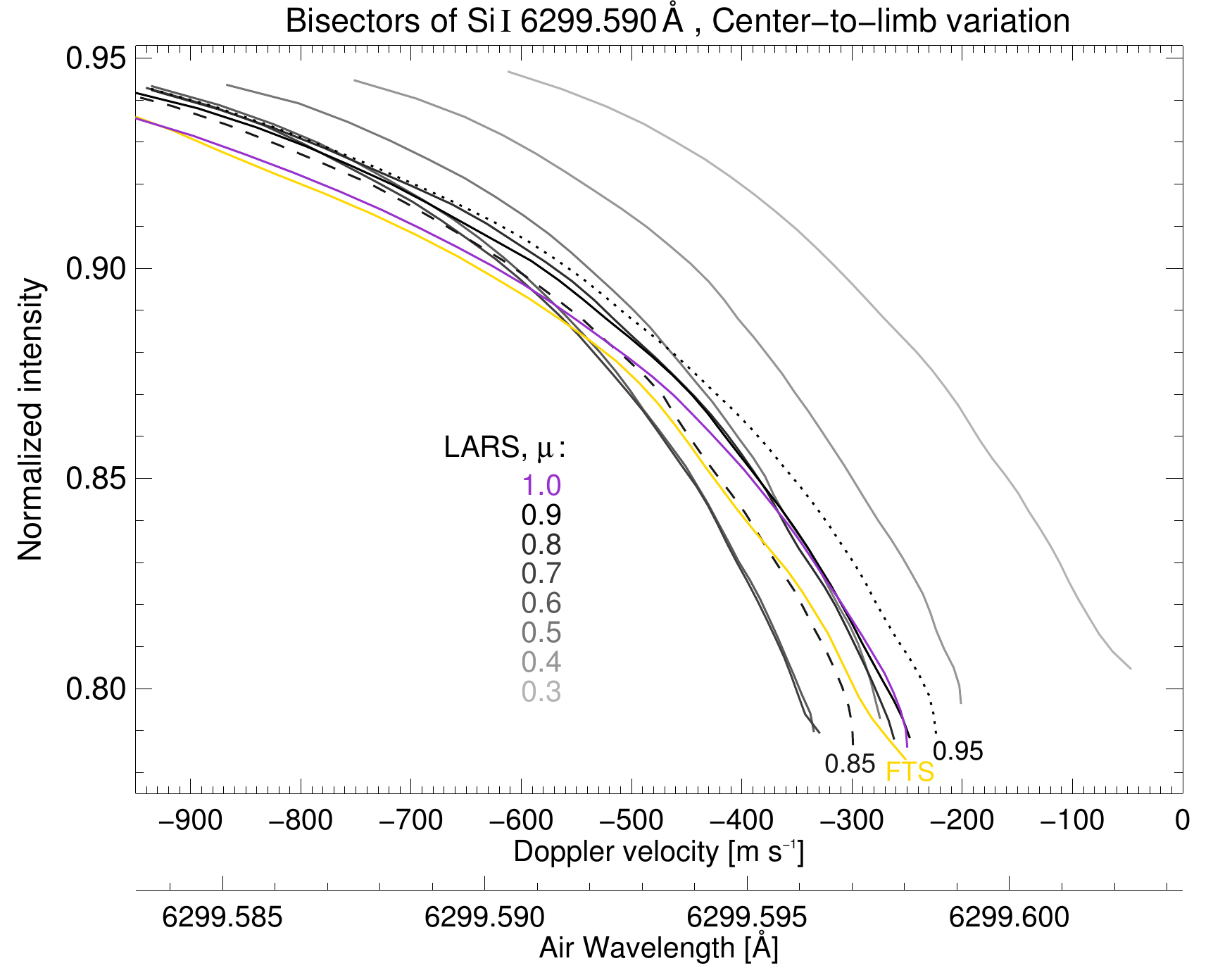}\\[0.1cm]
\includegraphics[width=0.96\columnwidth]{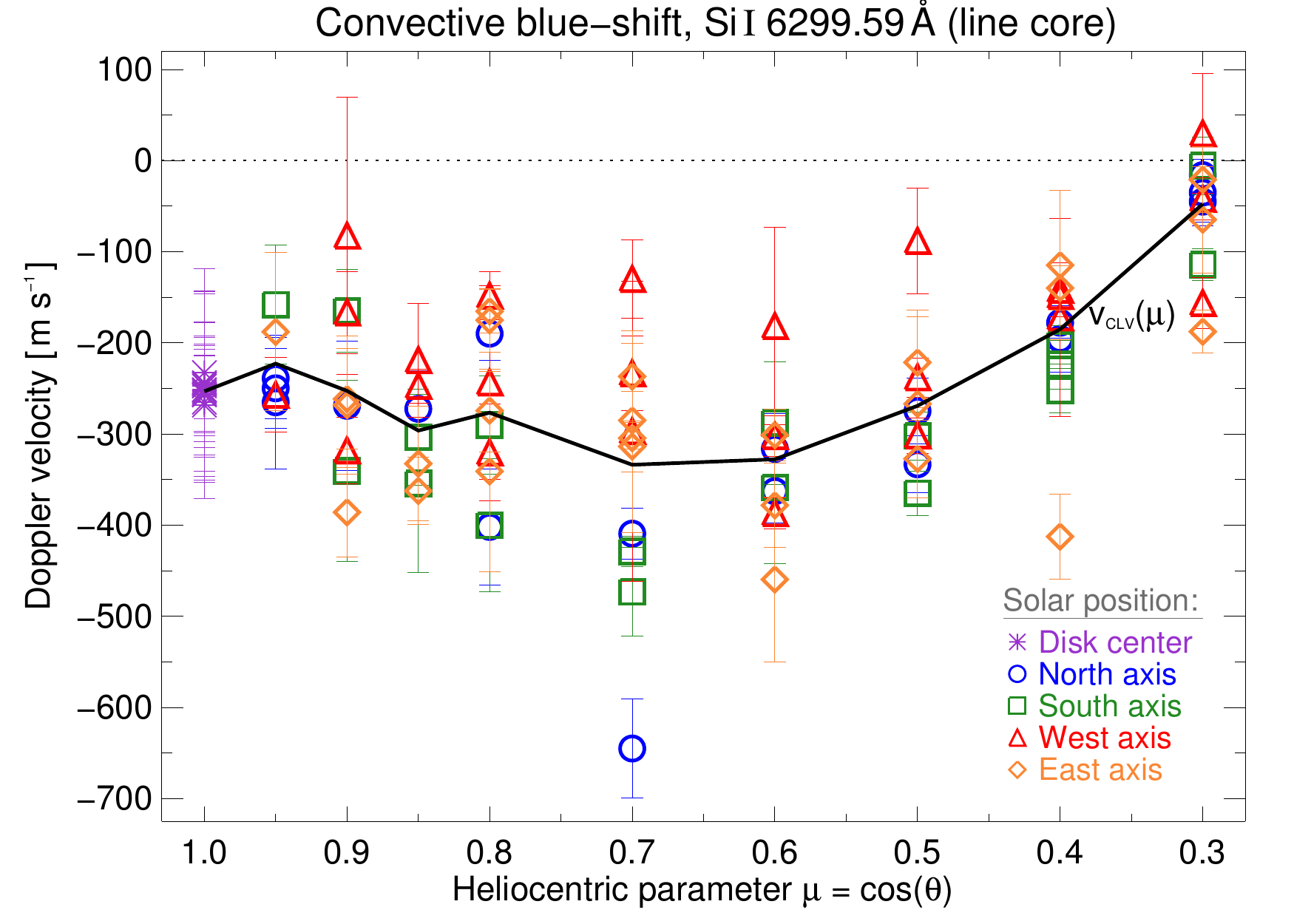}
\caption{Center-to-limb variation for the \ion{Si}{I}\,6299.6\,\AA\ line. Top and middle panel: Spectral profile and bisectors from disk center (purple) toward the solar limb (decreasing gray scale), with FTS atlas comparison (yellow). Intensities were normalized to the spectral continuum. Air wavelengths (in \AA) were translated into Doppler velocities (in ${\rm m\,s^{-1}}$). Bottom panel: Line core velocities plotted against the heliocentric position $\mu$, with colored symbols indicating the solar axis. The average trend is drawn in black as a thick, solid line (compare Fig.\,\ref{fig_sec3_clv_linecore}).}
\label{fig_secA_analysis_SiI62996}
\end{figure}

\begin{figure}[htpb]
\vspace{1.1cm}
\includegraphics[width=0.96\columnwidth]{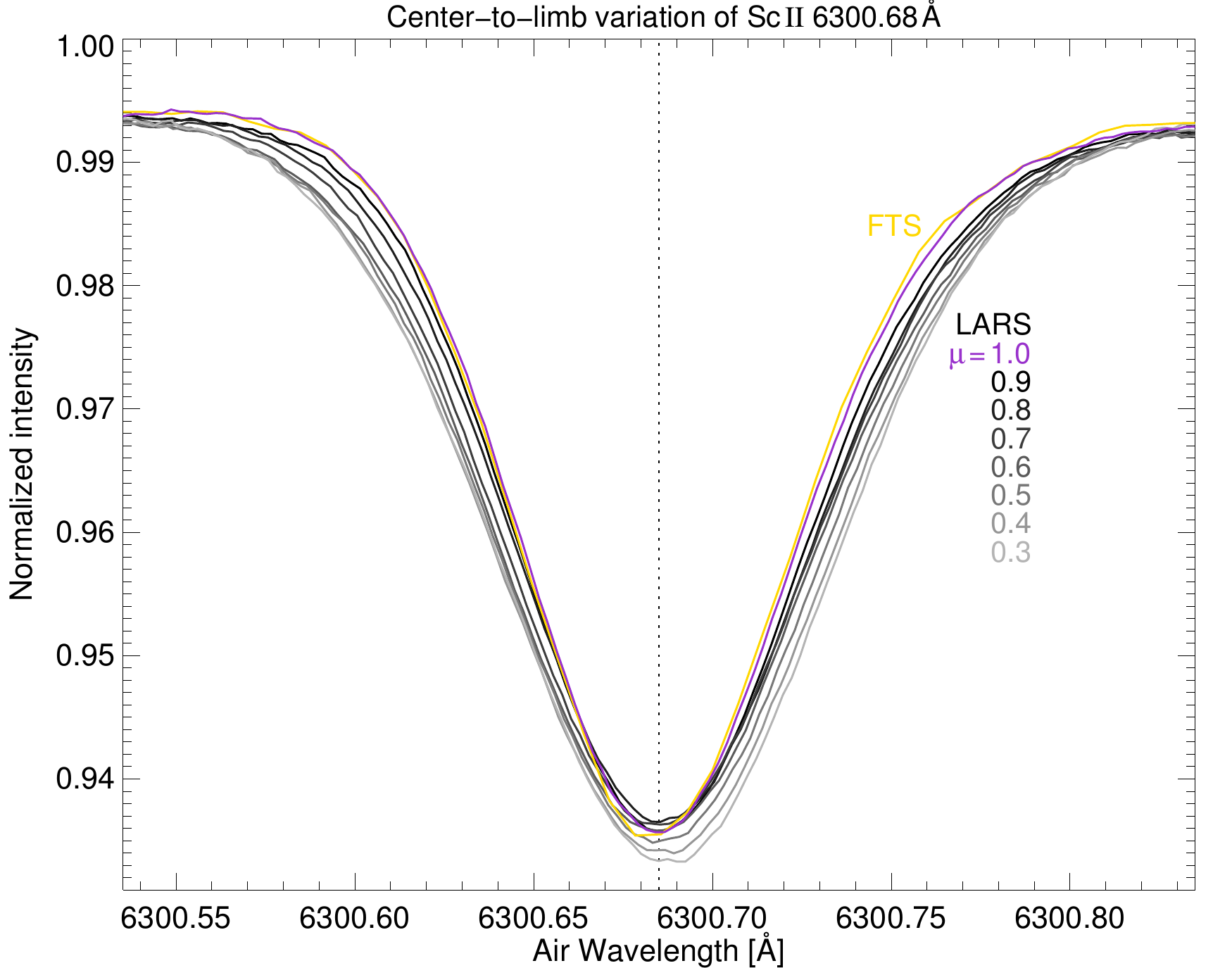}\\[0.1cm]
\includegraphics[width=0.96\columnwidth]{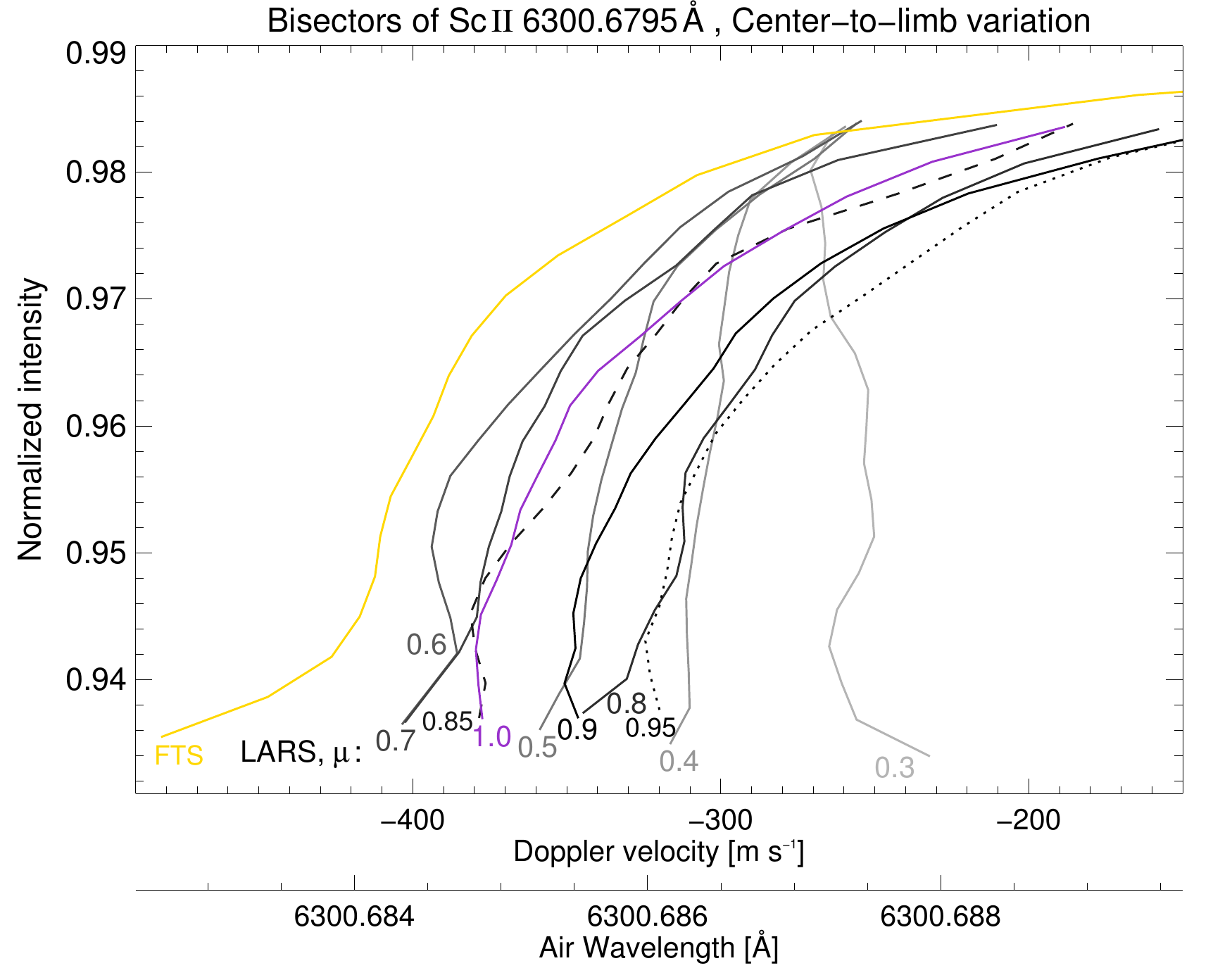}\\[0.1cm]
\includegraphics[width=0.96\columnwidth]{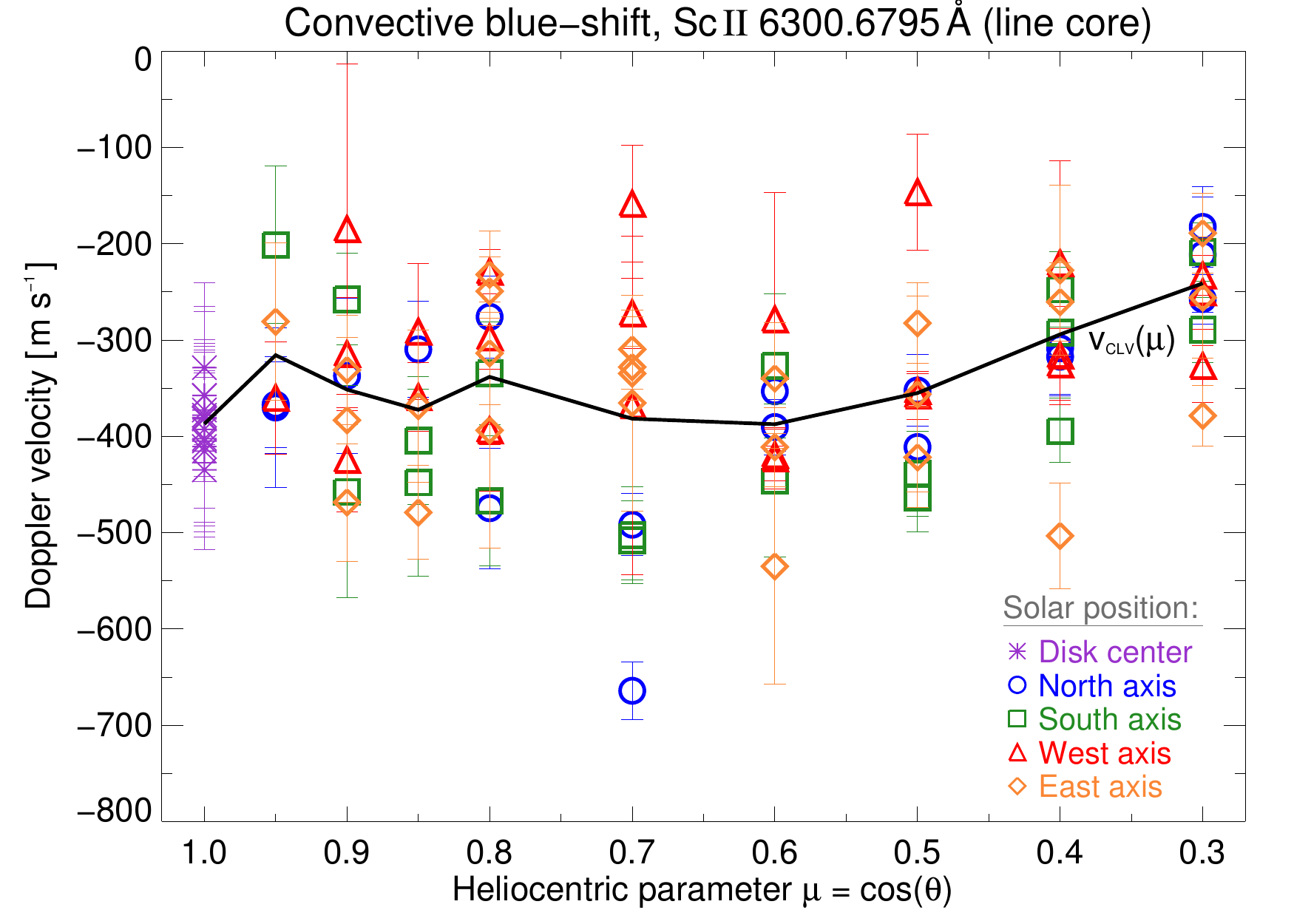}
\caption{Center-to-limb variation for the \ion{Sc}{II}\,6300.7\,\AA\ line. Top and middle panel: Spectral profile and bisectors from disk center (purple) toward the solar limb (decreasing gray scale), with FTS atlas comparison (yellow). Intensities were normalized to the spectral continuum. Air wavelengths (in \AA) were translated into Doppler velocities (in ${\rm m\,s^{-1}}$). Bottom panel: Line core velocities plotted against the heliocentric position $\mu$, with colored symbols indicating the solar axis. The average trend is drawn in black as a thick, solid line.}
\label{fig_secA_analysis_ScII63007}
\end{figure}

\begin{figure}[htpb]
\includegraphics[width=0.97\columnwidth]{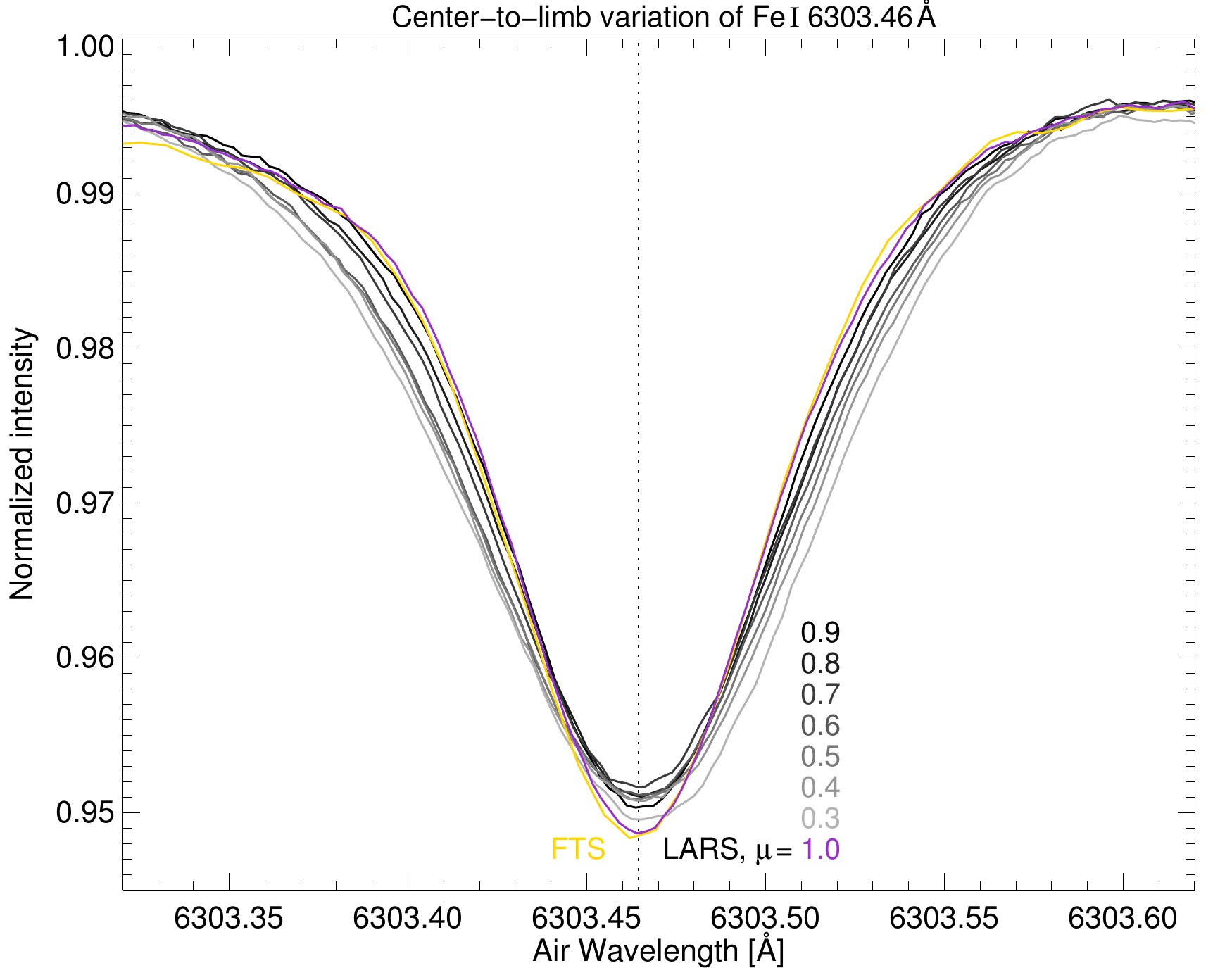}\\[0.1cm]
\includegraphics[width=0.97\columnwidth]{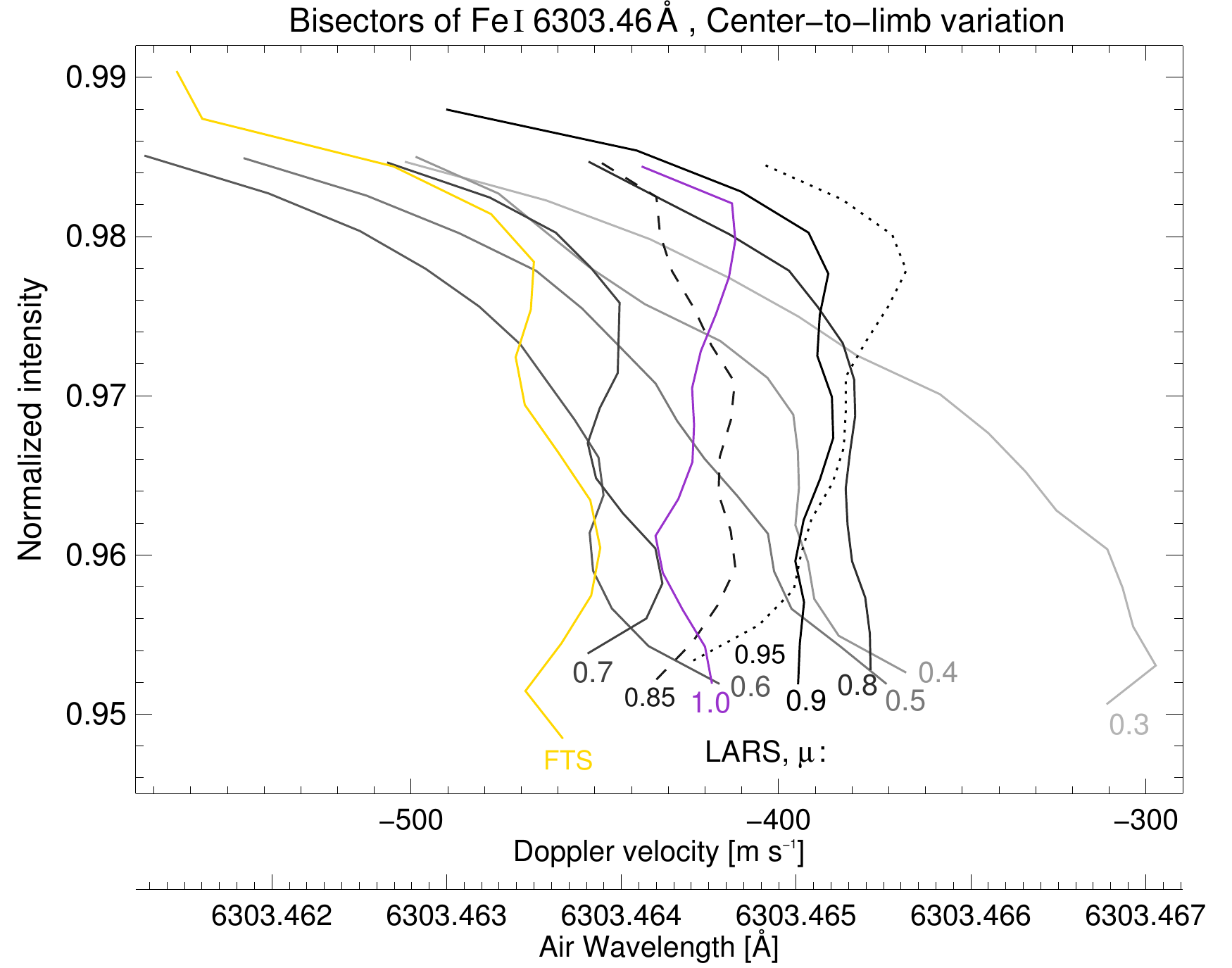}\\[0.1cm]
\includegraphics[width=0.97\columnwidth]{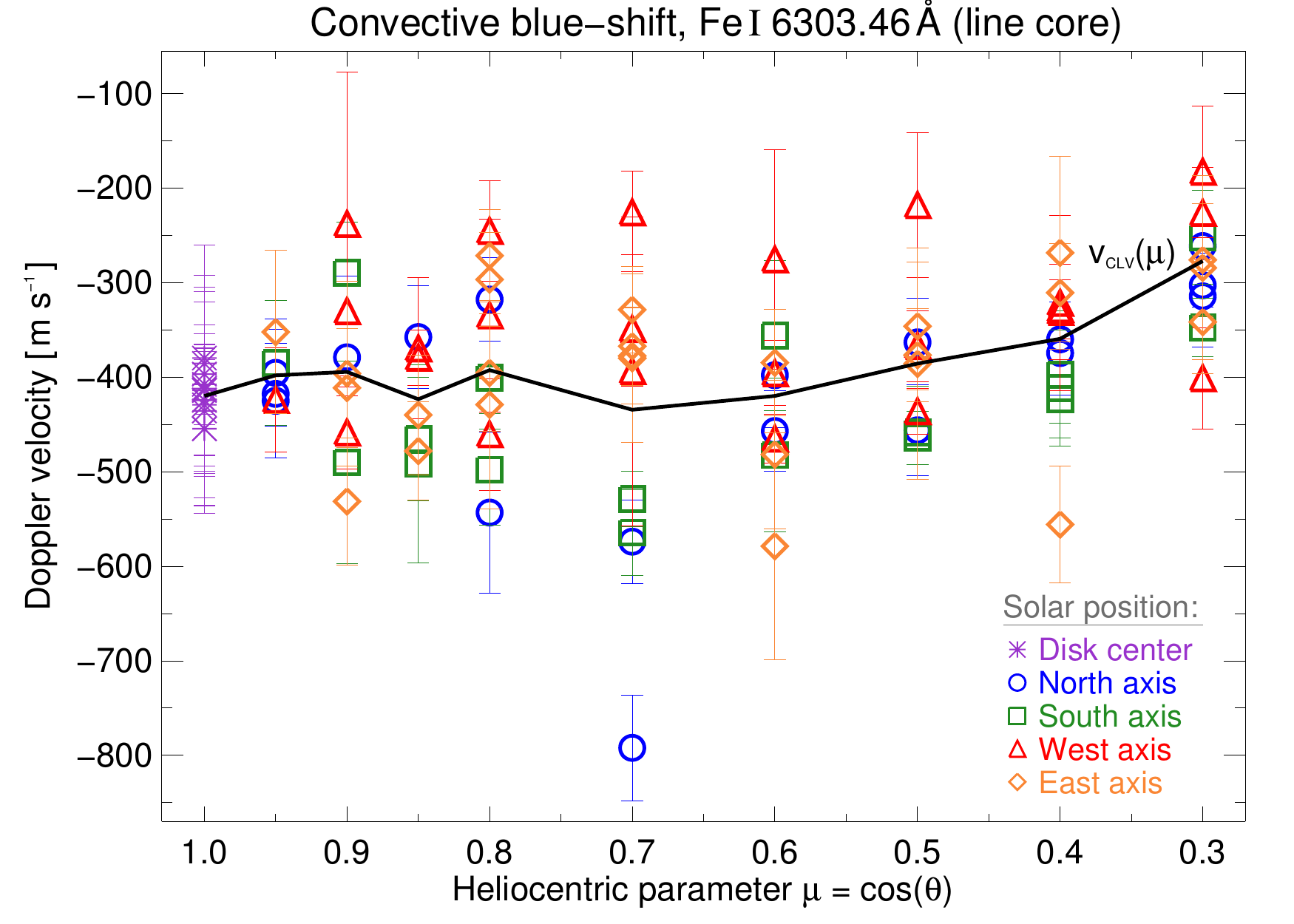}
\caption{Center-to-limb variation for the \ion{Fe}{I}\,6303.5\,\AA\ line. Top and middle panel: Spectral profile and bisectors from disk center (purple) toward the solar limb (decreasing gray scale), with FTS atlas comparison (yellow). Intensities were normalized to the spectral continuum. Air wavelengths (in \AA) were translated into Doppler velocities (in ${\rm m\,s^{-1}}$). Bottom panel: Line core velocities plotted against the heliocentric position $\mu$, with colored symbols indicating the solar axis. The average trend is drawn in black as a thick, solid line.}
\label{fig_secA_analysis_FeI63035}
\end{figure}

\begin{figure}[htpb]
\includegraphics[width=0.97\columnwidth]{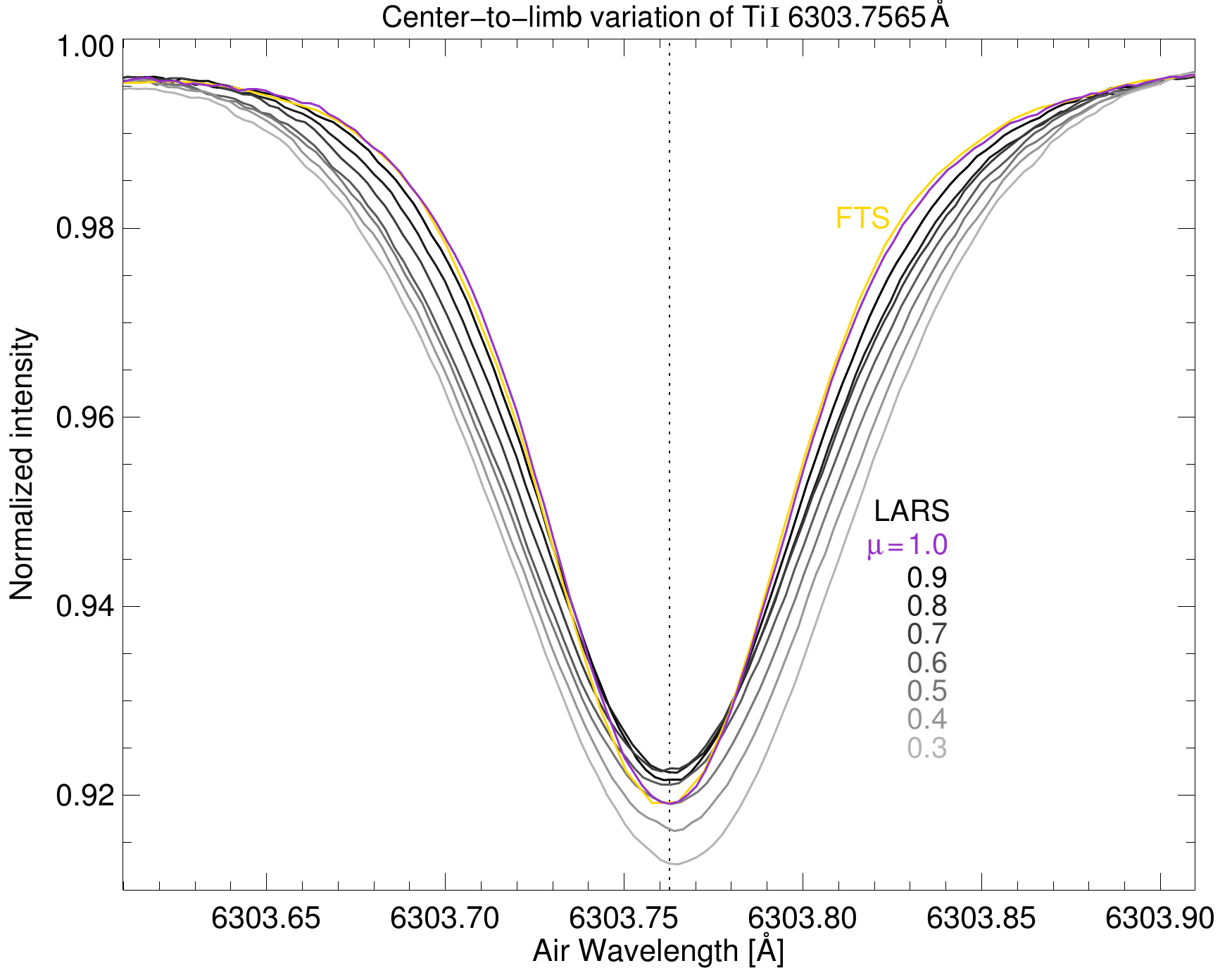}\\[0.1cm]
\includegraphics[width=0.97\columnwidth]{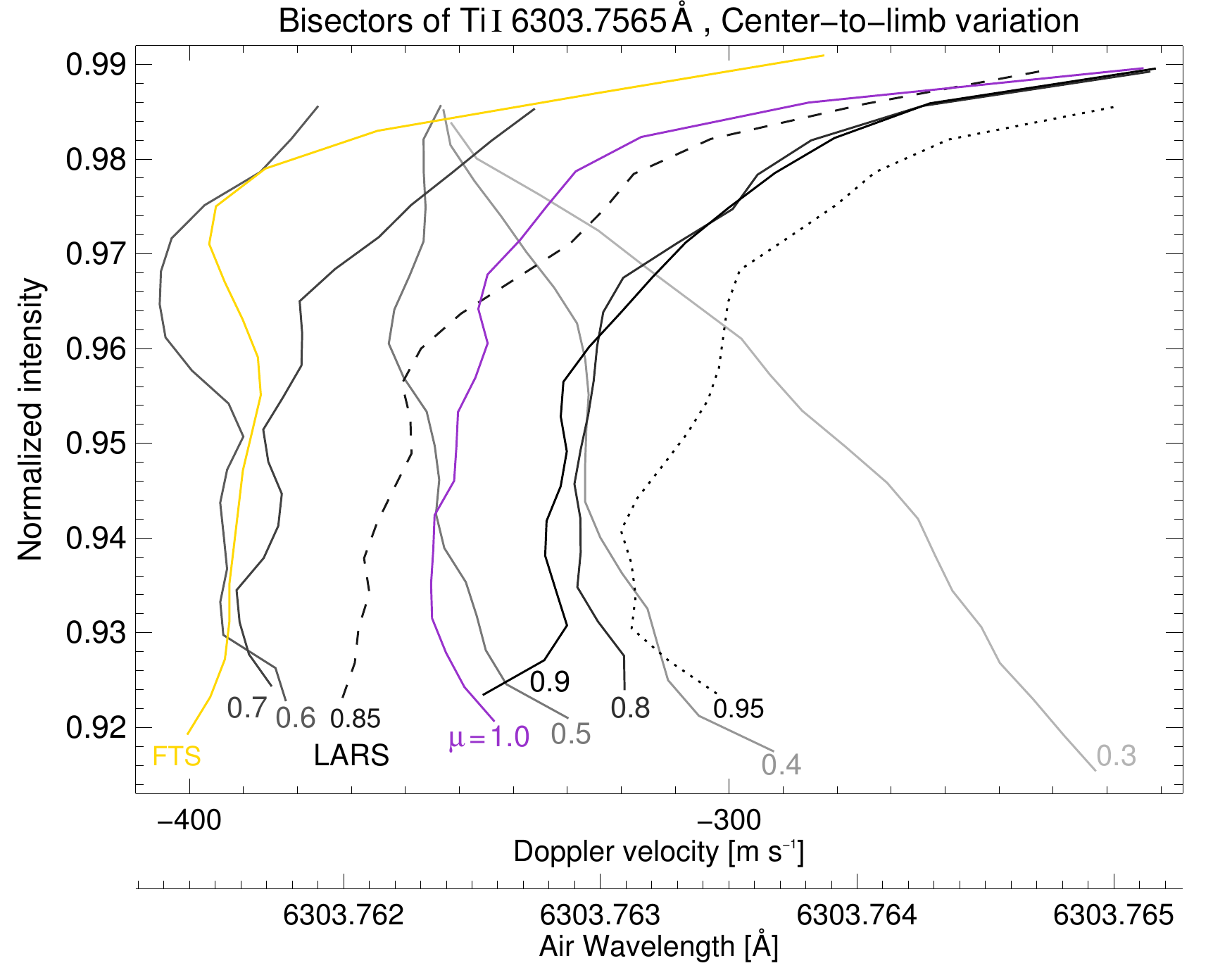}\\[0.1cm]
\includegraphics[width=0.97\columnwidth]{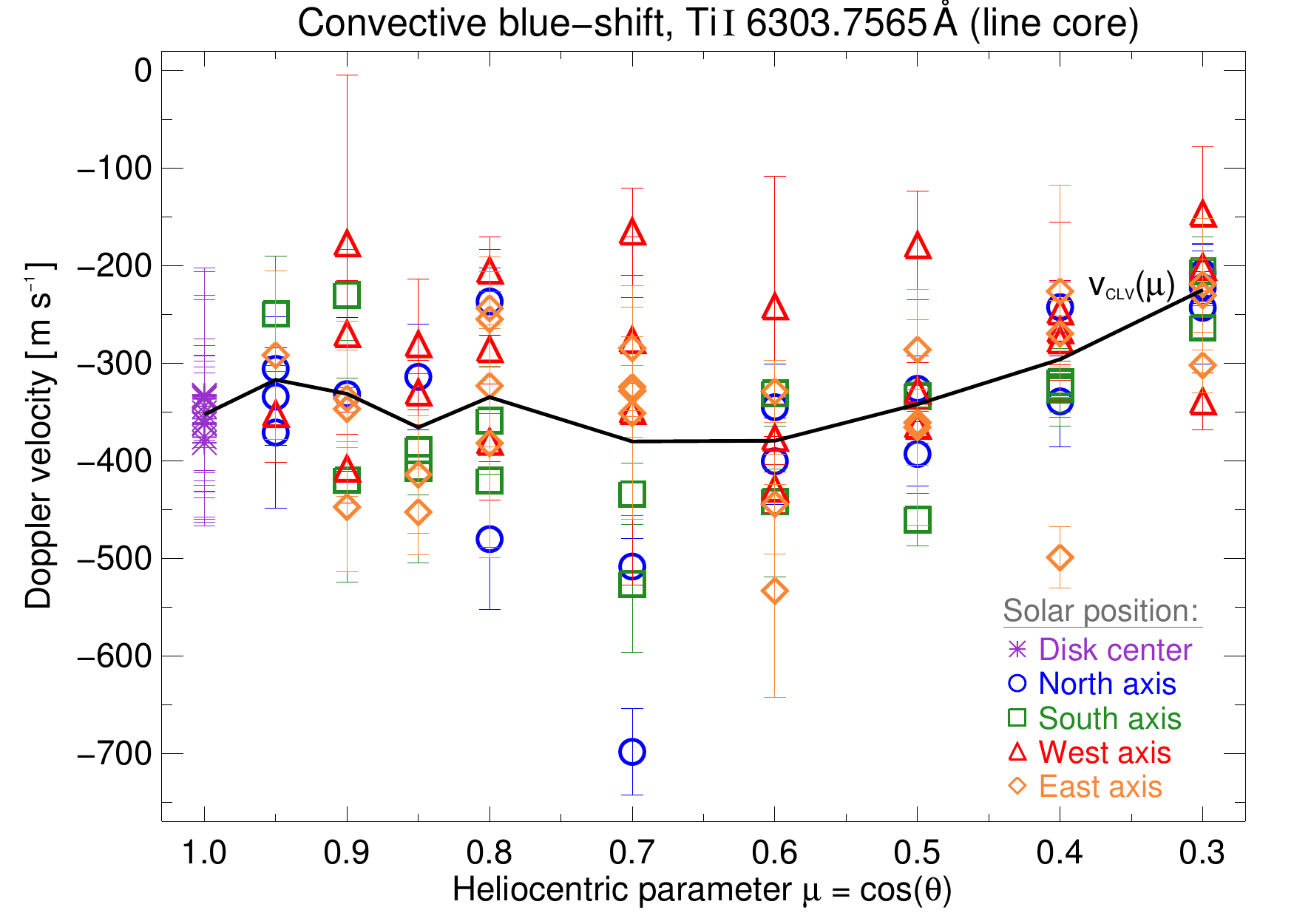}
\caption{Center-to-limb variation for the \ion{Ti}{I}\,6303.8\,\AA\ line. Top and middle panel: Spectral profile and bisectors from disk center (purple) toward the solar limb (decreasing gray scale), with FTS atlas comparison (yellow). Intensities were normalized to the spectral continuum. Air wavelengths (in \AA) were translated into Doppler velocities (in ${\rm m\,s^{-1}}$). Bottom panel: Line core velocities plotted against the heliocentric position $\mu$, with colored symbols indicating the solar axis. The average trend is drawn in black as a thick, solid line.}
\label{fig_secA_analysis_TiI63038}
\end{figure}


\begin{figure}[htpb]
\includegraphics[width=0.97\columnwidth]{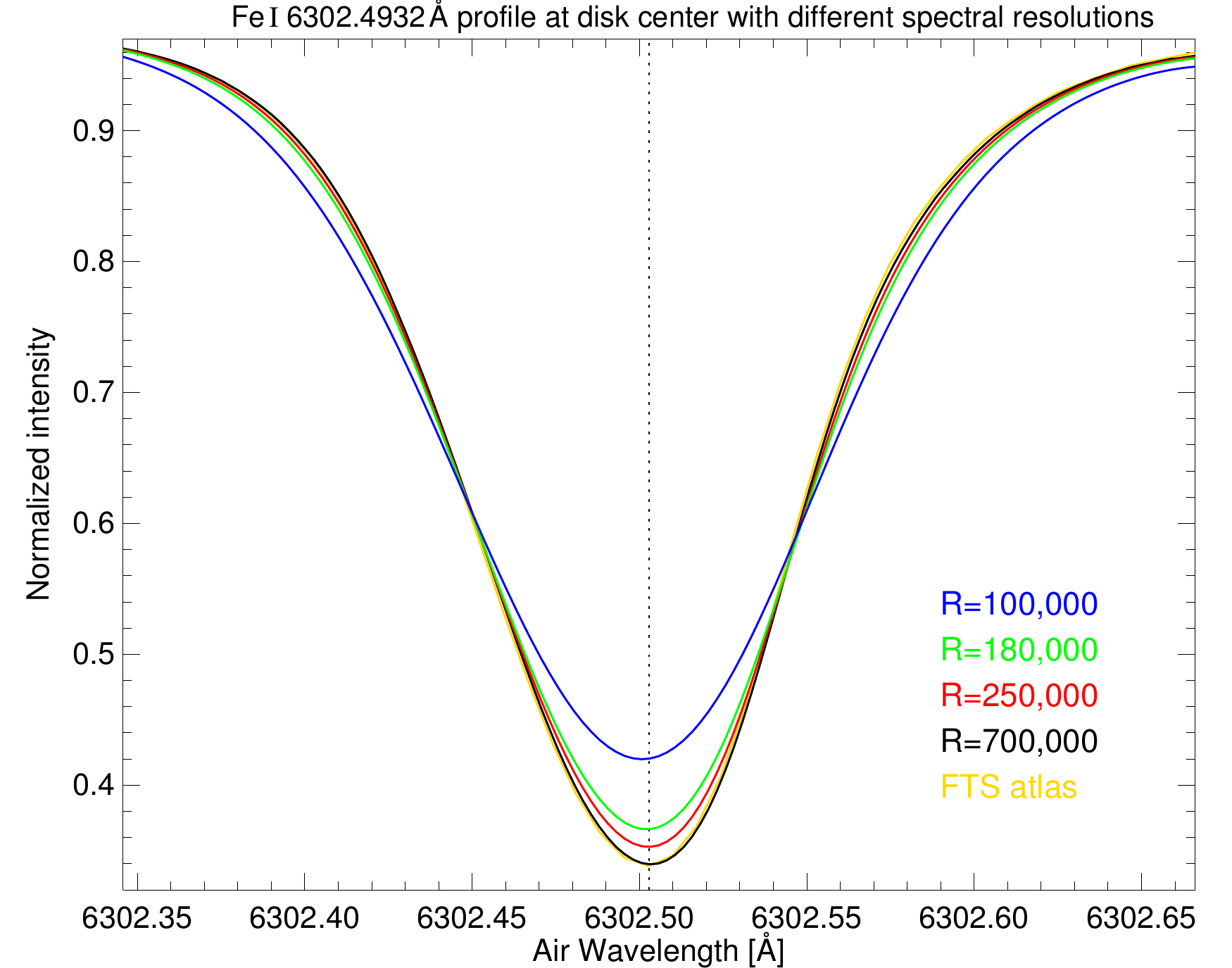}\\[0.1cm]
\includegraphics[width=0.97\columnwidth]{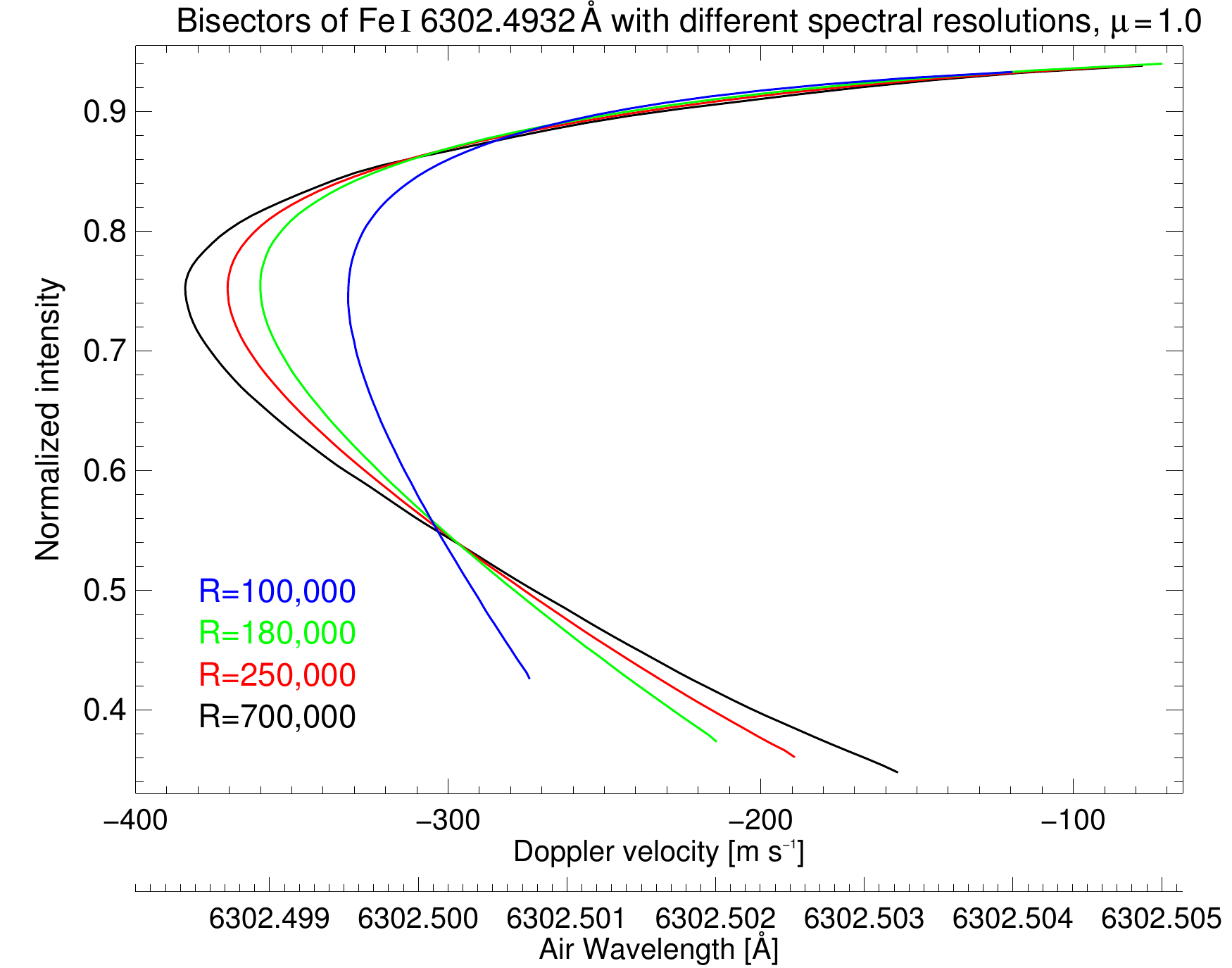}\\[0.15cm]
\includegraphics[width=0.95\columnwidth]{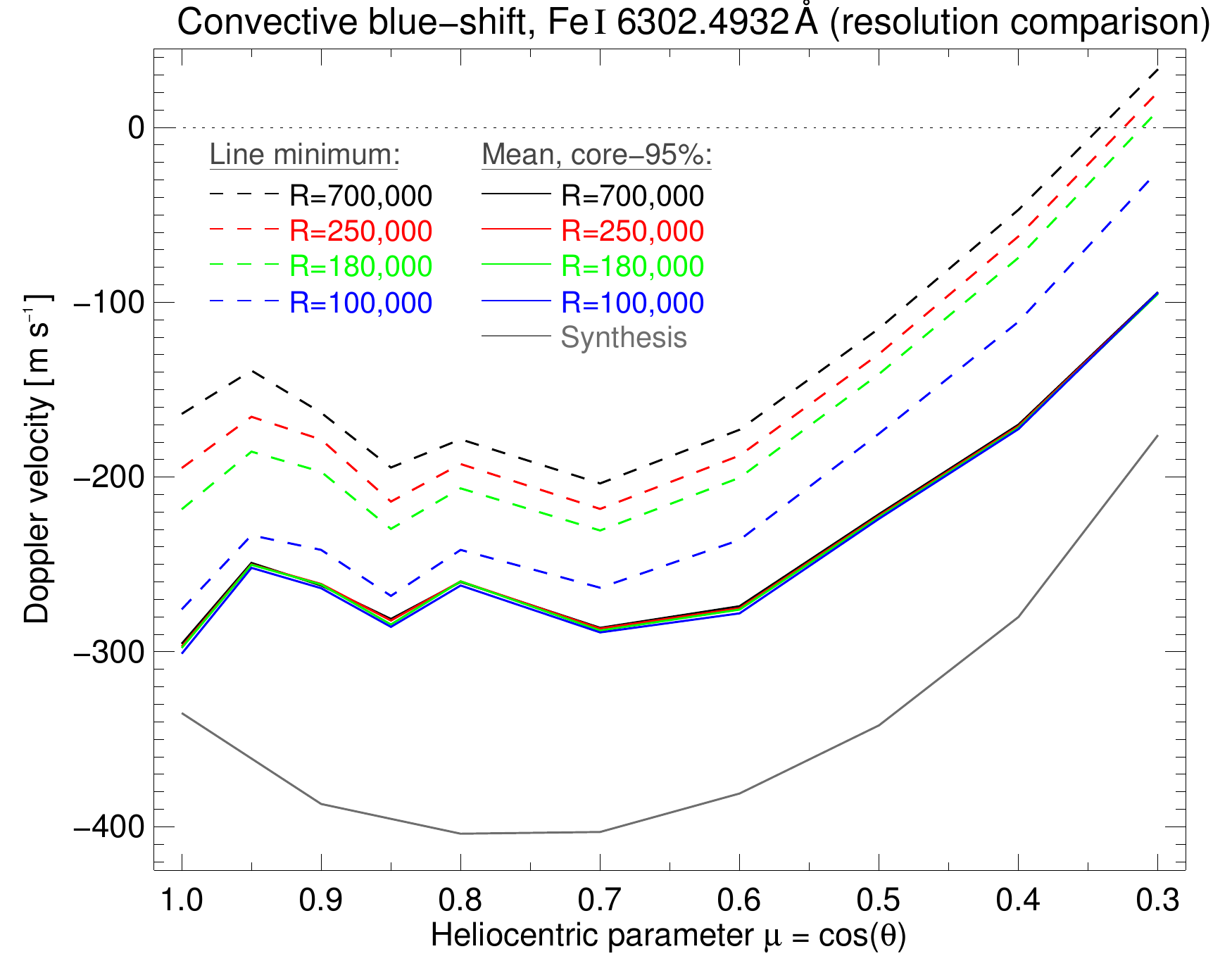}
\caption{Degradation of the \ion{Fe}{I}\,6302.5\,\AA\ line analysis to lower spectral resolutions, from 700\,000 (black) to 250\,000 (red), 180\,000 (green), and 100\,000 (blue). The three panels show the change of the spectral line profile at disk center (top panel), the corresponding bisectors (center panel), and the center-to-limb variation of the convective shift (bottom panel). The FTS atlas profile is added in yellow. Air wavelengths are given in \AA, translated Doppler velocities in ${\rm m\,s^{-1}}$. Intensities were normalized to the spectral continuum. The different center-to-limb blueshifts (lower panel) were either obtained for the line core (dashed lines) or the average bisector (solid lines).}
\label{fig_secA_resolution}
\end{figure}

\end{appendix}

\end{document}